\documentclass{jfm}
\usepackage{upmath}
\usepackage{amsmath}
\usepackage{amsfonts}
\usepackage{amssymb}
\usepackage{pgf,pgfarrows}
\usepackage{amsbsy}
\usepackage{wasysym}
\usepackage{graphicx}
\usepackage{subfig}

\ifCUPmtlplainloaded \else
  \checkfont{eurm10}
  \iffontfound
    \IfFileExists{upmath.sty}
      {\typeout{^^JFound AMS Euler Roman fonts on the system,
                   using the 'upmath' package.^^J}
}
      {\typeout{^^JFound AMS Euler Roman fonts on the system, but you
                   dont seem to have the}
       \typeout{'upmath' package installed. JFM.cls can take advantage
                 of these fonts,^^Jif you use 'upmath' package.^^J}
       
      }
  \else
    
  \fi
\fi
\ifCUPmtlplainloaded \else
  \checkfont{msam10}
  \iffontfound
    \IfFileExists{amssymb.sty}
      {\typeout{^^JFound AMS Symbol fonts on the system, using the
                'amssymb' package.^^J}
\let\le=\leqslant  \let\leq=\leqslant
       \let\ge=\geqslant  \let\geq=\geqslant
      }{}
  \fi
\fi
\ifCUPmtlplainloaded \else
  \IfFileExists{amsbsy.sty}
    {\typeout{^^JFound the 'amsbsy' package on the system, using it.^^J}
}
    {\providecommand\boldsymbol[1]{\mbox{\boldmath $##1$}}}
\fi

\def\dd{{\rm d}}
\def\ee{{\rm e}}
\def\ii{{\rm i}}

\affiliation{
$^1$Institute of Applied Mathematics, University
of British Columbia, 6356 Agricultural Road, Vancouver,
BC, V6T 1Z2, Canada.\\[\affilskip]
$^2$Department of Civil Engineering, University
of British Columbia, 2002-6250 Applied Science Lane, Vancouver,
BC, V6T 1Z4, Canada.\\[\affilskip]
$^3$Department of Mathematics, University of British Columbia,
1984 Mathematics Road, Vancouver, BC, V6T 1Z2, Canada.\\[\affilskip]
$^4$Department of Mechanical Engineering, University of
British Columbia, 2054-6250 Applied Science Lane, BC, V6T 1Z4, Canada.}
\pubyear{2009}
\volume{xxx}
\pagerange{xxx--yyy}
\begin{document}

\title[Stability of plane Couette-Poiseuille flow with uniform
cross-flow]{On the stability of plane Couette-Poiseuille flow with uniform
cross-flow}
\author[A.~Guha and I.A.~Frigaard]
{A\ls N\ls I\ls R\ls B\ls A\ls N\ns G\ls U\ls H\ls A$^{1,2}$ \and
\ns I\ls A\ls N\ns A.\ns F\ls R\ls I\ls G\ls A\ls A\ls R\ls D$^{3,4}$}
\date{?? and in revised form ??}
\maketitle

\begin{abstract}
We present a detailed study of the linear stability of plane Couette-Poiseuille flow in the presence of a cross-flow. The base flow is characterised by the cross flow Reynolds number, $R_{inj}$ and the dimensionless wall velocity, $k$. Squire's transformation may be applied to the linear stability equations and we therefore consider 2D (spanwise-independent) perturbations. Corresponding to each dimensionless wall velocity, $k\in[0,1]$, two ranges of $R_{inj}$ exist where unconditional stability is observed. In the lower range of $R_{inj}$, for modest $k$ we have a stabilisation of long wavelengths leading to a cut-off $R_{inj}$. This lower cut-off results from skewing of the velocity profile away from a Poiseuille profile, shifting of the critical layers and the gradual decrease of energy production. Cross-flow stabilisation and Couette stabilisation appear to act via very similar mechanisms in this range, leading to the potential for robust compensatory design of flow stabilisation using either mechanism.

As $R_{inj}$ is increased, we see first destabilisation and then stabilisation at very large $R_{inj}$. The instability is again a long wavelength mechanism. Analysis of the eigenspectrum suggests the cause of instability is due to resonant interactions of Tollmien-Schlichting waves. A linear energy analysis reveals that in this range the Reynolds stress becomes amplified, the critical layer is irrelevant and  viscous dissipation is completely dominated by the energy production/negation, which approximately balances at criticality. The stabilisation at very large $R_{inj}$ appears to be due to decay in energy production, which diminishes like $R_{inj}^{-1}$. Our study is limited to two dimensional, spanwise  independent perturbations.
\end{abstract}

\section{Introduction}

From the perspective of applications in technology, Poiseuille flow of viscous fluid along a duct is undoubtedly one of the most important flows studied as it underpins the field of hydraulics. Instability and subsequent transition from laminar flow marks a paradigm shift in the dominant transport mechanisms of mass, momentum and heat, and it is for this reason that the subject remains of enduring interest, even after more than 100 years of study. In this paper we focus on two methods for affecting the linear stability of plane Poiseuille (PP) flow. The first method consists of introducing a Couette component to the flow, by translation of one of the walls. The second method consists of introducing a cross-flow, e.g.~via injection through a porous wall. While both effects have been studied individually to some extent, there are fewer studies of the two effects combined, which is the main focus here.

We first summarise the effects of a Couette component on a plane Poiseuille flow. The main curiosity here stems from the observation that PP flow is linearly unstable when the critical Reynolds number exceeds $R_c \approx 5772$, (Reynolds number based on the center-line axial velocity and the half-width of the channel; see \cite{Orszag1971}), whereas the plane-Couette (PC) flow is absolutely stable with respect to infinitesimal amplitude disturbances, $R_c = \infty$; see \cite{Romanov1973}. Superimposing PP and PC flows, we may ask if a small Couette component can affect the stability of the PP flow. Stability of plane Couette-Poiseuille (PCP) flow was first studied by \cite{Potter1966} and later by \cite{Hains1967}, \cite{ReynoldsPotter1967} and \cite{CowleySmith1985}. The results are typically understood with respect to a Reynolds number that is based on the maximal velocity of the Poiseuille component, say $R_p$, and the ratio of wall velocity to maximal velocity of the Poiseuille component, denoted $k$. For small Couette components, $k$, it is possible to observe some destabilisation of the flow, (depending on the wavenumber), but as soon as $k > 0.3$ a strong stabilisation of the flow sets in. As the velocity ratio $k$ exceeds $0.7$, the  neutral stability curve completely vanishes and the flow becomes unconditionally linearly stable, i.e.~$R_c \to \infty$. The term ``cut-off'' velocity has been used to describe this stabilisation; see \cite{ReynoldsPotter1967}.

Although plane Couette flows are widely studied it is worth noting that they are actually difficult to produce, i.e.~outside of the computational and theoretical domain. In many duct flows axial translation of a wall is either not possible or is limited in terms of speed. High $R$ frequently means high velocities, lowering the range of achievable $k$ as the flow velocity increases. Therefore, the range of practical flows for which a sufficiently stabilizing Couette component can be introduced is limited and we know of no technological applications where this is used for stabilisation.

Annular Couette-Poiseuille (ACP) flows are more practically relevant and have also been studied extensively (\cite{Mott1968,Sadeghi1991}). For example, ACP flows occur when removing/inserting drillpipe or casing from vertical wellbores during an operation called ``tripping''. Sadeghi and Higgins studied the flow between two concentric cylinders, the outer being stationary while the inner is moved with a constant (dimensionless) velocity $k$ in the streamwise direction. They showed that varying the radius ratio ($\eta$) between the outer and inner cylinders can have a dramatic effect on the stability characteristics. The limit $\eta\to 1$ approximates PCP flow and is unconditionally linearly stable for $k \geq 0.7$, thereby confirming \cite{Potter1966}. By increasing $\eta$, the cut-off condition is attained for lower values of $k$ and the cut-off relation between $k$-$\eta$ is almost linear. Similar to \cite{Mott1968}, they argued that increasing $\eta$ increases the asymmetry of the base flow profile which in turn increases the stability with respect to axisymmetric disturbances. Their findings are very relevant to our work, since we later show that the stability achieved by increasing $\eta$ in ACP flows and that achieved by applying a small cross-flow in PCP flows are essentially similar.

Shear flows with cross-flow occur in a range of natural settings as well as in various technological applications. As examples, we cite studies in sediment-water interfaces over permeable seabeds (\cite{Goharzadeh2005});  fluid transport and consequent mass transfer at the walls of blood vessels, the lungs and kidneys (\cite{Maj2002}); and flow through the fractures of geological formations (\cite{Berkowitz2002}). In some technological applications cross-flow is an inherent part of the process, e.g.~dewatering of pulp suspensions in paper making, whereas in others it is introduced to affect the stability. An example of the latter is the use of wall suction to delay the transition to turbulence over the surface of an aircraft wing (\cite{Joslin1998}).

The stability of PP flow with cross-flow was first analysed by \cite{Hains1971} and \cite{Sheppard1972}, both of whom have shown that a modest amount of cross-flow produces significant increase  of the critical Reynolds number. These results are however slightly problematic to interpret in absolute terms, since at a fixed pressure gradient along the channel, increasing the cross-flow  decreases the velocity along the channel, (hence effectively the Reynolds number). This difficulty was noted by \cite{Fransson2003}, who used the maximal channel velocity as their velocity scale (instead of that based on the PP flow without cross-flow), and thus separated the effects of base velocity magnitude from those of the base velocity distribution. Using this velocity scale in their Reynolds number $R$, they showed regimes of both stabilisation and destabilisation as the cross-flow Reynolds number was increased. For example, for $R=6000$ and wavenumber $\alpha=1$, \cite{Fransson2003} have shown that the cross-flow was stabilising up to a cross-flow Reynolds number $R_{inj} \approx 3.4 $, and then starts destabilising before re-stabilising again at $R_{inj}\approx 635$. The initial regime of stabilisation is the one corresponding to the earlier results.

Although cross-flow affects the base velocity profile, the main change to the linear stability problem is to add an inertial cross-flow term to the Orr-Sommerfeld operator. One reason why addition of terms like the cross-flow term can destabilise an otherwise stable shear flow is suggested by the two-dimensional instability of Blasius boundary layer, as studied by \cite{Baines1996}. In such flows, the resulting growing disturbance is known as a Tollmien-Schlichting (T-S) wave. They showed that the interaction between two idealized modes, viz. an ``inviscid'' neutral mode at zero viscosity and a decaying viscous mode (or modes) existing at uniform shear undergo resonant interactions. The latter is forced by the former through the no-slip wall boundary conditions.

In the present study we focus on the combination of cross-flow and Couette component. Our motivation stems from a desire to understand how the two mechanisms interact, since in terms of technological application different mechanical configurations may be more or less amenable to cross-flow and/or wall motion. This means that there is value in knowing when one effect may compensate for the other in stabilizing (or destabilizing) a given flow. To the best of our knowledge, stability of PCP flow with cross-flow has only been studied in any generality by \cite{Hains1971}. In considering the base flow for PCP flow with cross-flow, (which is parameterized by $R_{inj}$ and $k$), the relation $kR_{inj} = 4$, defines an interesting paradigm in which the base velocity in the axial direction is linear. These Couette-like flows have been studied by \cite{Nicoud1997} for increasing $R_{inj}$. They found a critical value of $R_{inj} \approx 24$, below which which no instability occurs, (we have translated their critical value of $48$ into the $R_{inj}$ that we use). Therefore, we observe that understanding of cross-flow PCP flows is far from complete. We aim to contribute to this understanding.

The 3D linear stability of PCP flow with cross-flow is amenable to Squires transformation, so that the linear instability occurs first for 2D (spanwise-independent) perturbations. It is these perturbations that we study here. Our aim is to demarcate clearly in the $(R_{inj},k)$-plane, regions of unconditional stability, i.e.~where there is a cut-off wall velocity or injection velocity. We also wish to understand the underlying linear stability mechanisms as $R_{inj}$ and $k$ are varied.

Although the study of 2D perturbations is justified from the pure perspective of linear stability, it must be acknowledged that 3D and nonlinear effects are likely to be relevant in instabilities that are observed to grow, i.e. the actual transition. The past two decades have seen extensive study of transient growth mechanisms, due to non-normality of the operator associated with linearized Navier-Stokes equations. Algebraic growth of $O(Re^2)$ may occur for linearly stable disturbances that decay only slowly over a timescale of $O(Re)$. It has been proposed that this transient algebraic growth is responsible for subcritical transition in wall-bounded shear flows. For an overview of these developments we refer to \cite{Reddy1993,Schmid2001,Chapman2002,Schmid2007}.

At the same time as transient growth mechanisms have undergone extensive research, self sustaining nonlinear mechanisms were proposed by Waleffe and others, e.g.~\cite{Hamilton1995,Waleffe1997}. In this scenario energy from the mean flow can be fed back into streamwise vortices, thus resisting viscous decay. Self-sustained exact unstable solutions to the Navier-Stokes equations were found by \cite{Faisst2003} and by \cite{Wedin2004}. Much current effort is focused at understanding the link between these self-sustained unstable solutions and observed transitional phenomena, such as intermittency, streaks, puffs and slugs; see e.g.~\cite{Hof2004,Hof2005,Eckhardt2007,Kerswell2007}.

Our study does not deal with any of the complexities of transition mentioned above, and as such the relevance may be questioned. This is a fair criticism, but on the other hand we note that for other classical shear flows that are linearly stable at all $Re$, careful control of  apparatus imperfections and the level of flow perturbations can significantly retard the point at which transition is observed. For example, in Hagen-Poiseuille flow of Newtonian fluids one typically observes transition to turbulence starting for $Re \gtrsim 2000$. However, an experimental flow loop in Manchester UK produces stable laminar flows for $Re \approx 24,000$, \cite{Hof2004,Peixinho06}, and stable flows have even been reported up to $Re \approx 100,000$, \cite{Pfenniger61}. This all suggests that significantly enhanced stability may be achieved experimentally, where predicted by the linear theory.

The question of how to achieve a PCP flow with cross-flow in practice is also relevant. Evidently all Poiseuille flows occur in finite geometries with entrance effects, side walls and imperfections in the planar walls, so that the notion of a truly planar infinite flow is anyway flawed. Uniform base flows studied in hydrodynamic stability are invariably an approximation of experimental reality. Even in the absence of wall motion planar Poiseuille flow is difficult experimentally, due to spanwise perturbations and inflow non-uniformities. This said, a geometry with a uniformly translating channel walls is particularly difficult to achieve and as mentioned before, $k \approx 0.7$ is difficult for high $R$ flows where $R$ is increased via flow rate. Imposing a uniform cross-flow along with a streamwise pressure variation is more practically achievable in practice. As an example, see \cite{Vadi2001}. A uniform trans-membrane pressure cross-flow micro-filtration system is able to maintain uniform trans-membrane pressure with high cross-flow velocity ($\hat{V}_{inj}$) and improves the utilization of available filtration area. In the patent of \cite{Sand2001} the concept of operating a membrane filtration unit using UTMP has been proposed, such that pressure drop along the channel can be adjusted independent of the $\hat{V}_{inj}$. A different generic concept for achieving a uniform cross-flow over a finite length of a porous-walled channel, consists of injecting the fluid along a secondary channel behind one of the porous wall that has a linearly converging geometry. Although there are clear technical challenges, it is worth remarking that the injection velocities needed for stability are very modest by comparison with the wall velocities, and therefore as a target appear achievable.

An outline of our paper is as follows. We commence below in \S \ref{sec:base} by introducing the base flow and linear stability problem. We describe our numerical method and present benchmark results that illustrate typical effects of varying $R_{inj}$ and $k$. These results serve to motivate the presentation of results, which follows in the following 3 sections. Section \ref{sec:low} considers low $R_{inj}$ and significant $k$, where we see that long wavelengths dominate. In \S \ref{sec:medium} we characterise the flows for intermediate $R_{inj}$ and small $k$, where short wavelengths are the least stable. Finally, we consider large $R_{inj}$ in \S \ref{sec:high}, where we find both destabilisation and eventual stabilisation. The paper concludes in \S \ref{sec:discuss} with a summary of the principal results.

\section{Stability of plane Couette-Poiseuille flow with cross-flow}
\label{sec:base}

The base flow considered in this paper is a plane Couette-Poiseuille flow (PCP) with imposed uniform cross-flow. This flow is two-dimensional,
viscous, incompressible and fully developed in the streamwise direction, $\hat{x}$,(all dimensional variables are denoted with a ``hat'',
i.e.~$\hat{\cdot}$). The imposed base velocity in the $\hat{y}$-direction,~$\hat{v}$, is constant and equal to the injection/suction velocity $\hat{V}_{inj}$. Since $\hat{v}$ is constant, the $\hat{x}$-component of velocity,~$\hat{u}$ depends only on $\hat{y}$. The flow domain is bounded by walls at~$\hat{y}=\pm\hat{h}$, and is driven in the $\hat{x}$-direction by a constant pressure gradient and by translation of the upper wall, at speed $\hat{U}_{c}$. The $\hat{x}$-component of velocity, $\hat{u}(\hat{y})$, is found from the $\hat{x}$-momentum equation, which simplifies to:
\begin{equation}
\hat{V}_{inj}\frac{\partial\hat{u}}{\partial\hat{y}}=-\frac{1}{\hat{\rho}}\frac{\partial\hat{p}}{\partial\hat{x}}+\hat{\nu}\frac{\partial^{2}\hat{u}}{\partial\hat{y}^{2}}\label{eq:M2}\end{equation}
where~$\hat{\rho}$ is the density,~$\hat{\nu}=\hat{\mu}/\hat{\rho}$
is the kinematic viscosity, and $\hat{\mu}$ the dynamic viscosity.
The boundary conditions at $\hat{y}=\pm\hat{h}$ are:
\begin{equation}
\hat{u}(-\hat{h})=0,\,\,\,\,\hat{u}(\hat{h})=\hat{U}_{c}\label{eq:bcs}\end{equation}
To scale the problem we scale all lengths with $\hat{h}$, hence $(x,y)=(\hat{x}/\hat{h},\hat{y}/\hat{h})$.
For the velocity scale two choices are common. First, the imposed
pressure gradient defines a ``Poiseuille'' velocity scale:
\begin{equation}
\hat{U}_{p}=-\frac{\hat{h}^{2}}{2\hat{\mu}}\frac{\partial\hat{p}}{\partial\hat{x}},\label{eq:Updefine}\end{equation}
which is equivalent to the maximum velocity of the plane Poiseuille
flow, driven by the pressure gradient alone. Second, we may take the
maximum velocity, which we need to compute.~$\hat{U}_{p}$ is the choice of \cite{Potter1966}, and thus allows
one to compare directly with the studies of PCP flows. In the absence
of a cross-flow, the maximal velocity is not actually very sensitive
to the wall velocity $\hat{U}_{c}$, at least for $\hat{U}_{c}<\hat{U}_{p}$,
which covers the range over which the flow stabilises. However, in
the case of a strong cross-flow the $\hat{x}$-velocity is reduced
significantly below $\hat{U}_{p}$, which therefore loses its meaning.
Consequently we adopt the second choice and scale with the maximal
velocity, $\hat{U}_{max}$. This choice retains physical meaning in
the base velocity, but does introduce algebraic complexity.

The solution is found from \eqref{eq:M2}--\eqref{eq:bcs}~after
detailed but straightforward algebra:
\begin{eqnarray}
\hat{u}(\hat{y}) & = & \hat{U}_{p}\left[\frac{4\cosh R_{inj}-kR_{inj}\ee^{-R_{inj}}+[kR_{inj}-4]\ee^{R_{inj}\hat{y}/\hat{h}}}{2R_{inj}\sinh R_{inj}}+\frac{2\hat{y}}{R_{inj}\hat{h}}\right], \label{eq:baseflow-d}\end{eqnarray}
where $k$ and $R_{inj}$ are defined by:
\begin{eqnarray} k &=& \frac{\hat{U}_{c}}{\hat{U}_{p}}, \\ R_{inj} &=& \frac{\hat{V}_{inj}\hat{h}}{\hat{\nu}}. \end{eqnarray}
These two dimensionless parameters uniquely define the dimensionless
base flow. The parameter $k$ is the velocity ratio of Couette to
Poiseuille velocities, which is useful as it allows direct comparison
with earlier results on stabilisation of PCP flows without cross-flow.
The parameter $R_{inj}$ is simply a Reynolds number based on the
injection velocity. Primarily here we consider the ranges: $k\in[0,1]$
and $R_{inj}\ge0$.

For relatively weak crossflow velocities, the velocity component $u(y)$ has a single
maximum at a value of $y=y_{max}$ defined by:\begin{equation}
\ee^{R_{inj}y_{max}}=\frac{\sinh R_{inj}}{R_{inj}}\frac{4}{4-kR_{inj}}\end{equation}
The maximal velocity $\hat{U}_{max}$, is then evaluated from \eqref{eq:baseflow-d}.
Since,~$\sinh R_{inj}\ge R_{inj}$, we can see that $y_{max}>0$ for
$k\ge0$ and $R_{inj}>0$. Both the injection cross-flow and Couette
component act to skew the velocity profile towards the upper wall.
For stronger cross-flow velocities, (or sufficiently large $k$),
the maximal velocity occurs at the upper wall, i.e.~$\hat{U}_{max}=\hat{U}_{c}$.

The division between weak and strong cross-flows, taking into account
also the Couette component, is defined by the line
\begin{equation}
kR_{inj}=4\left[1-\frac{\sinh R_{inj}}{R_{inj}\ee^{R_{inj}}}\right]\end{equation}
The dimensionless base velocity is given by:
\begin{equation} u(y) = \left\{ \begin{array}{l} \displaystyle{\frac{4 \cosh R_{inj} - kR_{inj} \ee^{-R_{inj}} - [4-kR_{inj}]\ee^{R_{inj} y} +4 y \sinh R_{inj} }{4 \cosh R_{inj} - kR_{inj} \ee^{-R_{inj}} - [4-kR_{inj}]\ee^{R_{inj} y_{max}} +4 y_{max} \sinh R_{inj}} } , \\[1ex] \hspace{6cm} k R_{inj} \le 4\left[1 - \frac{\sinh R_{inj}}{R_{inj} \ee^{R_{inj}}} \right] ,\\[1ex] \displaystyle{\frac{1}{k} \left[ \frac{4 \cosh R_{inj} - kR_{inj} \ee^{-R_{inj}} - [4-kR_{inj}]\ee^{R_{inj}y}}{2 R_{inj}\sinh R_{inj}} +\frac{2 y}{R_{inj}} \right]}, \\[1ex] \hspace{6cm} k R_{inj} > 4\left[1 - \frac{\sinh R_{inj}}{R_{inj} \ee^{R_{inj}}} \right]. \end{array} \right. \label{eq:baseflow} \end{equation}
It can be verified that in the limit $R_{inj}\to0$, with $k$ fixed,
the classical form of PCP base velocity profile is retrieved:
\begin{equation} u(y) \sim \displaystyle{\frac{1-y^2 + \frac{k}{2}(1+y) + R_{inj}[ \frac{1}{3}(y-y^3) - \frac{k}{4}(1-y^2)]}{1+\frac{k}{2}+\frac{k^2}{16} - R_{inj}[\frac{k}{6} - \frac{k^3}{64} ]}} \end{equation}
as $R_{inj}\to0$, with $k\le4[1-\sinh R_{inj}/(R_{inj}\ee^{R_{inj}})]/R_{inj}\sim4[1+R_{inj}/3]$.

\begin{figure}
~~~~~~~~~~~~~~~~~~~~~~~~~~~~\includegraphics[scale=0.08]{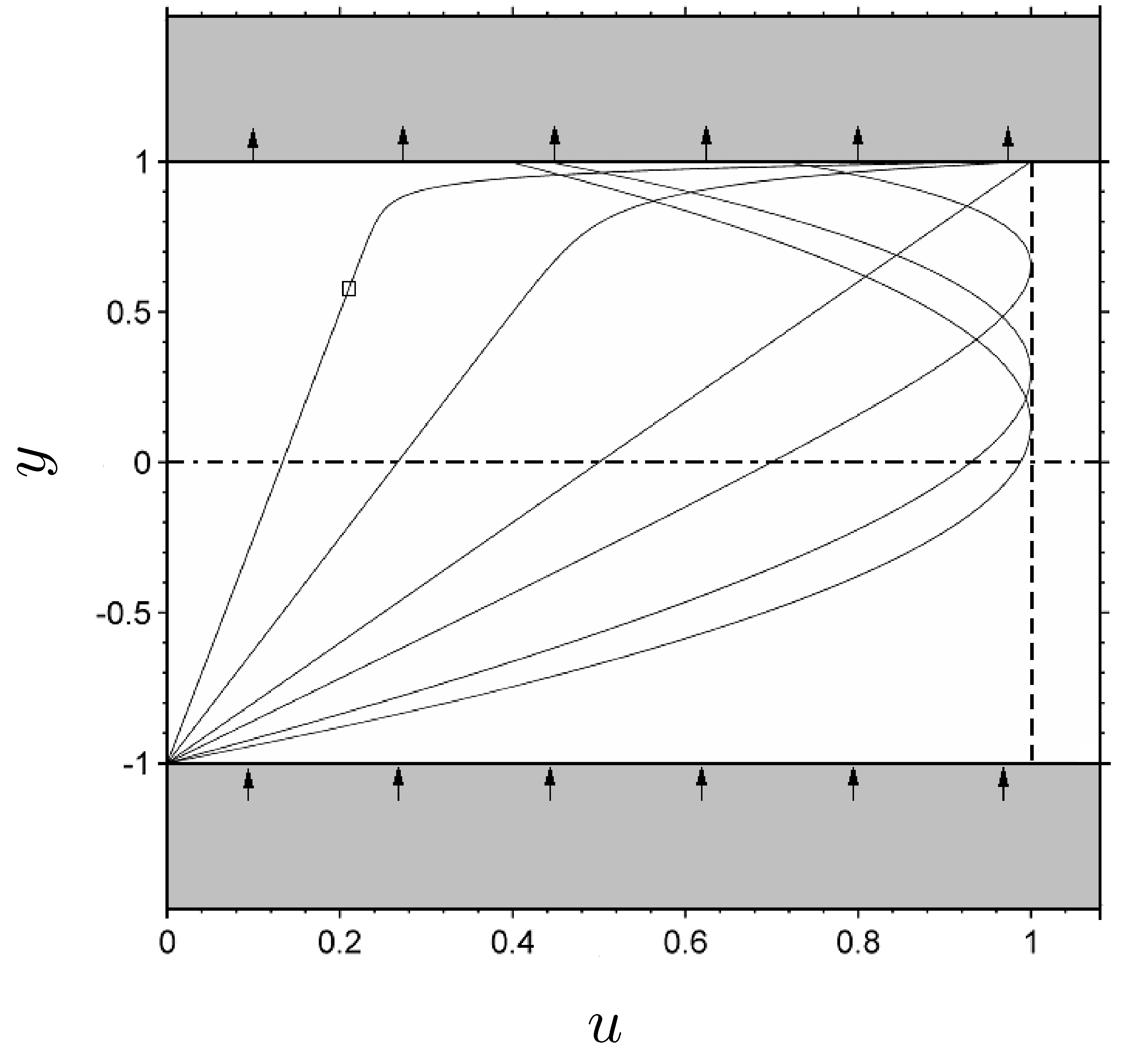}

\caption{Mean Velocity Distribution for $k=0.5$ and $R_{inj}$$=0$, $1$,
$4$, $8$, $15$ and $30$~($R_{inj}=30$ marked with a $\square$)\label{fig:1}}

\end{figure}

Examples of the base velocity profile are given in Fig.~\ref{fig:1},
for for $k=0.5$ and different values of $R_{inj}$. Observe that
for $R_{inj}=8$, when $kR_{inj}=4$, the velocity profile is linear.
This flow has been termed a ``generalised Couette'' flow by \cite{Nicoud1997}.

\begin{figure}
\subfloat[]{\includegraphics[scale=0.08]{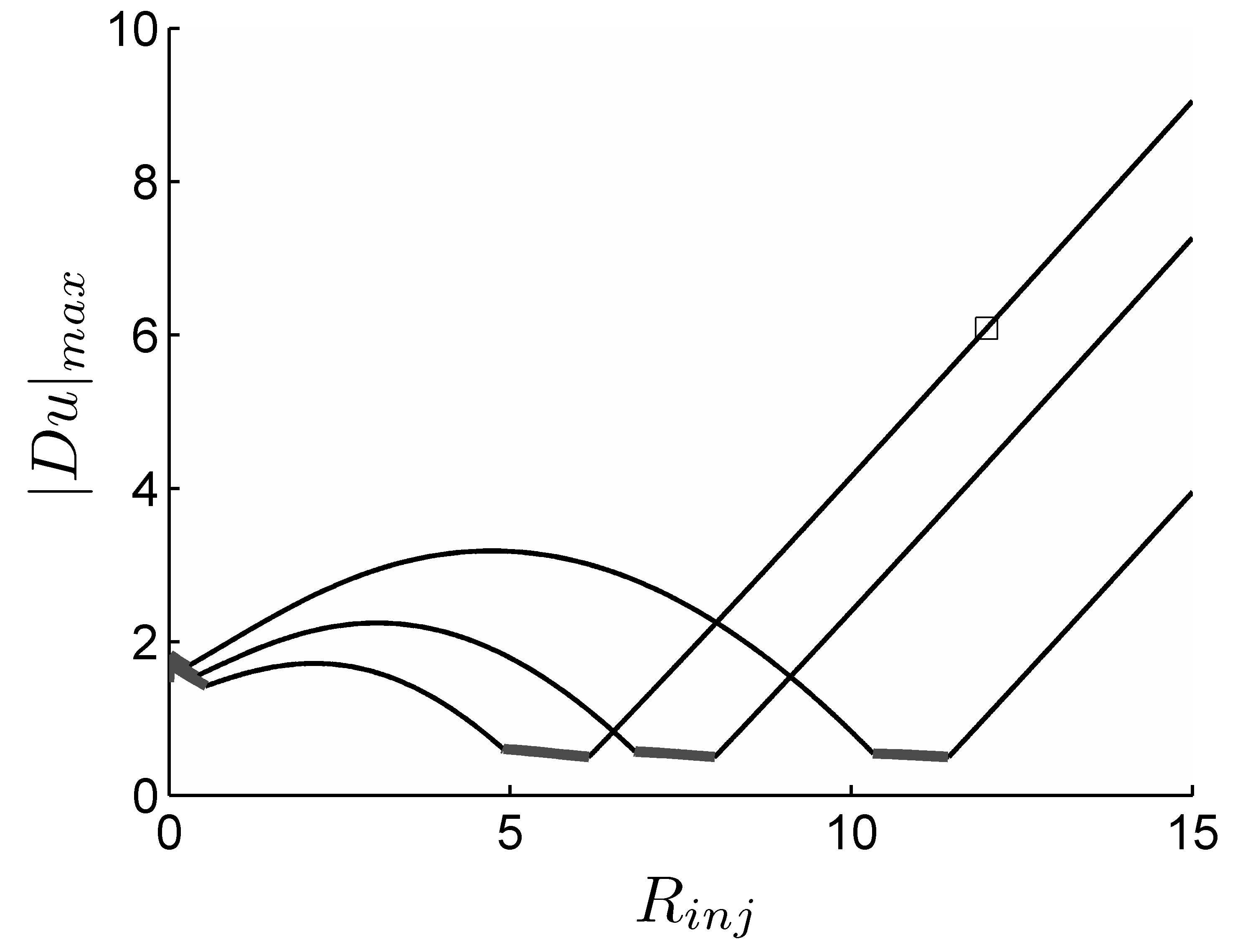}

}~\subfloat[]{\includegraphics[scale=0.08]{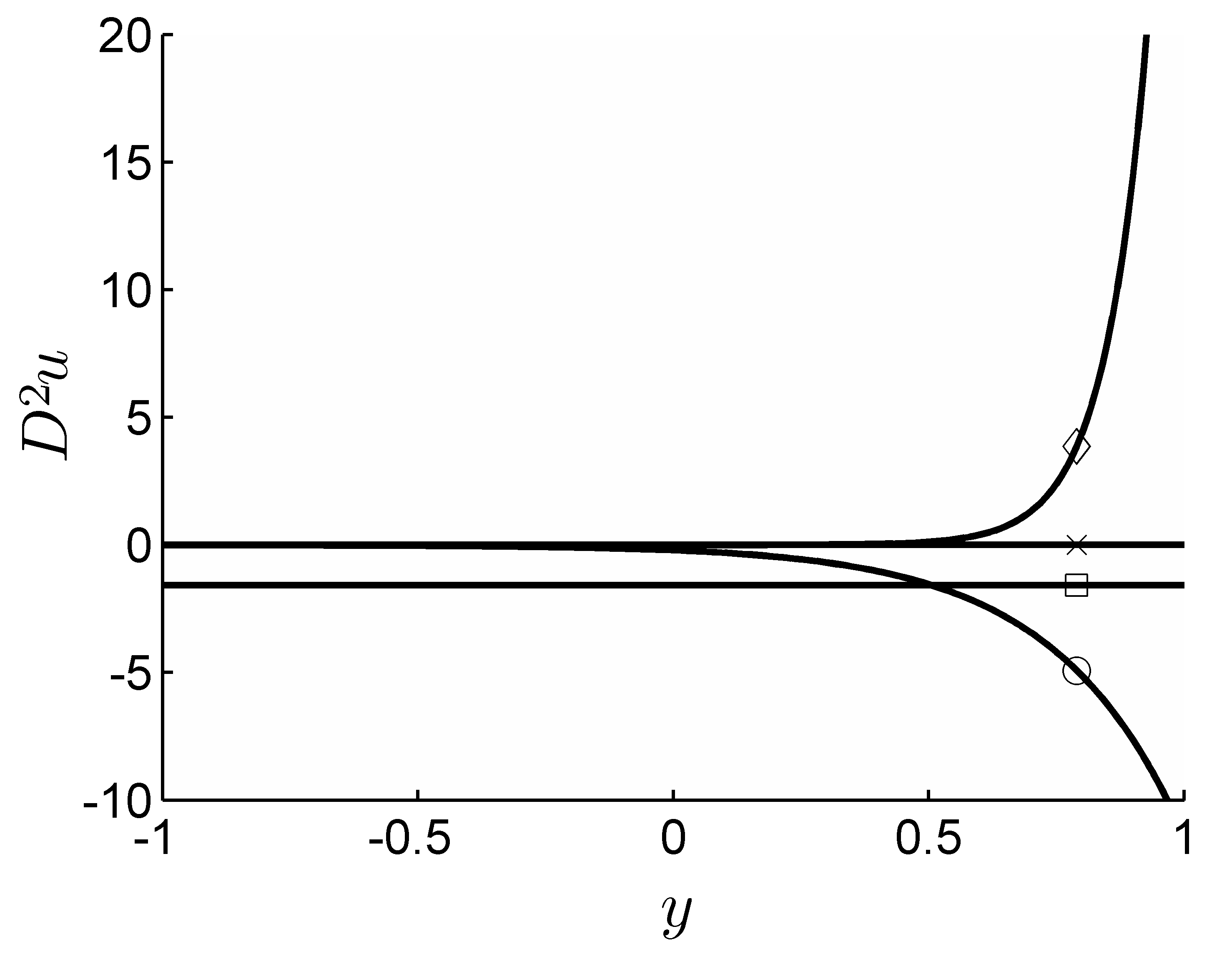}

}\caption{(a)~Maximal velocity gradient, $|Du|_{max}$, plotted against $R_{inj}$
for $k=0.35$,~$0.5$,~$0.65$, ($k=0.65$ marked with a $\square$).
The thick line indicates where the maximum is attained at $y=-1$;
otherwise at $y=1$.~(b)~Variation of $D^{2}u$ with $y$ for $k=0.5$
for : $R_{inj}=0$,~($\square$); $R_{inj}=4$,~($\circ$); $R_{inj}=8$,~($\times$);
$R_{inj}=12$, ($\diamond$).\label{fig:2}}

\end{figure}

We shall denote differentiation with respect to $y$ by the operator
$D$. The first and second derivatives of the base flow, $Du$ and
$D^{2}u$ respectively, influence the stability of the flow. We find
that $D^{2}u$ has sign determined by $(kR_{inj}-4)$, and increases
in absolute value exponentially towards the upper wall. For $kR_{inj}<4$,
the velocity is concave, and is convex otherwise. Since $D^{2}u$
does not change sign, the maximal absolute value of the first derivative
is found either at the upper or lower wall, $y=\pm1$. The maximal
velocity gradients are found at the lower wall for small $R_{inj}$
and also for a range of $R_{inj}$ close to $kR_{inj}=4$, but otherwise
are found at $y=1$; see Fig.~\ref{fig:2}(a). At large $R_{inj}$
the maximal velocity increases almost linearly:
\begin{eqnarray} |Du|_{max} & = & |Du(y=1)| = \frac{1}{k} \left[ \frac{[kR_{inj}-4]\ee^{R_{inj}}}{2 \sinh R_{inj}} +\frac{2}{R_{inj}} \right] \nonumber \\ & \sim & R_{inj} - \frac{4}{k} + \frac{2}{k R_{inj} } + O(\ee^{-2R_{inj}}).
\label{Dumax}
\end{eqnarray}

Figure~\ref{fig:2}(b) shows examples of the profiles of $D^{2}u$. We observe
that $D^{2}u\approx0$ over a large range of $y$, close to the lower
wall, whenever a significant amount of cross-flow is present, i.e.~$R_{inj}\gtrsim1$.

\subsection{Dimensionless groups}

\begin{figure}
\subfloat[]{\includegraphics[scale=0.08]{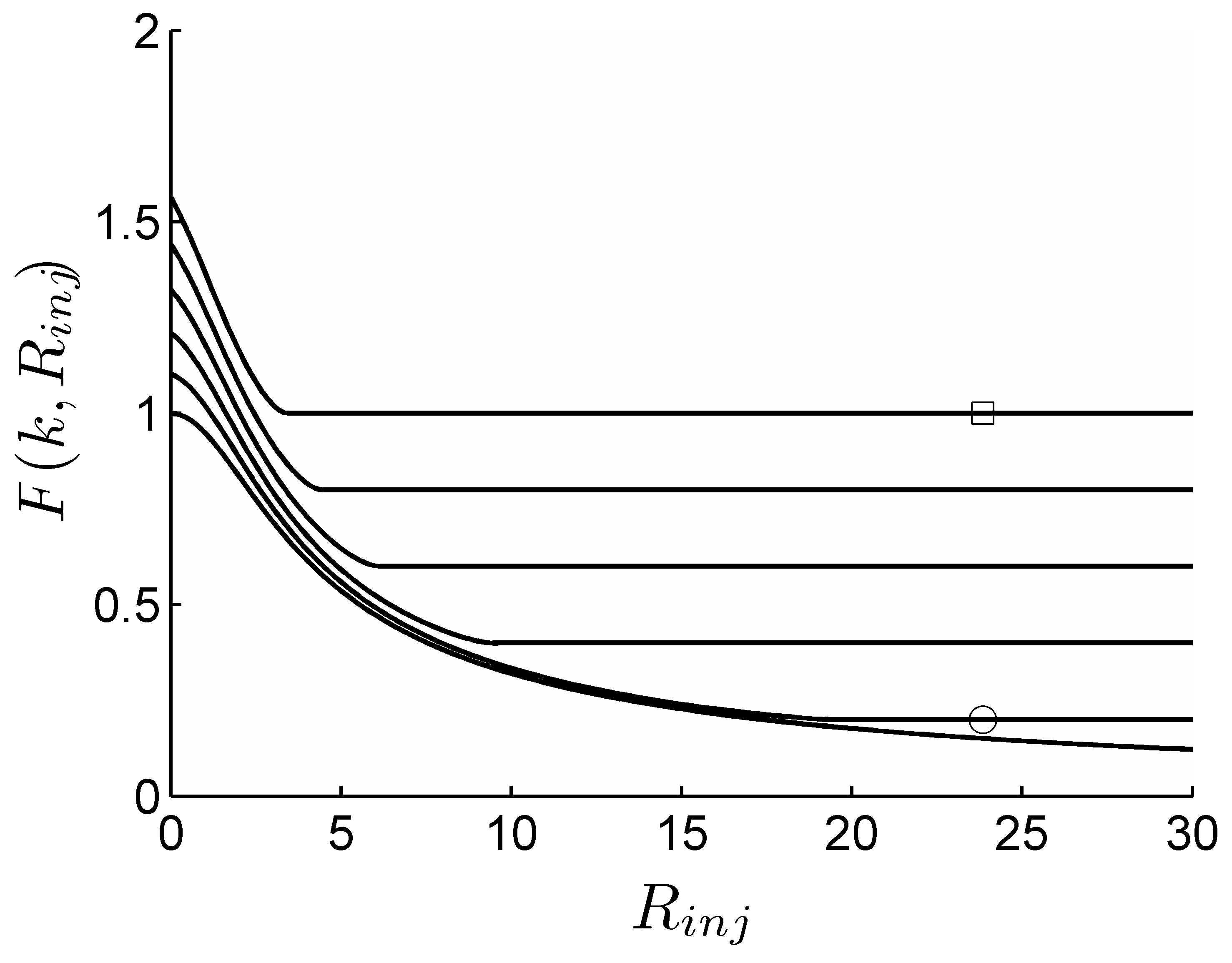}

}~\subfloat[]{\includegraphics[scale=0.08]{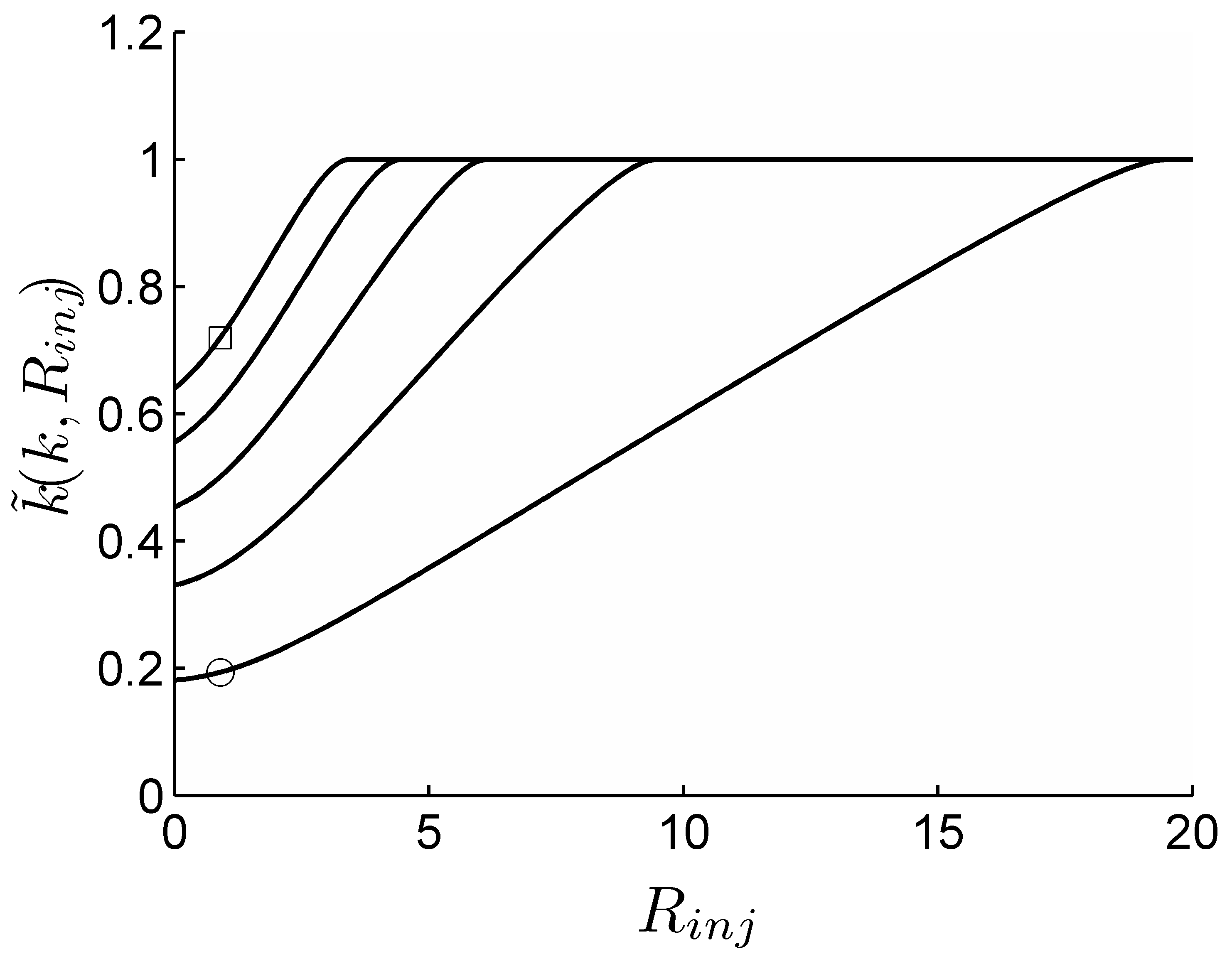}

}

\caption{(a)~$R/R_{p}=F(k,R_{inj})$ for $k=0,\,0.2,\,0.4,\,0.6,\,0.8,\,1$;~(b)~dimensionless
wall speed $\tilde{k}(k,R_{inj})$ for $k=0.2,\,0.4,\,0.6,\,0.8,\,1$.
In both plots, $k=0.2$ is marked with $\circ$ and $k=1$ is marked
with $\square$.\label{fig:figbase3}}

\end{figure}

The base base flow is fully defined by the parameters $k$ and $R_{inj}$,
as discussed above. In addition, the transient flow and associated
stability problem will depend on the streamwise Reynolds number, $R$,
which we define in terms of $\hat{U}_{max}$, i.e. \begin{eqnarray} R &=& \frac{\hat{U}_{max} \hat{h} }{ \hat{\nu} }. \end{eqnarray}
To aid the reader in interpreting our results in terms of those previously
published, it is helpful to consider also a Reynolds number based
on the Poiseuille velocity, $\hat{U}_{p}$, say $R_{p}=\hat{U}_{p}\hat{h}/\hat{\nu}$.
Straightforwardly, we find $R=R_{p}F(k,R_{inj})$:
\begin{equation}
F(k,R_{inj})=\left\{ \begin{array}{l}
{\displaystyle {\frac{4\cosh R_{inj}-kR_{inj}\ee^{-R_{inj}}+[kR_{inj}-4]\ee^{R_{inj}y_{max}}+4y_{max}\sinh R_{inj}}{2R_{inj}\sinh R_{inj}}},}\\[1ex]
\hspace{4.3cm}kR_{inj}\le4\left[1-\frac{\sinh R_{inj}}{R_{inj}\ee^{R_{inj}}}\right],\\
k,\hspace{4cm}kR_{inj}>4\left[1-\frac{\sinh R_{inj}}{R_{inj}\ee^{R_{inj}}}\right].\end{array}\right.\label{eq:Rep}\end{equation}
Note that $F(k,R_{inj})=\hat{U}_{max}/\hat{U}_{p}$, which is fixed
by the parameters $k$ and $R_{inj}$. Thus, for fixed $k$ and $R_{inj}$
an increase in $R$ is interpreted as an increase in $R_{p}$, and
vice versa. It is also useful to know the ratio of upper wall velocity
to the maximal velocity, i.e.~$\hat{U}_{c}/\hat{U}_{max}$, which
we shall denote $\tilde{k}$, given simply by the ratio $k/F(k,R_{inj})$:
\begin{equation}
\tilde{k}(k,R_{inj})=\left\{ \begin{array}{l}
{\displaystyle {\frac{2kR_{inj}\sinh R_{inj}}{4\cosh R_{inj}-kR_{inj}\ee^{-R_{inj}}+[kR_{inj}-4]\ee^{R_{inj}y_{max}}+4y_{max}\sinh R_{inj}}},}\\[1ex]
\hspace{4.3cm}kR_{inj}\le4\left[1-\frac{\sinh R_{inj}}{R_{inj}\ee^{R_{inj}}}\right],\\
1,\hspace{4cm}kR_{inj}>4\left[1-\frac{\sinh R_{inj}}{R_{inj}\ee^{R_{inj}}}\right].\end{array}\right.\label{eq:wallspeed}\end{equation}
The ratio $R/R_{p}$ and the upper wall speed $\tilde{k}$ are illustrated
in Fig.~\ref{fig:figbase3} for convenience.

\subsection{The stability problem}

The base flow is two-dimensional, but since $v=R_{inj}$ is constant,
the 3D linear stability equations are only modified by the addition
of a constant convective term: \[
R_{inj}\frac{\partial}{\partial y}\mathbf{u}', \]
where $\mathbf{u}'=(u',v',w')$ denotes the linear perturbation. The
classical Squire transformation can therefore be applied to the temporal
problem, showing that for any unstable 3D linear disturbance there
exists an unstable 2D linear disturbance at lower $R$; see \cite{Squire1933}.

It suffices to consider only 2D disturbances and we adopt the usual
normal mode approach to linear spatially periodic perturbations, introducing
a stream function which we represent in modal form as: \begin{eqnarray}
\hat{\psi}(x,y,t) & = & \phi(y)\ee^{[\ii\alpha(x-ct)]},\label{eq:5}\end{eqnarray}
 with $u'=D\phi(y)\ee^{[\ii\alpha(x-ct)]}$, $v'=-\ii\alpha\phi(y)\ee^{[\ii\alpha(x-ct)]}$.
Thus, $\alpha$ is real, denoting the wavenumber, $c$ denotes the
complex wave speed, ($c=c_{r}+\ii c_{i},~\ii=\sqrt{-1}$), and $\phi(y)$
denotes the amplitude of the stream function perturbation. The modified
Orr-Sommerfeld (O-S) equation for the flow is: \begin{eqnarray}
\ii\alpha R[(c-u)(\alpha^{2}-D^{2})-D^{2}u]\phi-R_{inj}D(\alpha^{2}-D^{2})\phi & = & (\alpha^{2}-D^{2})^{2}\phi,\label{eq:4}\end{eqnarray}
 and the boundary conditions are \begin{eqnarray}
\phi(\pm1) & = & D\phi(\pm1)=0.\label{eq:6}\end{eqnarray}
 The inclusion of the injection cross-flow results in
an additional 3rd order derivatives in the inertial terms, i.e. $R_{inj}D(\alpha^{2}-D^{2})\phi$. Note that $R_{inj}$ also influences stability via the base velocity profile $u(y)$. Equations (\ref{eq:4})--(\ref{eq:6}) constitute the eigenvalue problem. The eigenvalue $c$ is parameterised by the 4 dimensionless groups $(\alpha,R,R_{inj},k)$ and the condition of marginal stability is:
\begin{eqnarray}
c_{i}(\alpha,R,R_{inj},k) & = & 0\label{eq:7}\end{eqnarray}
 We attempt to characterise the stability of (\ref{eq:4})--(\ref{eq:6})
for positive $(\alpha,R,R_{inj},k)$. We may note that the limit $R\to\infty$
for finite $R_{inj}$, reduces (\ref{eq:4}) to the Rayleigh
equation. Since $D^{2}u$ is of one sign only, there are no inflection
points and hence no purely inviscid instability. This suggests that
the instabilities of (\ref{eq:4})--(\ref{eq:6}) will be viscous
in nature.

Addition of the constant cross-flow terms does not fundamentally alter
the O-S problem, and we expect a discrete spectrum. To find the spectrum
of (\ref{eq:4})--(\ref{eq:6}) we use a spectral approach, representing
$\phi$ by a truncated sum of Chebyshev polynomials: \begin{eqnarray}
\phi & = & \sum_{n=0}^{N}a_{n}T_{n}\left(y\right)\,\,\mbox{for}\,\, y\in\left[-1,1\right]\label{eq:8}\end{eqnarray}
 where $N$ is the order of the truncated polynomial, $a_{n}$ is
the coefficient of the $n$-th Chebyshev polynomial, $T_{n}\left(y\right)$.
This method is described for example in \cite{Schmid2001} and is widely
used. The discretised problem is coded and solved in Matlab. The accuracy
of the code has been checked against the results of \cite{Mack1976}
for the Blasius boundary layer, with various results for PP flow in
\cite{Schmid2001}, with the PCP flow results of \cite{Potter1966},
and finally against results for PP flow with cross-flow; see {\cite{Sheppard1972}}, \cite{Fransson2003}. The
results are accurate up to  three, four and five significant places when validated against \cite{Potter1966}, \cite{Mack1976} and \cite{Fransson2003}, respectively. All the numerical results given below have been computed with $N=120$. On using $200$ collocation points, the growth rates changed only in the fourth significant place in the worst case.

\subsection{Characteristic effects of varying $k$ and $R_{inj}$}
\label{sec:preliminary}

Before starting a systematic analysis of (\ref{eq:4})--(\ref{eq:6}), we briefly show some example results that illustrate the characteristic effects of varying the Couette component, $k$, and the cross-flow component, $R_{inj}$. These examples also serve to establish the framework of analysis used later in the paper. With reference to PP flow, \cite{Potter1966} first observed that the stability is increased by adding a Couette component while \cite{Fransson2003} showed that cross-flow can stabilise or destabilise PP flow.

\subsubsection{Eigenspectra}

\begin{figure}
~~~~~~~~~~~~~~~~~~~~~\subfloat[]{\includegraphics[scale=0.08]{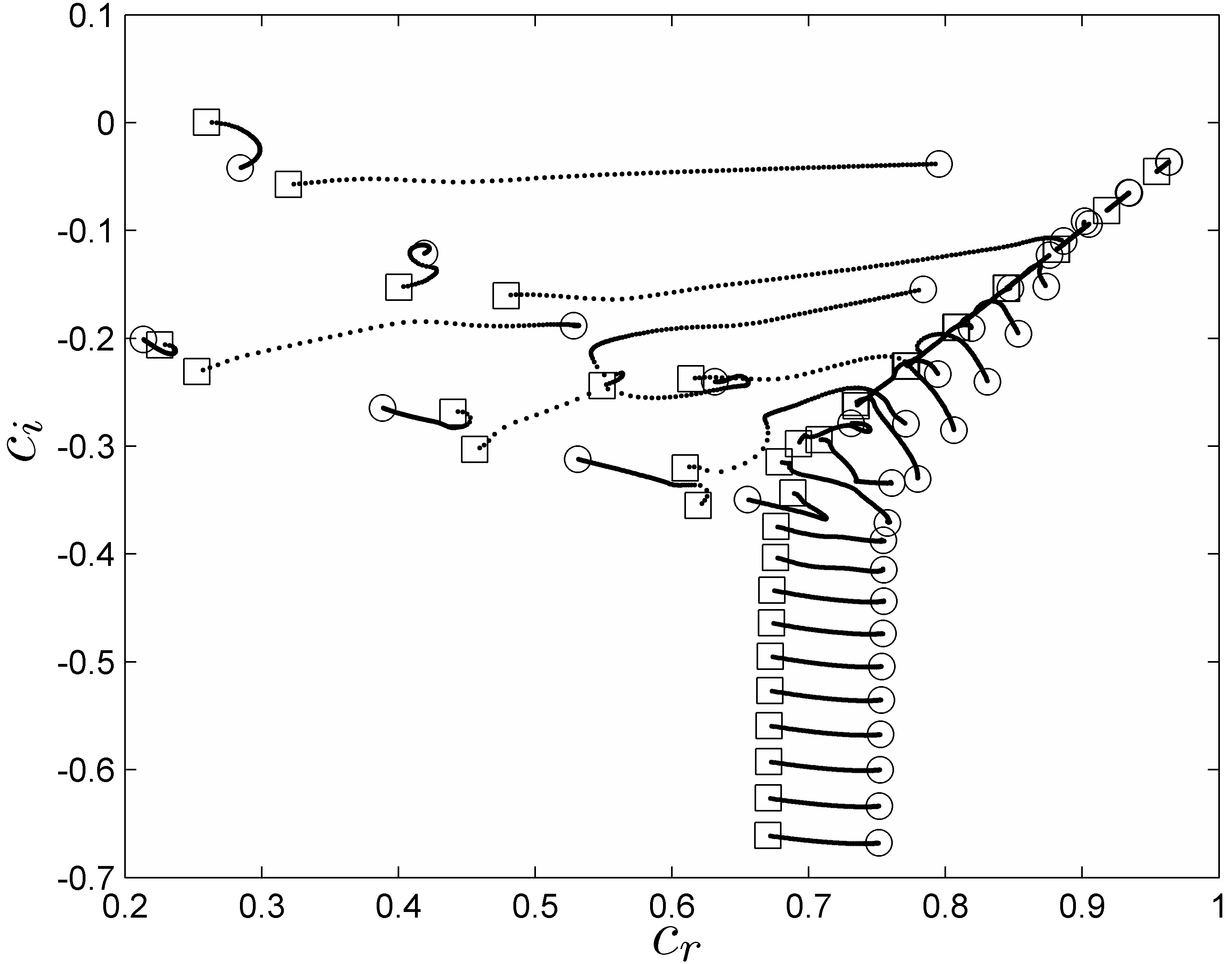}

}\\\subfloat[]{~~~~~~~~~~~~~~~~~~~~~\includegraphics[scale=0.08]{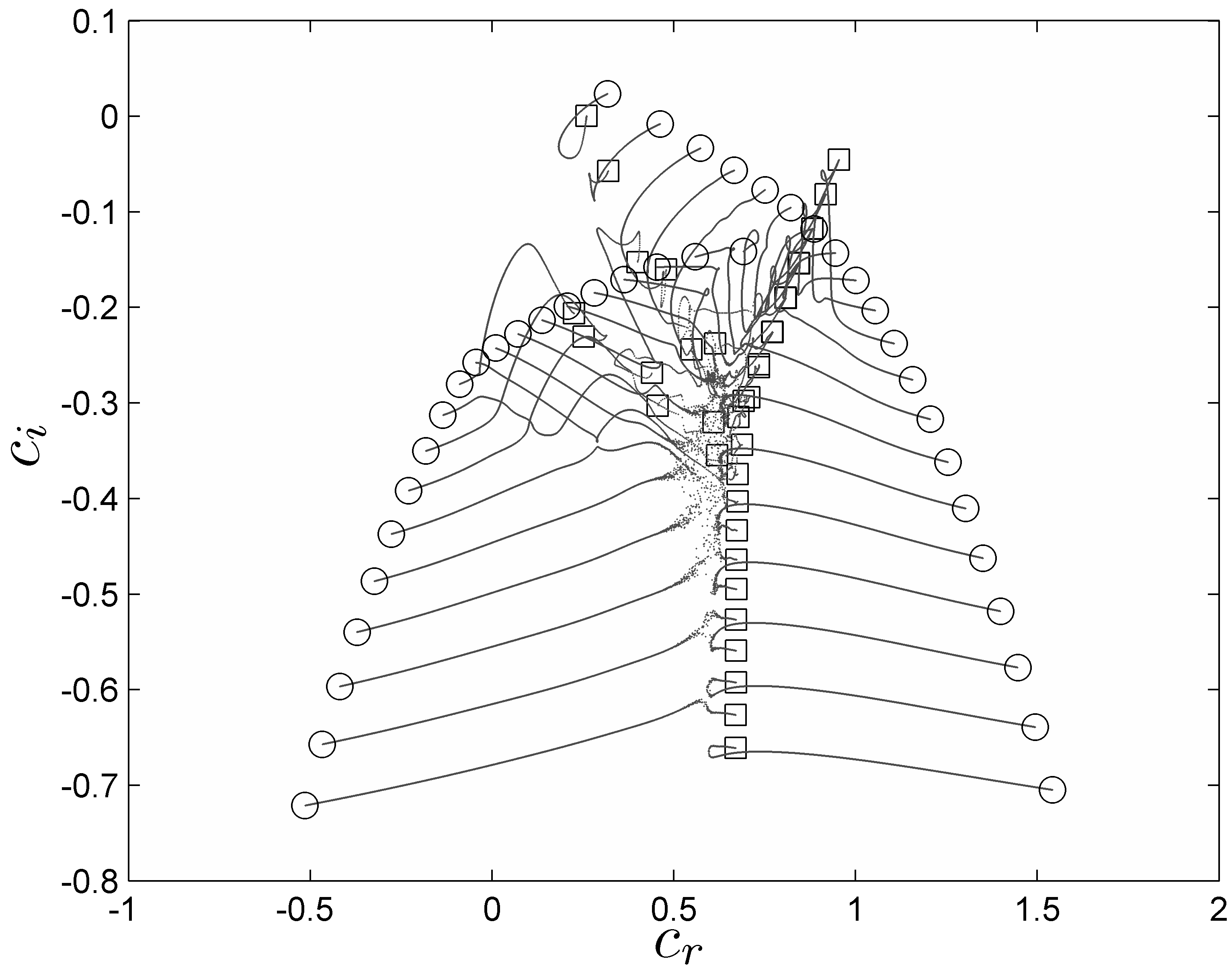}

}\caption{~Eigenspectrum of $(\alpha,R)=(1,6000)$ by varying $k$ and $R_{inj}$.~$40$ least stable modes are considered.  (a)~Effect of increasing $k$ from $0$ to $1$ in steps of $0.01$, keeping $R_{inj}=0$.
(b)~Effect of increasing $R_{inj}$ from $0$ to $100$ in steps of $0.05$, keeping $k=0$~(PP flow).~The symbols in (a) and (b) are similar and are denoted as follows: $k=0$~or~$R_{inj}=0$~by~$(\square)$,~$k=1$~or~$R_{inj}=100$~by~$(\circ)$ and intermediate $k$~or~$R_{inj}$ by ~(.).            Note that the PP flow spectrum is represented by the $\square$ in both figures, and shows the vertical family of S-modes, the branch of A-modes (diagonally upwards from centre to left) and branch of P-modes  (diagonally upwards from centre to right).
\label{fig:BasicSol}}
\end{figure}

Setting $(\alpha,R)=(1,6000)$, we investigate variations in the eigenspectrum of (\ref{eq:4})--(\ref{eq:6}).
According to a classification proposed by \cite{Mack1976}, the spectrum of PP flow spectra may be divided into 3 distinct families: A, P and S. Family A exhibits low phase velocity and corresponds to the modes concentrated near the fixed walls. Family P represents phase velocities, $c_{r}$, close to the maximum velocity in the channel. Family S corresponds to the mean modes and has phase velocity $c_{r}$ close to the mean velocity. In Figs~\ref{fig:BasicSol}(a) \& (b), we track the eigenmodes as  $k$ and $R_{inj}$, respectively, are varied from zero. The initial condition (denoted by $\square$) represents the PP flow.

Referring to Fig.~\ref{fig:BasicSol}(a), (where $R_{inj} = 0$), addition of the Couette component increases the mean velocity: the S modes shift from $c_{r}=0.6667$ at $k=0$~(PP flow) to  $c_{r}=0.7513$ at $k=1$. The family of P modes is also shifted to the right. The A modes are associated with both walls and as $k$ increases we see a splitting of the family, with the upper wall modes moving to the right as $k$ is increased. The least stable mode is a wall mode associated with the lower wall, which we observe stabilises monotonically as $k$ is increased. Figure~\ref{fig:BasicSol}(b) shows the effects of increasing $R_{inj}$, (holding $k=0$). The least stable A mode of PP flow initially stabilises and then destabilises with increasing $R_{inj}$. This behavior has also been observed by \cite{Fransson2003}. For large $R_{inj}$ the A, P, and S families have disappeared, instead leaving two distinct families of modes. It appears that each of the A, P, and S families splits, with some modes entering each of the two families (this alternate splitting is most evident for the S modes). As observed by \cite{Nicoud1997}, the phase speed no longer lies in the range of the axial velocity. This does not violate the conditions on $c_r$, given by \cite{Joseph1968} and \cite{Joseph1969}, since these conditions are derived for parallel flows only.

\subsubsection{Increasing $R_{inj}$}

\begin{figure}
\subfloat[]{\includegraphics[scale=0.055]{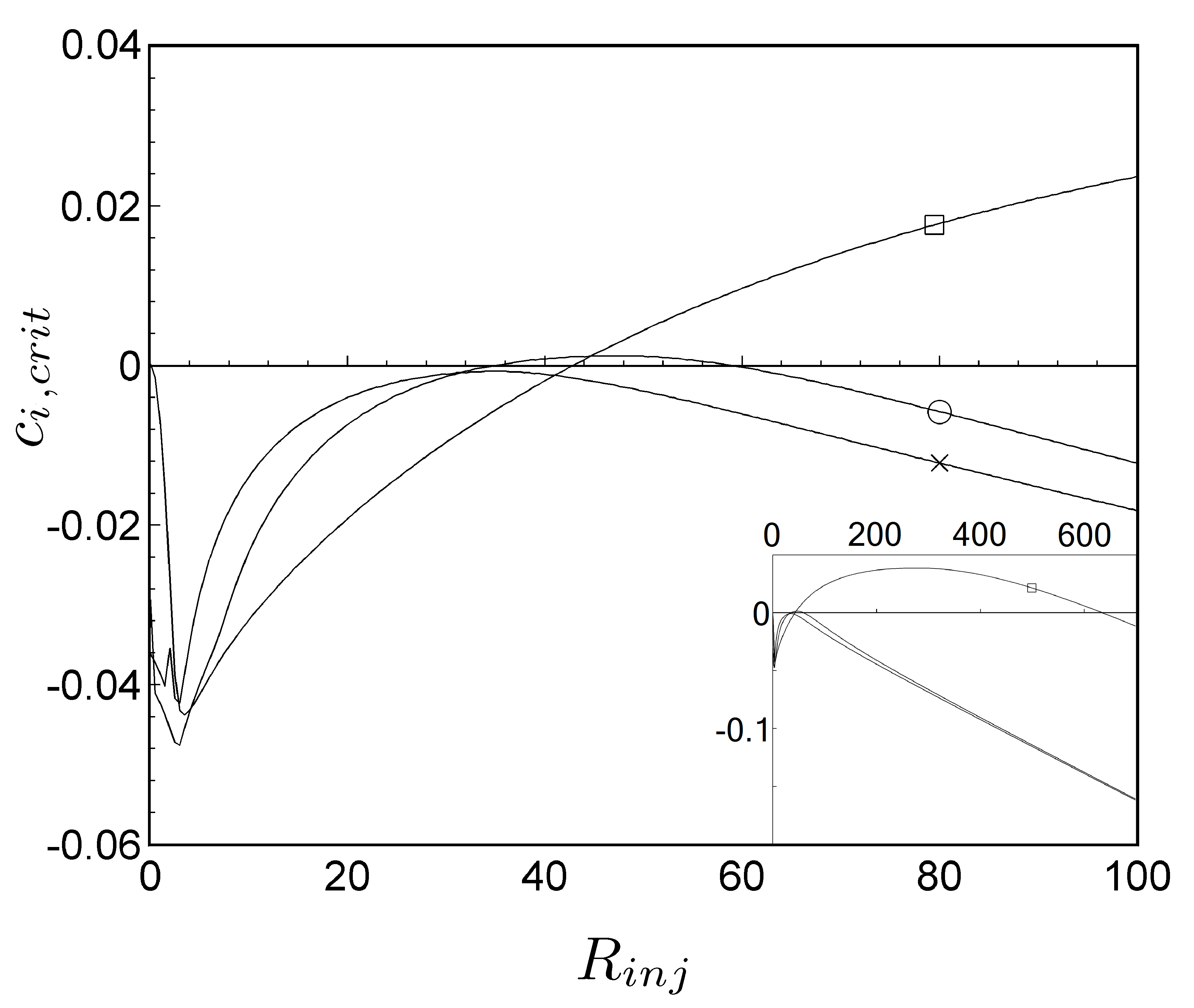}

}~\subfloat[]{\includegraphics[scale=0.06]{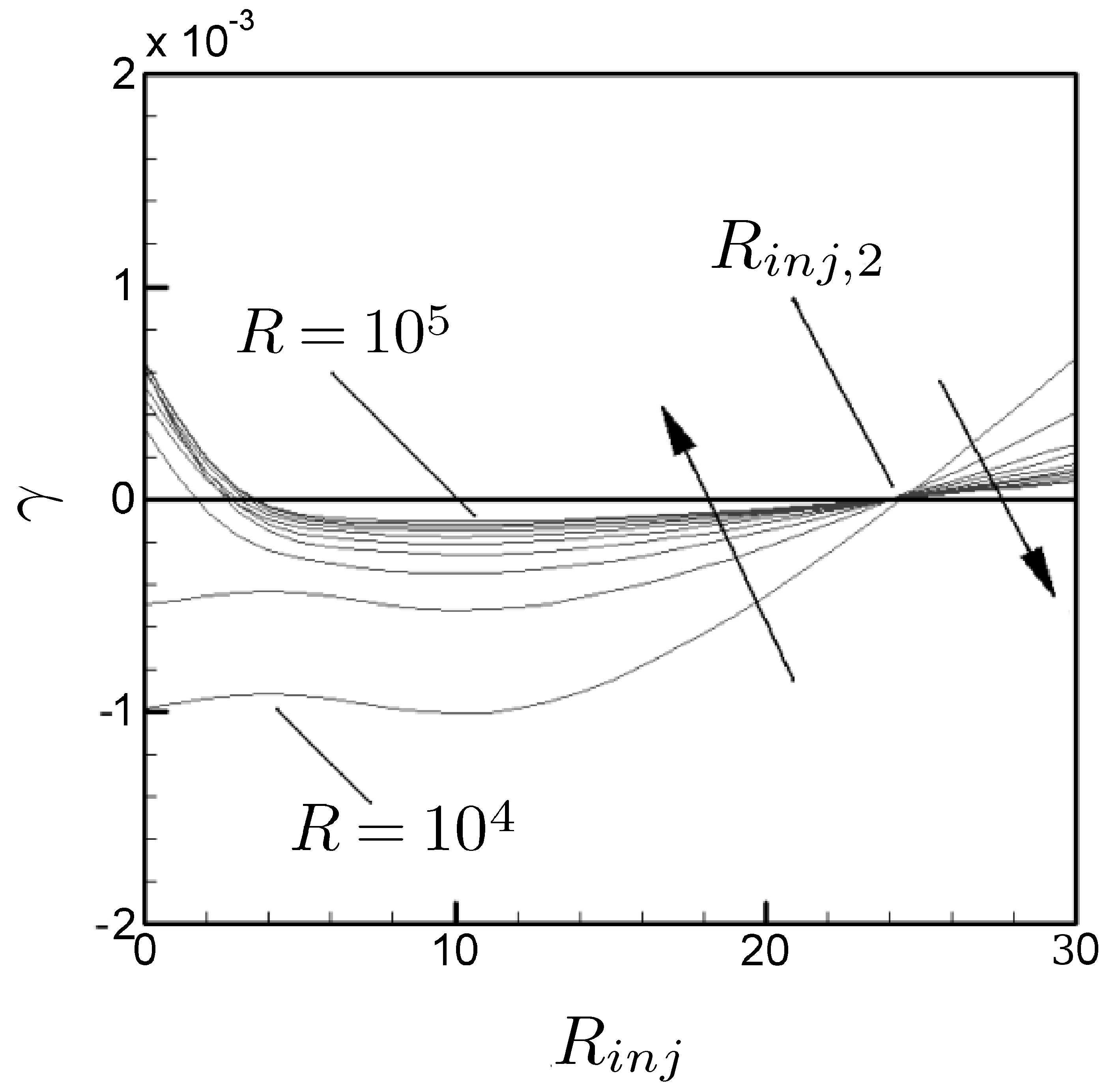}

}\caption{(a)~Effect of increasing $R_{inj}$ on the stability of PCP flow,
for $(\alpha,R)=(1,6000)$ and different values of $k=0\,(\square),\,0.5\,(\circ),\,1\,(\times)$.~(b)~Maximal
growth rate for increasing $R_{inj}$ at different $R$, ($k=0.5$
and the step in values of $R$ between curves is $10^{4}$).\label{fig:4}}

\end{figure}

Next, we illustrate the qualitative effects of increasing $R_{inj}$ at
fixed $(\alpha,R,k)$, in Fig.~\ref{fig:4}(a). We again  fix $\alpha=1$
and $R=6000$, and show the variation of the least stable eigenvalue,
for $k=0,\,0.5,\,1$. Our results for $k=0$ (PP flow) may be compared
directly with those of \cite{Fransson2003}. We observe that as $R_{inj}$
increases we have an initial range of stabilisation ($c_{i,crit}$ decreasing),
followed by a range of destabilisation ($c_{i,crit}$ increasing), and
finally again stabilisation at large $R_{inj}$, ($c_{i,crit}$ decreasing).
Qualitatively, we have observed these same three ranges of decreasing/increasing
$c_{i,crit}$, as $R_{inj}$ increases, for all numerical results that
we have computed, and this provides a convenient framework within
which to describe our results.

For fixed $(\alpha,R,k)$, the case $R_{inj}=0$ may either be stable
or unstable, in which cases there are respectively two or three marginal
stability values of $R_{inj}$. We denote these marginal values of
$R_{inj}$ by: $R_{inj,1}$, $R_{inj,2}$, $R_{inj,3}$, noting that
in the case that $R_{inj}=0$ is stable $R_{inj,1}$ is absent. More
clearly, $R_{inj,2}$ will always represent a transition from stable
to unstable, while $R_{inj,1}$~\&~$R_{inj,3}$ denote transitions from
unstable to stable. The PCP flows for $k=0.5$ and $1$ are stable
for $(\alpha,R)=(1,6000)$ in the absence of cross-flow, $R_{inj}=0$.
For a larger $R$,~$k=0.5$ is unstable at $R_{inj}=0$, but $k=1$
remains stable for all $(\alpha,R)$.

Figure~\ref{fig:4}(b) shows the maximal growth rate $\gamma$, for
increasing $R_{inj}$ at different $R$, with $k=0.5$. The maximal
growth rate is computed over wavenumbers $\alpha\in[0,1]$:
\begin{equation}
\gamma=\max_{\alpha\in[0,1]}\{\alpha c_{i}\},\end{equation}
which often captures the largest growth rates over all $\alpha$.
We observe that the first marginal value $R_{inj,1}$ increases with
$R$, but appears to converge towards a finite value as $R\to\infty$.
The second marginal value of $R_{inj,2}$ appears independent of
$R$, (at least numerically). For $k=0.5$ we have $R_{inj,2}\approx24.7$.~\cite{Nicoud1997}
have observed a similar behaviour in studying the generalised Couette
flow, (for which the constraint, $kR_{inj}=4$, is always satisfied).
They have found $R_{inj,2}\approx24$ (note that Nicoud and Angilella
use the full channel width as their length-scale, and therefore report $R_{inj,2}\approx48$,
in their variables). In contrast, the third marginal value, $R_{inj,3}$,
is strongly dependent on $R$. For example, for $k=0.5$, the values
corresponding to $R=10000$ and $100000$ are $R_{inj,3}\approx83$
and $R_{inj,3}\approx287$ respectively.

\subsubsection{Increasing $k$}
\label{sec:IncreasingK}

\begin{figure}
\subfloat[]{\includegraphics[scale=0.076]{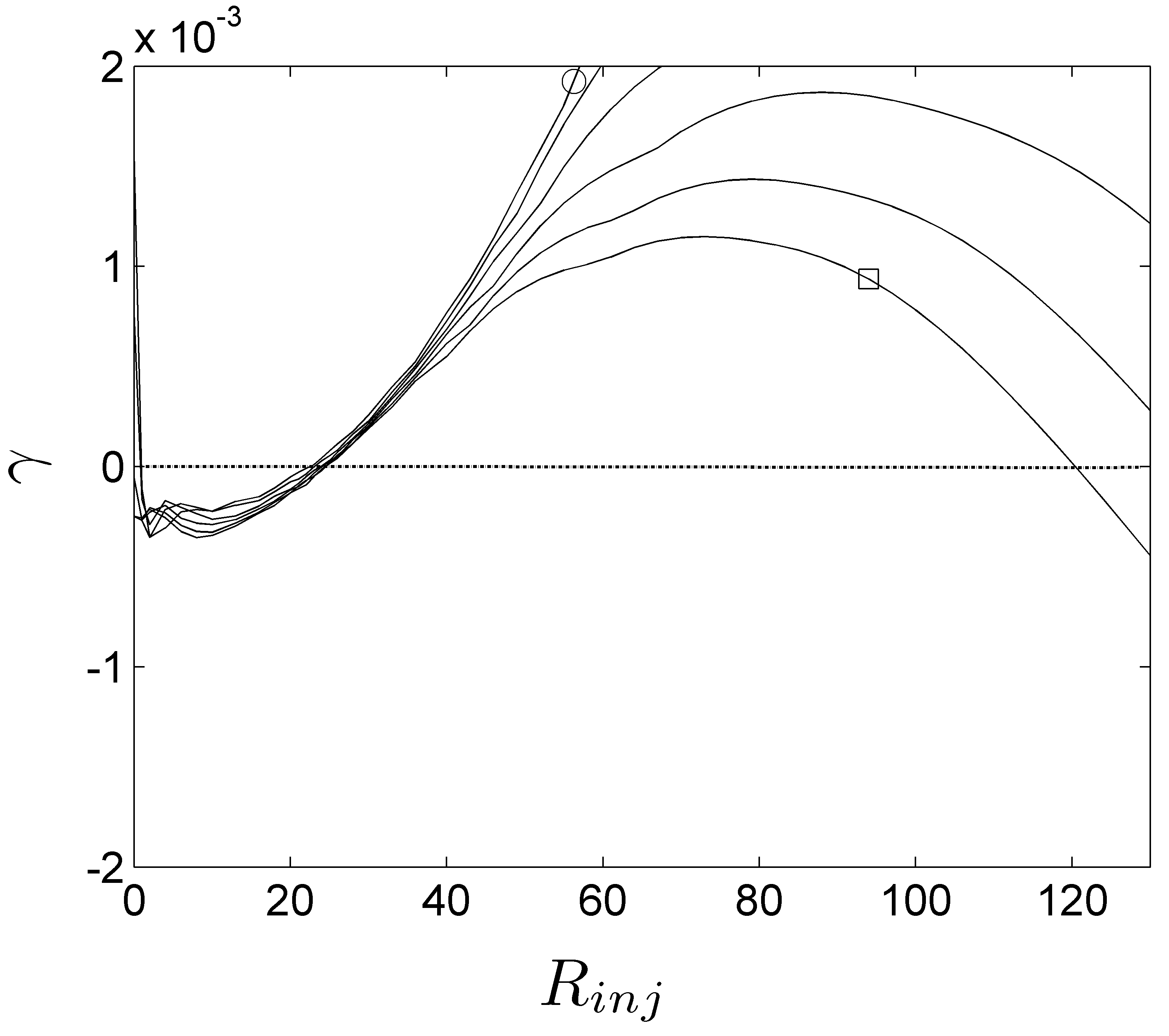}}~\subfloat[]{\includegraphics[scale=0.076]{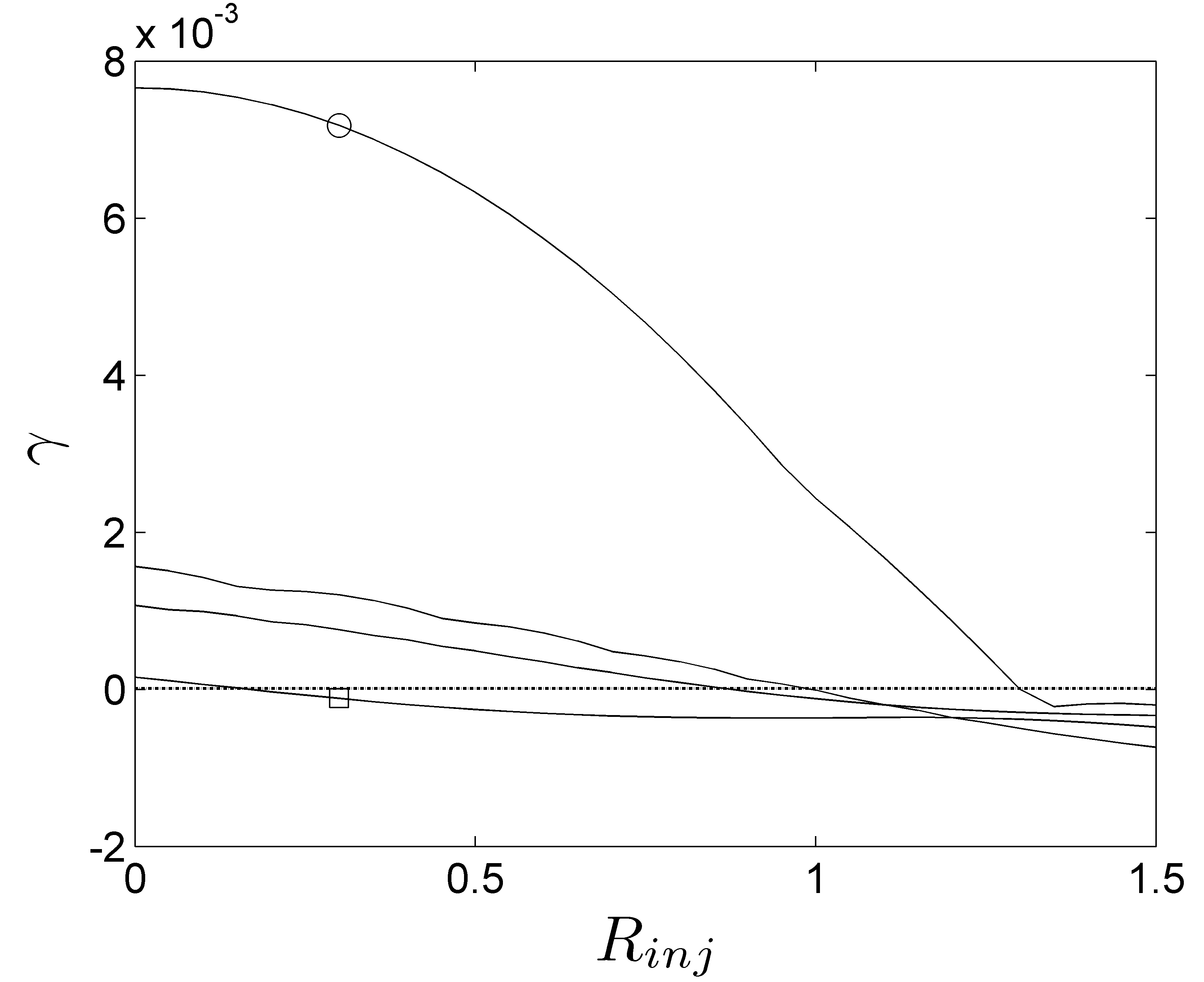}

}\caption{Maximal growth rate versus $R_{inj}$ at $R=40000$: (a) $R_{inj,2}$
\& $R_{inj,3}$ for $k=0\,(\circ)$ to {$1\,(\square)$};
(b) $R_{inj,1}$ for $k=0\,(\circ)$ to {$0.6\,(\square)$}.
Step size is $0.2$ in both figures.\label{fig:Maximal-expansion-rate}}

\end{figure}

Figure~\ref{fig:Maximal-expansion-rate} explores the effects of increasing the Couette component $k$, on $\gamma$
and on the marginal values of $R_{inj}$. Figure~\ref{fig:Maximal-expansion-rate}(a)
indicates that the sensitivity of $R_{inj,2}$ to $k$ is also not
extreme: we have found that this transition occurs within the range
$\sim22-25$ for $k\in[0,1]$. For each value of $k$ examined, we
also observe numerically a similar independence of $R_{inj,2}$ to
$R$ as seen earlier in Fig.~\ref{fig:4}(b) for $k=0.5$. The
3rd marginal value, $R_{inj,3}$, is strongly dependent on $k$. For
example, at $R=40000$, $R_{inj,3}(k=1)\approx120$ and $R_{inj,3}(k=0.8)\approx135$.
In general, increasing $k$ shifts $R_{inj,3}$ to the left, thereby
decreasing the span of the unstable region. Increasing $k$ also decreases
the maximum value of $\gamma$.

Figure~\ref{fig:Maximal-expansion-rate}(b) looks at the first transition, $R_{inj,1}$ at $R=40000$. \cite{Potter1966} was the first to observe that
for PCP flows (i.e. $R_{inj}=0$), a gradual increase in the wall velocity results in crossing a ``cut-off'' value of $k$, say $k_{1}$, such that
for $k>k_{1}$ the flow is unconditionally linearly stable.  It has already been pointed out from the results of Fig.~\ref{fig:4}(b)that $R_{inj,1}$ is  finite  as $R\to\infty$. In addition, the results in Fig.~\ref{fig:Maximal-expansion-rate}(b) indicate that $R_{inj,1}$ decreases with $k$ at a finite $R$. Hence, it can be inferred that as $R\to\infty$, the cut-off wall velocity, $k_{1}=k_{1}(R_{inj})$ must decrease with $R_{inj}$.

\section{PCP flows and the effects of small $R_{inj}$}
\label{sec:low}

Having developed a broad picture of the different transitions occurring in the flow, we now focus in depth at each range of $R_{inj}$, to understand the stability mechanisms in play. We start with the range of small $R_{inj}$.

PCP flows without cross-flow are stable to inviscid modes, but viscosity
admits additional modes, i.e.~the Tollmien-Schlichting (TS) waves,
which may destabilise, according to the value of $k$. When $\alpha R\gg1$ with $c\sim O(1)$, viscous
effects occur in thin oscillatory layers: (i) adjacent to the walls,
(of thickness $\sim(\alpha R)^{-1/2}$), and (ii) close to the critical
point(s), $y_{c}$, where $u(y_{c})=c_{r,crit}$ are found, (of thickness
$\sim(\alpha R)^{-1/3}$). It is in the critical layers that we see peaks
in the distribution of energy production, implying transfer from the
base flow. \cite{Potter1966} put forward the argument
that for a dimensionless wall velocity that exceeds $c_{r,crit}$,
the critical layer near the moving wall will vanish and there remains
only one critical layer, near the fixed lower wall. The thickness
of this second layer increases with wall velocity, thereby favouring
stabilisation.

This mechanism appears to correctly describe the long wavelength perturbations,
(at $R_{inj}=0$), which are found to be the least stable for $k\sim O(1)$.
Indeed \cite{CowleySmith1985} developed a long wavelength
analysis ($\alpha\sim R$), in order to predict the cut-off value
$k_{1}(R_{inj}=0)\approx0.7$. For values $k\sim O(1)$, PCP flows
have only a single neutral stability curve (NSC). However, \cite{CowleySmith1985}
noted that for smaller $k$, multiple neutral stability curves could
exist, and at shorter wavelengths. For example, when $0\leq k\leq R^{-2/7}$
there is one NSC, when $R^{-2/7}\leq k\leq R^{-2/13}$ there are three
NSC's, and when $R^{-2/13}\leq k\ll1$ there are two NSC's; see \cite{CowleySmith1985}.
Thus to understand the effect of cross-flow in PCP flows, the different
regimes of $k$ need to be considered separately.

\begin{figure}
\subfloat[]{\includegraphics[scale=0.059]{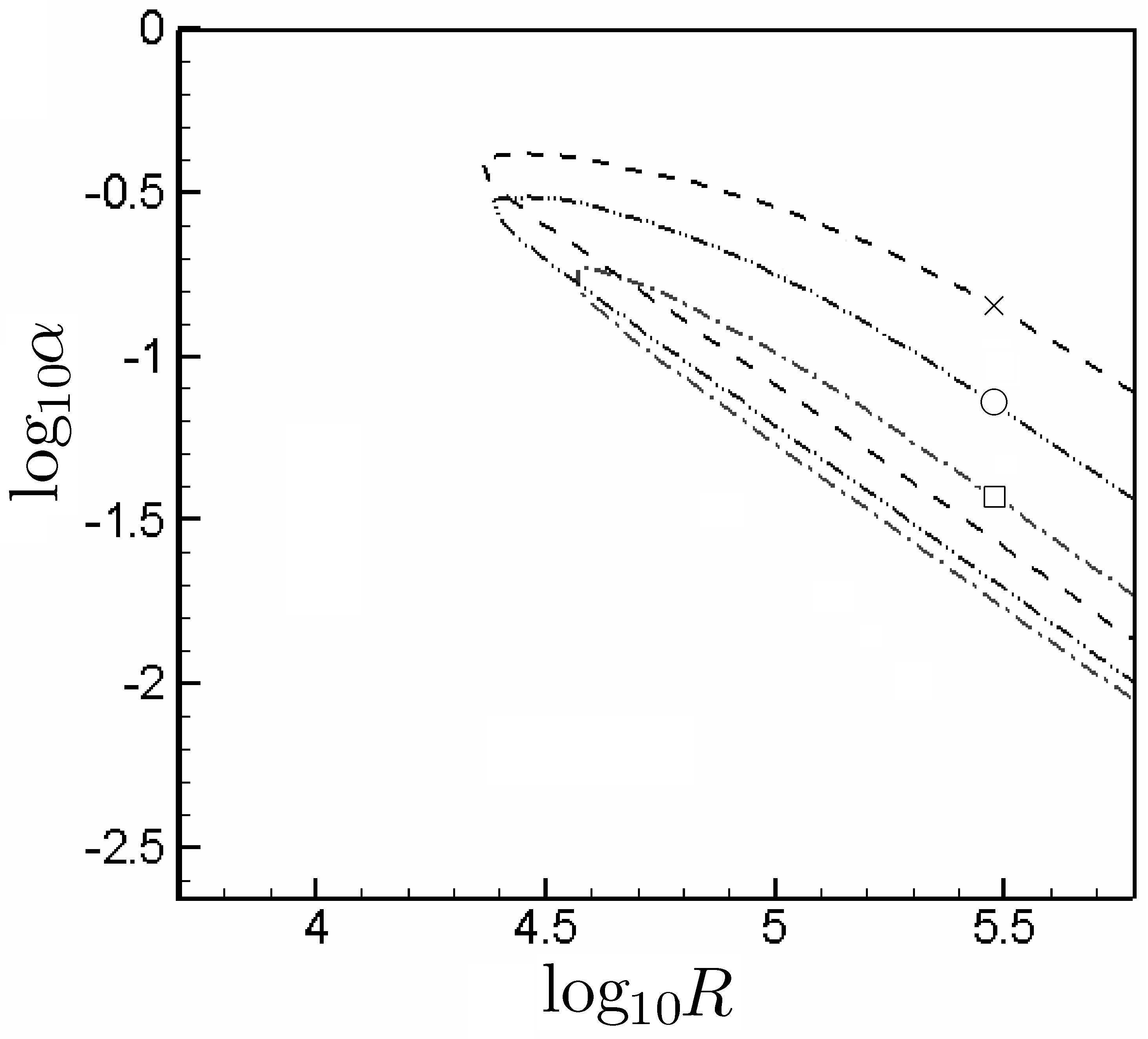}

}~\subfloat[]{\includegraphics[scale=0.08]{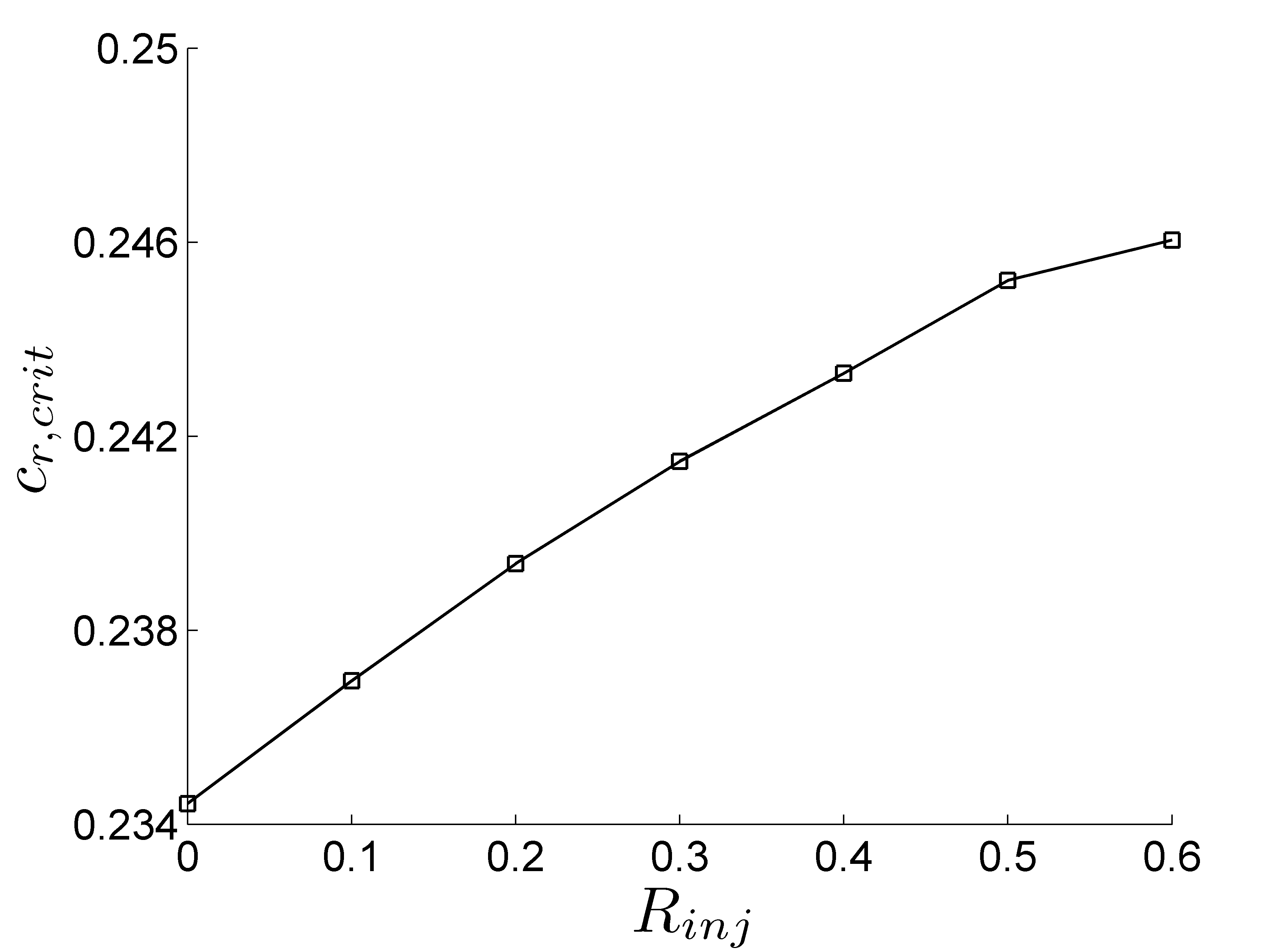}

}\caption{Critical values for $k=0.5$: (a) neutral stability curves for $R_{inj}=0\,(\times),\,0.3\,(\circ)$
and $0.53\,(\square)$; (b) variation in $c_{r,crit}$ with $R_{inj}$.\label{fig:k0.5}}

\end{figure}

For $R_{inj}\approx0$, we expect the stability behaviour to be close to that of the PCP flow without cross-flow.
Intuitively we expect cross-flow to stabilise, and so study the range
$c_{r,crit}<k\leq k_{1}(R_{inj}=0)$. We examine the NSC's obtained
from the O-S equation corresponding to $k=0.5$, under different values
of $R_{inj}$; see Fig.~\ref{fig:k0.5}(a). As expected, increasing
$R_{inj}$ results in a progressively larger critical $R=R_{crit}$.
We also observe that both the upper and the lower branches are oriented
at an angle of $45$ degrees, (i.e.~$\alpha\sim R^{-1}$), at high
values of $R$. On fixing $R_{inj}$ and increasing $k$ we have found
that for successively large $k$ the upper and lower branches move
together as $R_{crit}$ increases, eventually coalescing at $k=k_{1}(R_{inj})$.
This mechanism is identical with that observed by \cite{CowleySmith1985},
suggesting the applicability of a long wavelength approximation in
order to predict $k_{1}(R_{inj})$. Figure~\ref{fig:k0.5}(b) plots
the values of $c_{r}$ at criticality, as $R_{inj}$ is varied, also
for $k=0.5$. The critical values are tabulated in Table~\ref{tab:at-criticality}.
The dependence is initially linear. We observe that $k>c_{r,crit}$
over the computed range.

\begin{table}
~~~~~~~~~~~~~~~~~~~~~~~~~~~~~~~~~~~~~~~~~~\begin{tabular}{|c|c|c|c|}
\hline
$R_{inj}$ & $\alpha_{crit}$ & $R_{crit}$ & $c_{r,crit}$\tabularnewline
\hline
\hline
$0$ & $0.3851$ & $22600$ & $0.2344$\tabularnewline
\hline
$0.1$ & $0.3576$ & $22538$ & $0.2370$\tabularnewline
\hline
$0.2$ & $0.3275$ & $22924$ & $0.2394$\tabularnewline
\hline
$0.3$ & $0.2950$ & $23986$ & $0.2415$\tabularnewline
\hline
$0.4$ & $0.2550$ & $26321$ & $0.2433$\tabularnewline
\hline
$0.5$ & $0.2000$ & $31656$ & $0.2452$\tabularnewline
\hline
$0.6$ & $0.1200$ & $51115$ & $0.2461$\tabularnewline
\hline
\end{tabular}\caption{Critical values for $k=0.5$ and increasing $R_{inj}$.\label{tab:at-criticality}}

\end{table}

\subsection{Long wavelength approximation}

We follow the long wavelength distinguished limit approach of \cite{CowleySmith1985},
taking $\alpha\to0$ and $R\to\infty$, with $\lambda=(\alpha R)^{-1}$
fixed. The product $\alpha R$ is fixed along the upper and lower
branches of the NSC. Thus, as the two branches of the NSC coalesce,
in the $(k,\lambda)$-plane we observe $k\to k_{1}(R_{inj})$. In
the long wavelength limit, equation~\ref{eq:4} becomes:

\begin{equation}
i\lambda\left[D^{4}-R_{inj}D^{3}\right]\phi+(u-c)\, D^{2}\phi-(D^{2}u)\phi=0,\label{eq:9}\end{equation}
with boundary conditions (\ref{eq:6}).

\begin{figure}
\subfloat[]{\includegraphics[scale=0.057]{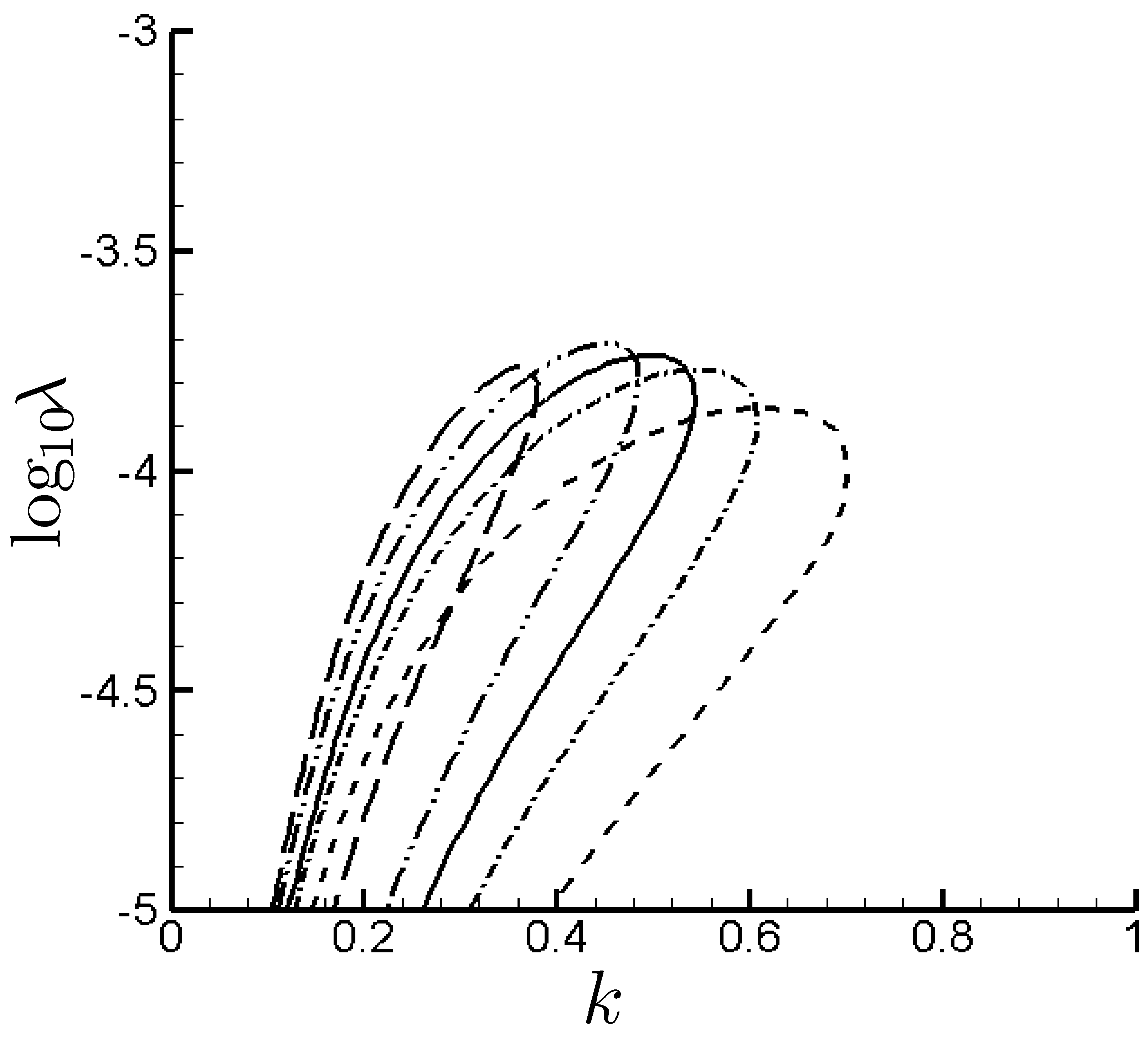}

}~\subfloat[]{\includegraphics[scale=0.06]{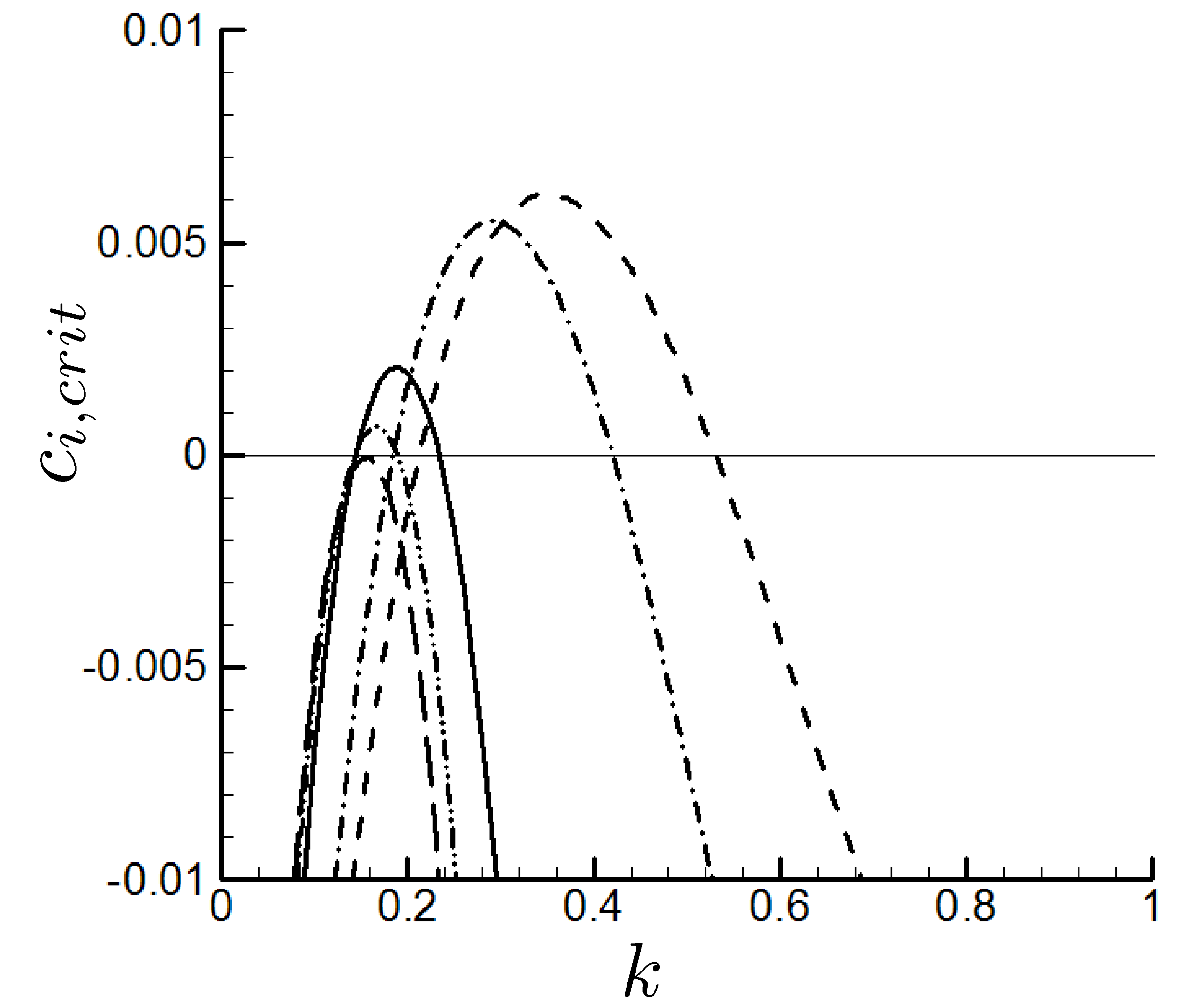}

}

\caption{
(a)~Long wave NSC's showing the dependence of $\lambda$ on $k$
for $R_{inj}=$$0$~(dashed line),~$0.3$~(dash-dot),~$0.5$~(solid),~$0.7$~(dash-dot-dot)
and $1$~(long dash); (b) $c_{i,crit}$ versus $k$ for $\lambda=2.5\times10^{-5}$,
and $R_{inj}=$$0$~(dashed line),~$0.3$~(dash-dot),~$1$~(solid),~$1.2$~(dash-dot-dot)
and $1.3$~(long dash). \label{fig:Long-wave-results}
}

\end{figure}

Figure~\ref{fig:Long-wave-results}(a) shows the NSC obtained from
(\ref{eq:9}), plotted in the $(k,\lambda)$-plane for various $R_{inj}$.
The cut-off value $k_{1}(R_{inj})$ is the maximal value of $k$ on
each of these curves. These values are listed in Table~\ref{tab:1}.
We also list the dimensionless wall speeds at cut-off, i.e.~$\tilde{k}(k_{1},R_{inj})$.
We observe that the cut-off wall speed decreases with $R_{inj}$. This is in agreement with the concluding remarks of \S \ref{sec:IncreasingK}.

\begin{table}
~~~~~~~~~~~~~~~~~~~~~~~~~~~~~~~~~~~~~~~~~~\begin{tabular}{|c|c|c|c|}
\hline
$R_{inj,1}$ & $k_{1}$ & $c_{r,crit}$ & $\tilde{k}(k_{1},R_{inj,1})$\tabularnewline
\hline
\hline
$0$ & $0.70$ & $0.2331$ & $0.5070$\tabularnewline
\hline
$0.3$ & $0.60$ & $0.2431$ & $0.4657$\tabularnewline
\hline
$0.5$ & $0.54$ & $0.2455$ & $0.4386$\tabularnewline
\hline
$0.7$ & $0.48$ & $0.2472$ & $0.4085$\tabularnewline
\hline
$1.0$ & $0.38$ & $0.2358$ & $0.3489$\tabularnewline
\hline
$1.29$ & $0.19$ & $0.1556$ & $0.1939$\tabularnewline
\hline
\end{tabular}\caption{Cut-off values, $k_{1}$, and wavespeed $c_{r,crit}$, for increasing $R_{inj}$.\label{tab:1}}

\end{table}

Figure~\ref{fig:Long-wave-results}(b) shows $c_i$ for the least stable eigenvalue of the long wavelength problem,
for fixed $\lambda=2.5\times10^{-5}$ and different values of $R_{inj}$, as $k$ is varied. When $R_{inj}\geq1.3$, we find that $c_{i,crit}\leq0,~\forall ~k\in(c_{r,crit},k_{1}(0)]$, implying that there are no neutral or unstable long wavelength perturbations in this range of $k$, (i.e.~at least until we approach the second transition at $R_{inj,2}$). Thus, in this initial range of say $R_{inj}\lesssim1.3$, provided that $k>c_{r,crit}$, we can talk equally of a cut-off value for $k$ or for $R_{inj}$.

\subsection{Effects of asymmetry of the velocity profile}

\begin{figure}
\subfloat[]{\includegraphics[scale=0.06]{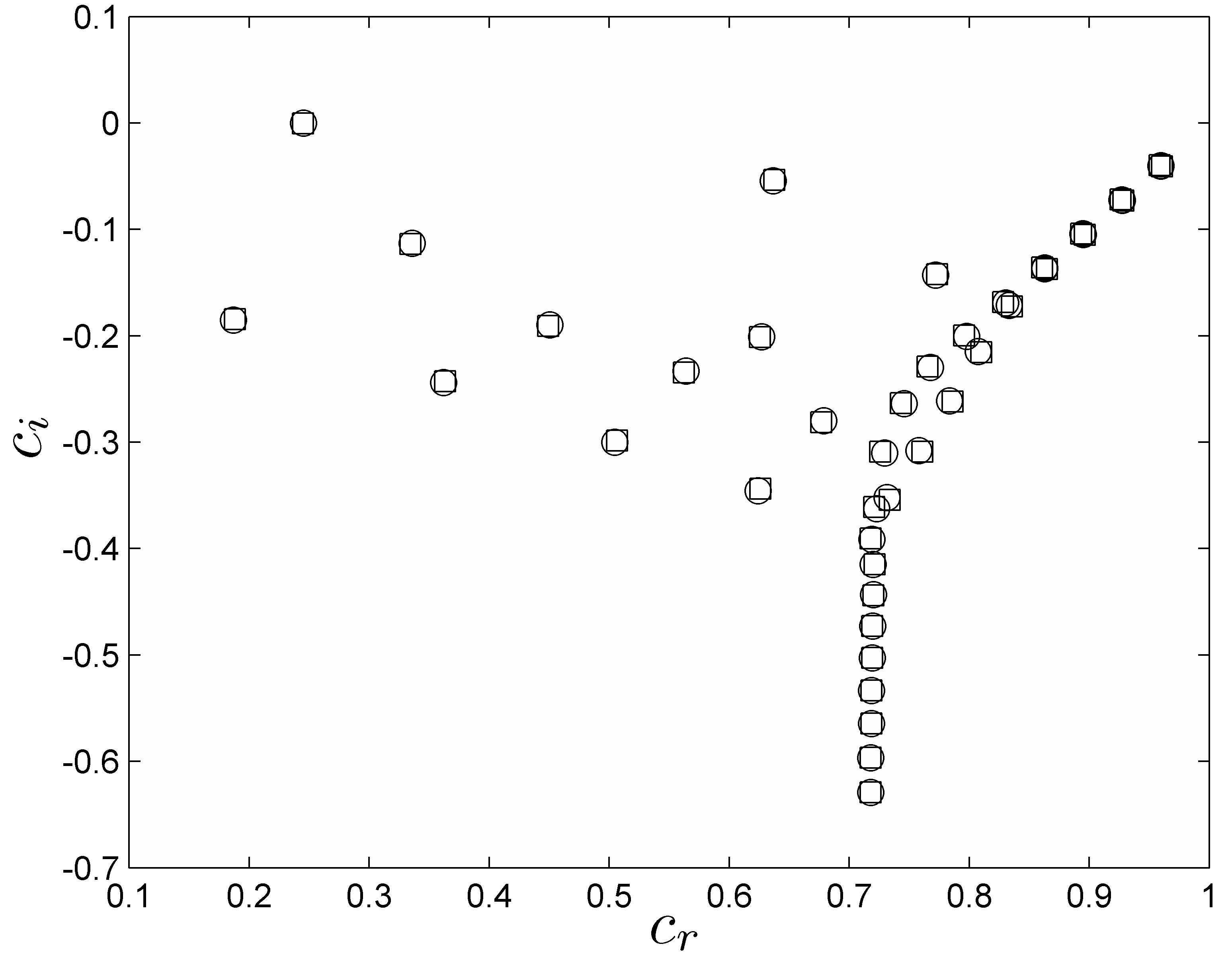}

}~\subfloat[]{\includegraphics[scale=0.06]{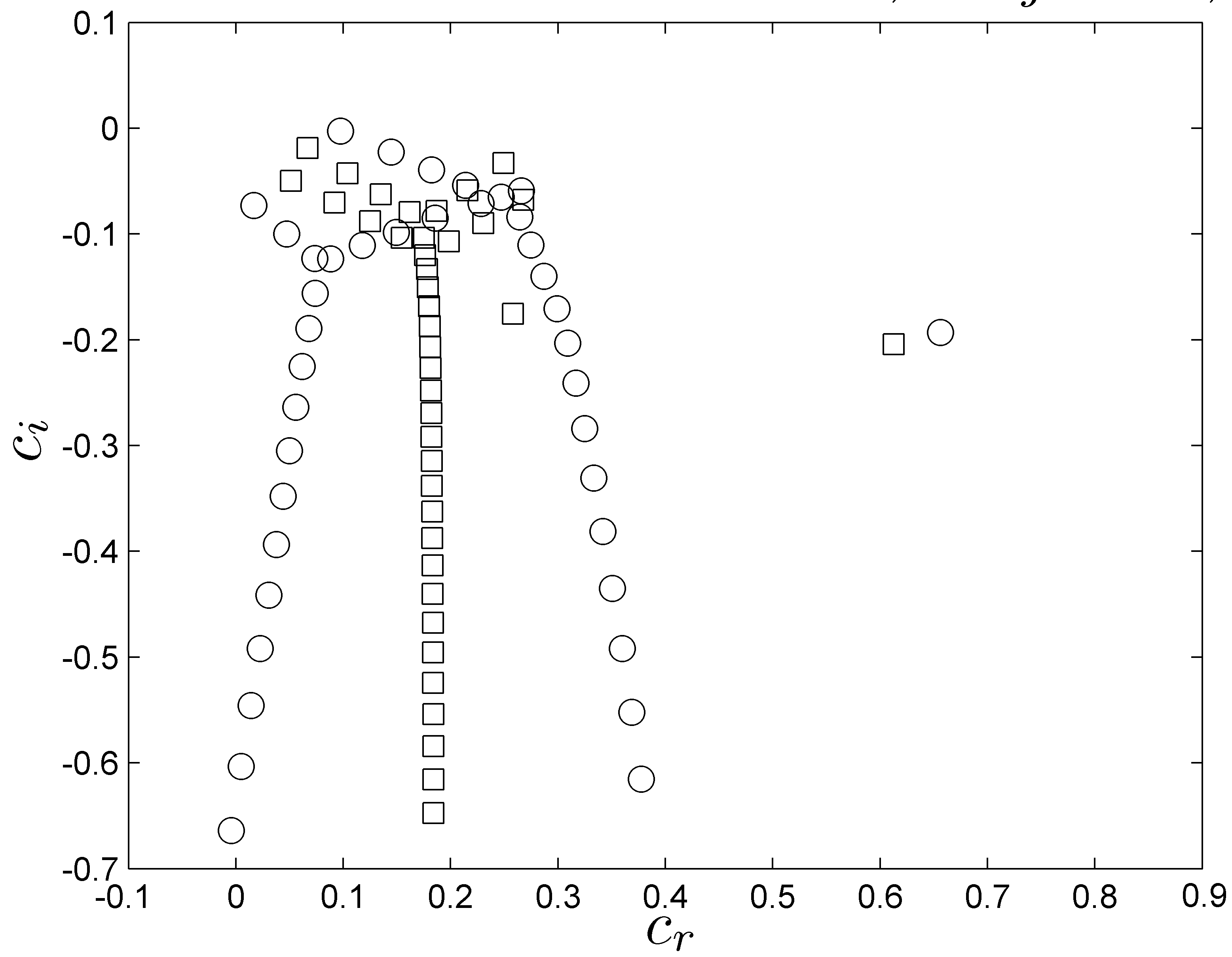}

}

\caption{~Eigenspectrum for $(k,\,\alpha,\, R)=(0.5,\,0.2,\,31656)$
(a)~$R_{inj}=0.5$~(Critical Conditions) and (b)~$R_{inj}=23.5$.~Symbol $\circ$ indicates the eigenspectrum from the O-S equation while $\square$ indicates the spectrum obtained by neglecting the additional cross-flow inertial term.
\label{fig:Comparison}
}
\end{figure}

We observe that $R_{inj}$ enters the stability problem in two distinct ways. The first one represents the direct contribution of the additional third order inertial term, $R_{inj}D(\alpha^{2}-D^{2})\phi$, in the O-S equation (\ref{eq:4}). For the second one, $R_{inj}$ influences the base velocity profile. To explore which of these effects is dominant, we show in Fig.~\ref{fig:Comparison} the spectra of (\ref{eq:4})--(\ref{eq:6}) obtained with and without the term, $R_{inj}D(\alpha^{2}-D^{2})\phi$, included in the computation.
The critical parameters corresponding to $R_{inj}=0.5$, in Table~\ref{tab:at-criticality}, are chosen and fixed for this  comparison. Figure~\ref{fig:Comparison}(a) shows the two spectra at $R_{inj}=0.5$, which are near identical, completely overlapping on the figure. This suggests that at smaller $R_{inj}$, the effects of cross-flow manifest completely via the base flow velocity profile.
Figure \ref{fig:Comparison}(b) shows a similar comparative study at a larger value of $R_{inj}$, closer to $R_{inj,2}$. In this figure we see a distinct difference between the spectra. The additional third order term is apparently responsible for the splitting of the A, P, and S families, illustrated earlier in Fig.~\ref{fig:BasicSol}(b).

\begin{figure}
~~~~~~~~~~~~~~~~~~\includegraphics[scale=0.075]{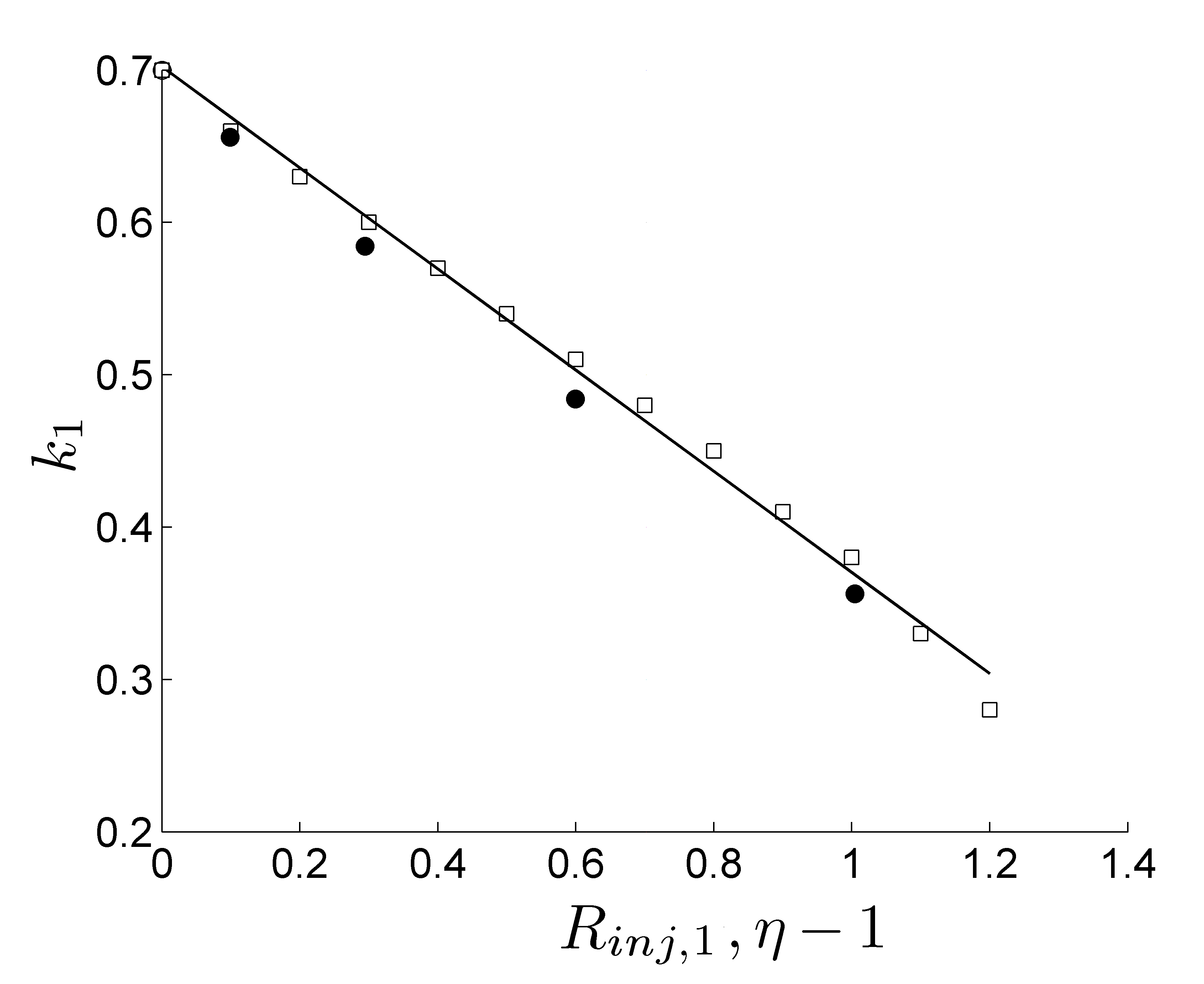}
\caption{~$k_{1}$ as a function of $R_{inj,1}$ (shown by $\square$) as well as the
radius ratio,~$\eta$ (shown by \CIRCLE{}) in ACP flow~(\cite{Sadeghi1991})}
\label{fig:Cut-off-Velocity}
\end{figure}

In Fig.~\ref{fig:Cut-off-Velocity}, we plot $k_{1}$ against $R_{inj} (= R_{inj,1})$. A linear dependence is evident. The slope of the line is approximately $-1/3$. The flow is unconditionally linearly stable above the line and conditionally unstable otherwise. For small values of $R_{inj}$, we have seen in Fig.~\ref{fig:1} that the principal effect is to skew the velocity profile towards the upper wall. A similar asymmetric skewing of the velocity profile is also induced in an annular Couette-Poiseuille (ACP) flow, through geometric means by varying the radius ratio, $\eta$, (defined as the radius of the outer stationary cylinder to the radius of inner moving cylinder). ACP flow has been studied extensively by~\cite{Sadeghi1991}, and we superimpose their results on ours, in Fig.~\ref{fig:Cut-off-Velocity}. The comparison is striking. We believe there are 2 features of Fig.~\ref{fig:Cut-off-Velocity} that are unusual and worthy of note. Unsurprising is of course the identical limits $R_{inj} = 0 = (\eta - 1)$. Note that $R_{inj} \to 0$ is the PCP flow, and $\eta \to 1$ represents the narrow gap limit of ACP, which is also the PCP flow.

The first feature is the very similar linear decay in critical $k = k_1(R_{inj})$, from the PCP values. It can be argued along the lines of \cite{Mott1968}, that for a fixed Couette component $(k)$, increasing the cross-flow for the PCP flow,  or the radius ratio in the ACP flow of \cite{Sadeghi1991}, skews the velocity profile more towards the moving boundary, thus increasing asymmetry and thereby stability. Since it has been already observed in Fig.~\ref{fig:Comparison}(a) that for small $R_{inj}$ the influence of injection on the eigenspectrum is through the velocity profile only, we do expect stabilisation. However, when $(\eta - 1)$ and $R_{inj}$ are of $O(1)$, we can see no obvious quantitative relation between these flows and even the stability operators are quite different.

The second noteworthy feature of Fig.~\ref{fig:Cut-off-Velocity} is that there is a minimum value of $k_{1}$~($k_{1,min}$) below which it is not possible to produce unconditional stability by applying (modest) cross-flow. This minimum value is found when $k_{1}\rightarrow c_{r,crit}$. We have found approximately that $k_{1,min}=0.19$ and the corresponding $R_{inj,1}=1.29$. This is very similar to \cite{Sadeghi1991}, who found that the critical layer near the moving wall of ACP flows remained up to $c_{r,crit} \approx 0.18$.

\subsubsection{Linear energy budget considerations}

The strong analogy with the ACP results of \cite{Sadeghi1991} suggests that a similar mechanism may be responsible for the stabilisation and cut-off behaviour. To investigate this we examine the linear energy equation, derived in modal form from the Reynolds-Orr energy equation. This yields the following two identities:
\begin{eqnarray}
c_i
&=&
\frac{\langle (\phi_r D\phi_i - \phi_i D\phi_r ) Du \rangle - \displaystyle{\frac{1}{\alpha R}}[ I_2^2 + 2 \alpha^2 I_1^2 + \alpha^4 I_0^2 ] }{I_1^2 + \alpha^2 I_0^2 } , \label{ci-identity} \\
c_r &=& \frac{\langle (\alpha^2 |\phi|^2 + |D\phi|^2 ) u \rangle + \displaystyle{\frac{R_{inj}}{\alpha R}}
\langle \alpha^2 (\phi_r D\phi_i - \phi_i D\phi_r ) + (D\phi_r D^2\phi_i - D\phi_i D^2\phi_r ) \rangle}
{I_1^2 + \alpha^2 I_0^2 } \nonumber \\[-1ex] \label{cr-identity}
\end{eqnarray}
where $I_k = I_k(\phi)$ is the semi-norm defined by:
\[ I_k = \left[ \int_{-1}^{1} |D^k \phi|^2~\dd y \right]^{1/2}, ~~k=0,1,2, \]
and where
\[ \langle f \rangle = \int_{-1}^{1} f(y)~\dd y . \]
Before proceeding further, we note that $R_{inj}$ only appears indirectly in (\ref{ci-identity}), reinforcing the assertion that for order unity $R_{inj}$, the principle contribution to stability of  injection is via the mean flow. Indeed, in the long wavelength limits of cut-off $k$ that we have studied, we have found values $\lambda = (\alpha R)^{-1} \lesssim 10^{-4}$ for instability. Thus, in (\ref{cr-identity}) the term directly involving $R_{inj}$ has minimal effect on $c_r$, explaining the
observations in Fig.~\ref{fig:Comparison}(a).

The identity (\ref{ci-identity}) can also be interpreted as an energy equation, in form:
\begin{equation}
\frac{d}{dt}\left\langle T_{1}\right\rangle =\left\langle T_{2}\right\rangle -\frac{1}{R}\left\langle T_{3}\right\rangle \label{eq:ReynoldsOrr}
\end{equation}
where
\begin{eqnarray}
T_{1} & = & 0.5 \left(\left|D\phi\right|^{2}+\alpha^{2}\left|\phi\right|^{2}\right), ~~~~\frac{\dd}{\dd t}T_{1} = \alpha c_{i} T_{1} ,\\
T_{2} & = & 0.5 \alpha \tau Du, ~~~~ \tau = \phi_r D\phi_i - \phi_i D\phi_r, \\
T_{3} & = &0.5(\left|D^{2}\phi\right|^{2}+2\alpha^{2}\left|D\phi\right|^{2}+\alpha^{4}\left|\phi\right|^{2}) .
\end{eqnarray}
The left-hand side of (\ref{eq:ReynoldsOrr}) represents the temporal variation of the spatially
averaged (one wavelength) kinetic energy. The first term on the right-hand side of (\ref{eq:ReynoldsOrr}) is the exchange of energy between the base flow and the disturbance. The last term, $\left(\frac{1}{R}\left\langle T_{3}\right\rangle \right)$, represents the rate of viscous dissipation. At criticality, the two terms on the right-hand side balances each other, but the spatial distributions of $T_2$ and $T_3/R$ indicate where the energy is generated and dissipated in the channel.

\begin{figure}
\subfloat[]{\includegraphics[scale=0.0515]{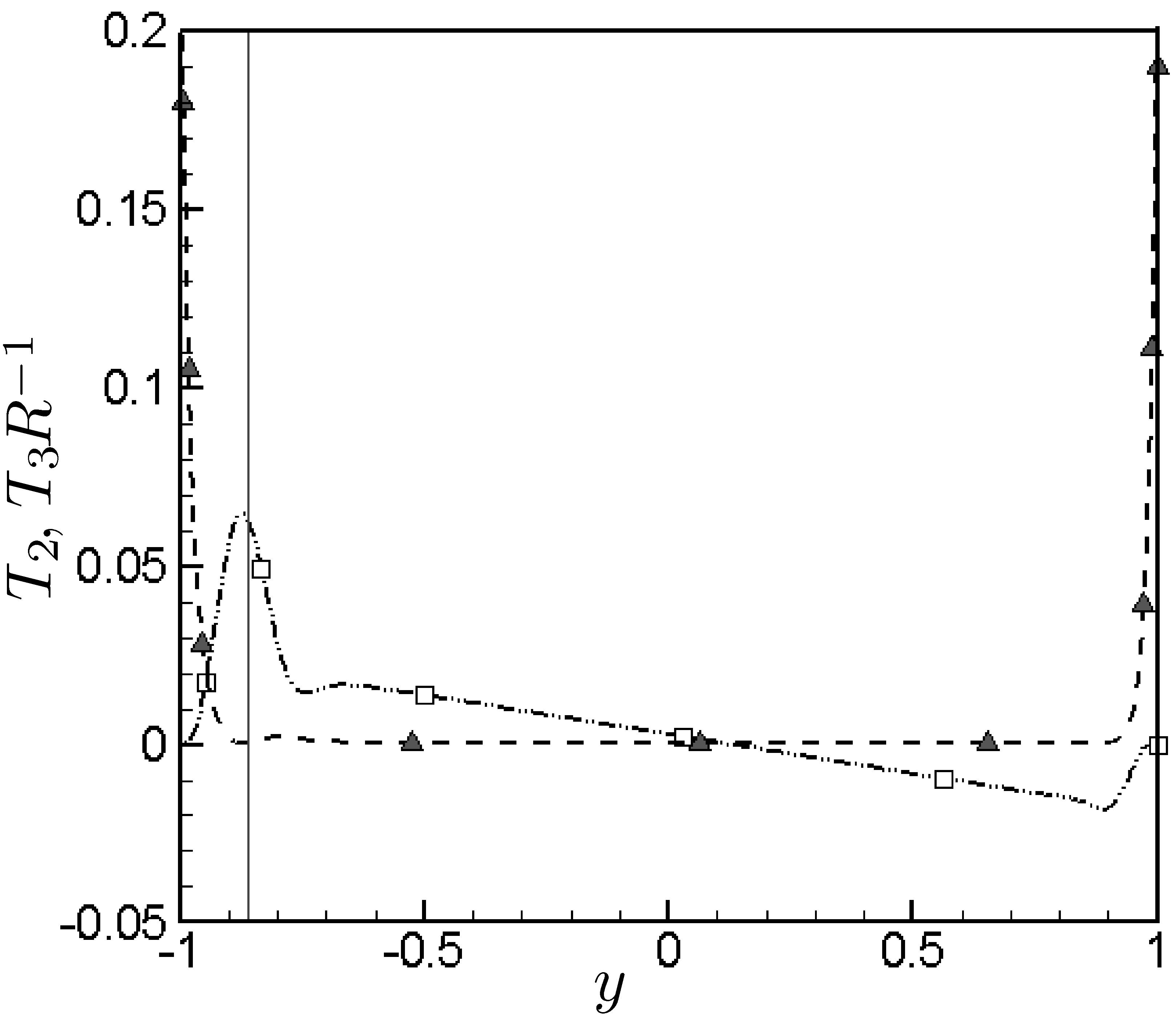}
}~\subfloat[]{\includegraphics[scale=0.05]{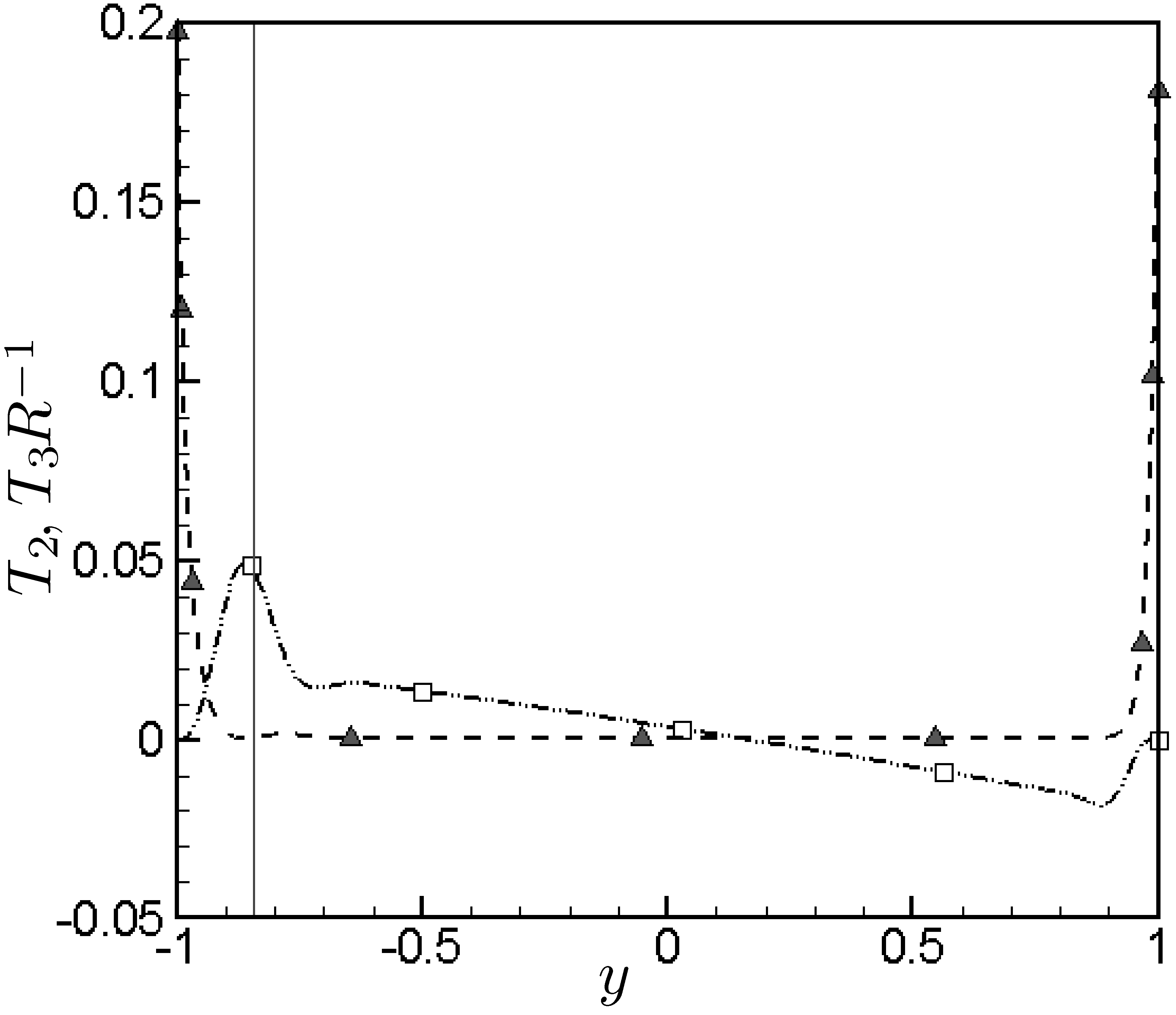}
}\\\subfloat[]{\includegraphics[scale=0.05]{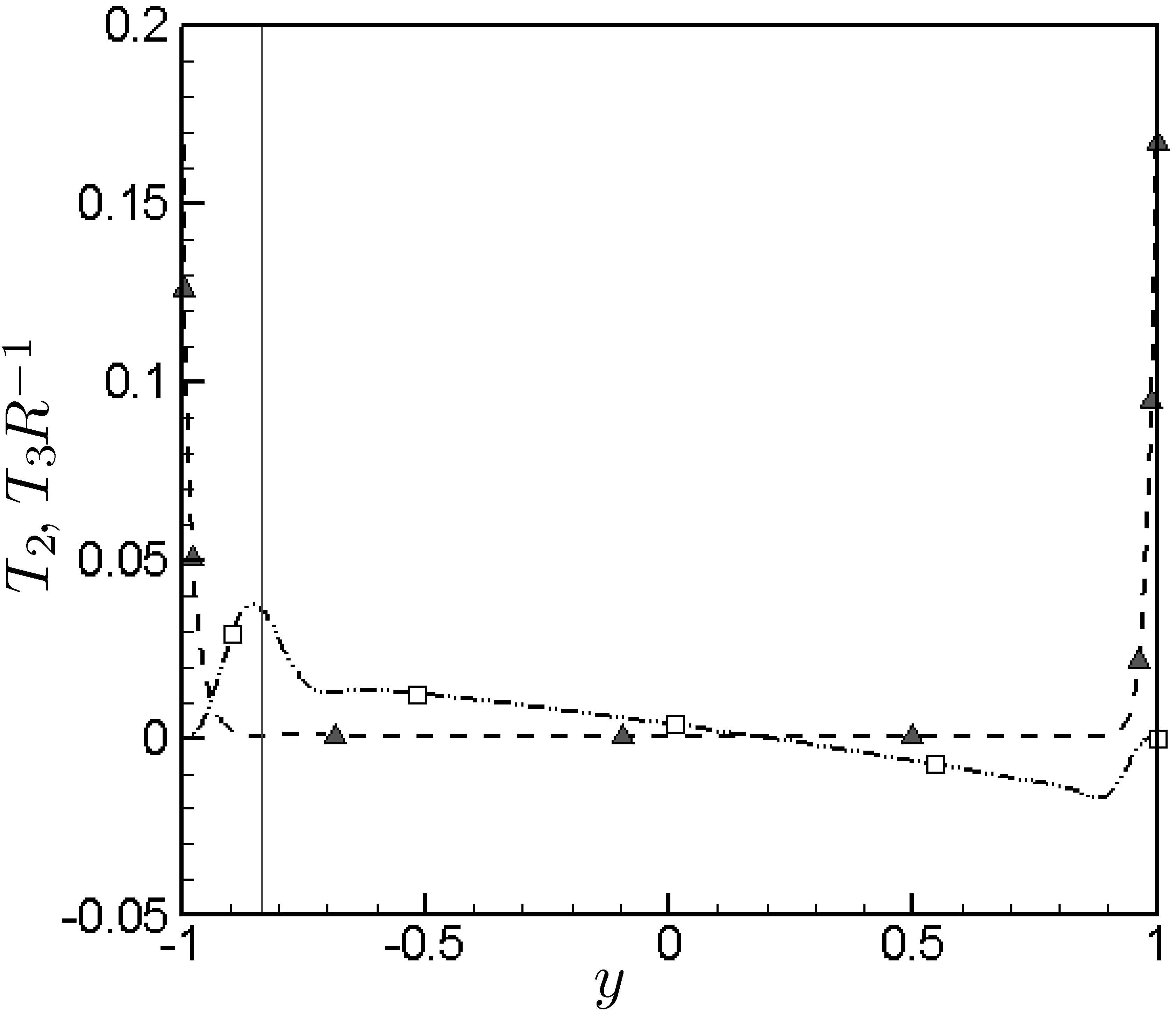}
}~\subfloat[]{\includegraphics[scale=0.05]{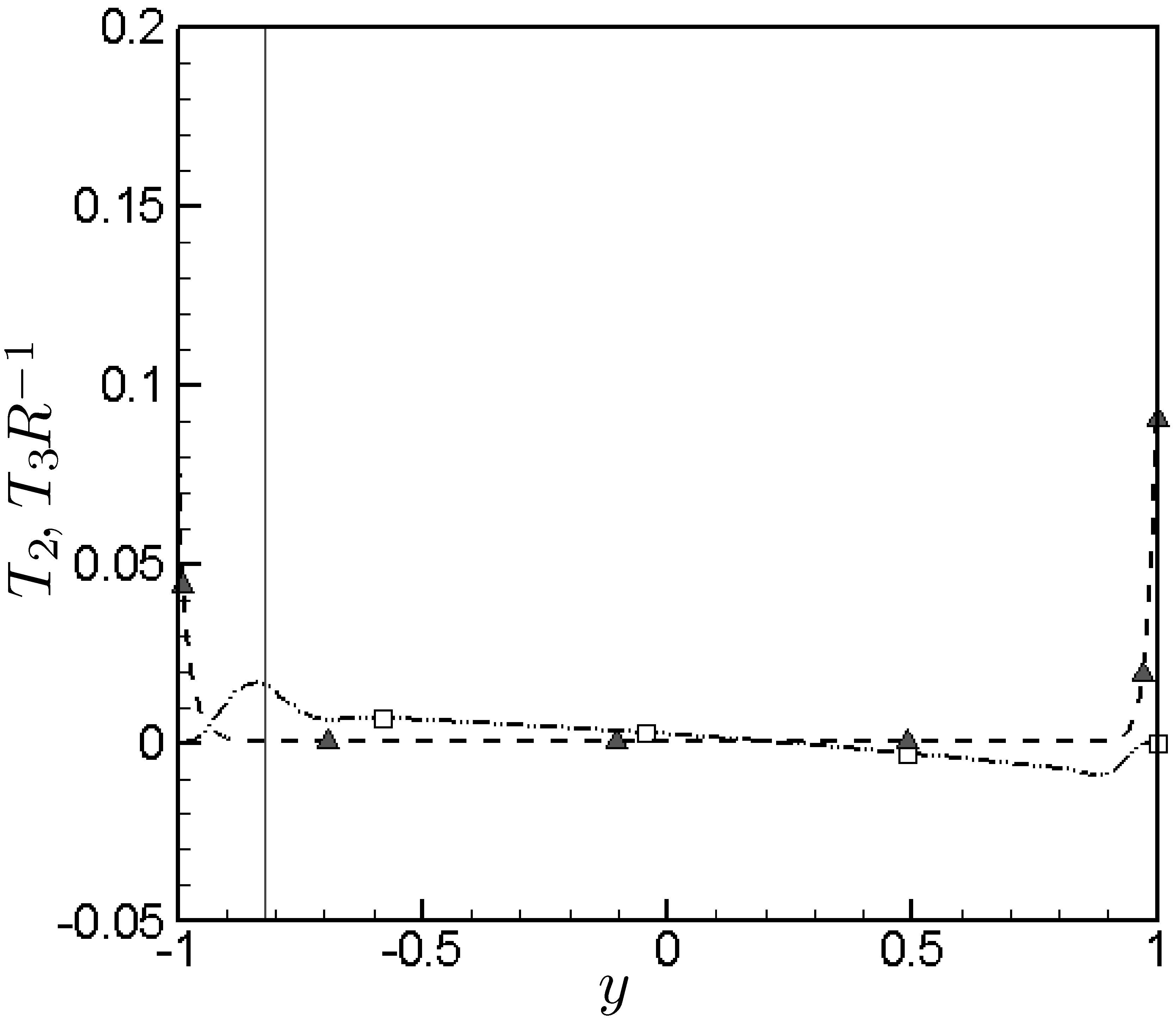}
}
\caption{Distribution of energy production~$(T_{2})$~and dissipation~$(\frac{1}{R}T_{3})$~terms
across the domain corresponding to criticality at  $R_{inj}$=~(a)~$0$,~(b)~$0.2$,~(c)~$0.4$~and~(d)~$0.6$.
In all the cases, $k=0.5$. Dash-dot-dot line with symbol $\square$ represents $T_{2}$,~dashed line with  filled $\triangle$ represents~$\frac{1}{R}T_{3}$
and solid vertical line represents the location of the critical layer.}
\label{fig:Critenergy}
\end{figure}

\cite{Sadeghi1991} extensively utilised this linear energy approach in studying the effect of $k$ on stability of ACP flow. They found that increase in the value of $k-c_{r,crit}$ decreases the Reynolds stress $(\tau)$ near the moving wall until it becomes negative, hence stabilising. The critical layer near the moving wall vanishes for $k > c_{r,crit}$ and as $k$ increases the Reynolds stress becomes progressively negative within the critical layer at the fixed wall, but this behavior is destabilising since the velocity gradient is negative there for ACP flow.

Figures~\ref{fig:Critenergy}(a)-(d) examine the distribution of $T_2$ and $T_3/R$ for the least stable eigenmode for the parameters listed in Table~\ref{tab:at-criticality}, i.e.~we fix $k=0.5$ and increase $R_{inj}$ up to $R_{inj}=R_{inj,1} \approx 0.6$. The critical layer is marked with a vertical line. We observe that both the rate of energy transfer and the rate of viscous dissipation decrease with the cross-flow. Without cross-flow, $T_{2}$ is positive and negative respectively in the lower (injection) and upper (suction) halves of the domain. Increasing the cross-flow decreases both the positive (near injection wall) and negative (near suction wall) peaks. The location of the critical layer also moves away from the injection wall due to the skewing of the velocity profile. When $R_{inj} \approx R_{inj,1}$, $\left\langle T_{2}\right\rangle $ and $\frac{1}{R}\left\langle T_{3}\right\rangle $ not only equalize but (since $\phi$ has been normalised), will have magnitudes $O(\alpha^{-1})$ since $\alpha R=$ constant at cut-off; (see also \cite{Sadeghi1991}). This reduced energy budget as $R_{inj} \approx R_{inj,1}$. This is the primary reason for the cut-off.

\subsection{Summary}

For the range of small to order unity $R_{inj}$ with $k \geq c_{r,crit}$, the flow instability is dominated by long wavelength perturbations. This instability mchanism exhibits a cut-off phenomenon characterised by a near linear boundary in the $(R_{inj},k)$-plane. The initial cut-off mechanism is very similar to that for ACP, as studied by \cite{Sadeghi1991}, combining skewing of the velocity profile, shifting of the critical layer and decay of the net perturbation energy.

\section{Intermediate $R_{inj}$ and short wavelength instabilities}
\label{sec:medium}

\begin{figure}
\subfloat[]{\includegraphics[scale=0.06]{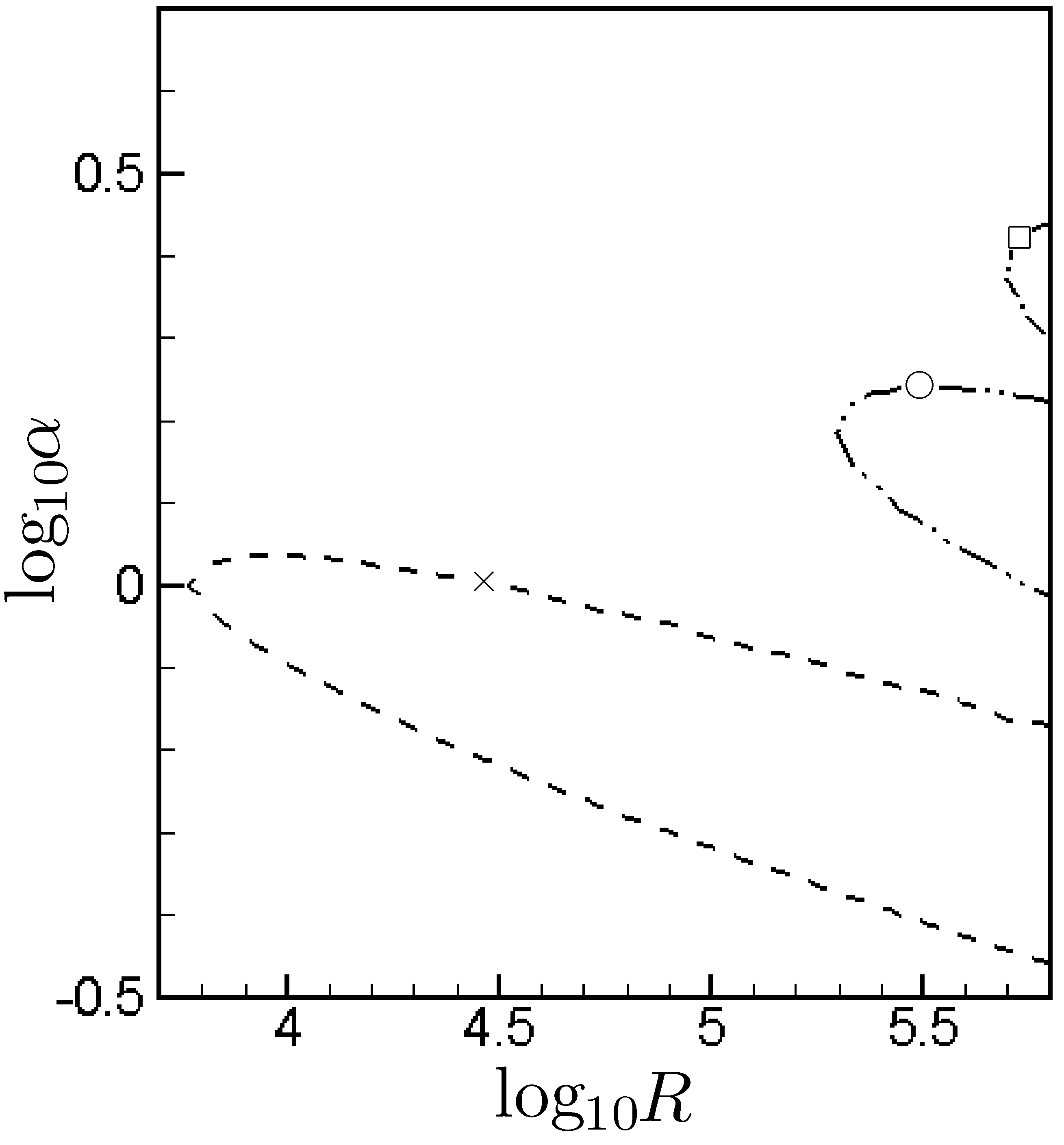}

}~~~~~~\subfloat[]{\includegraphics[scale=0.059]{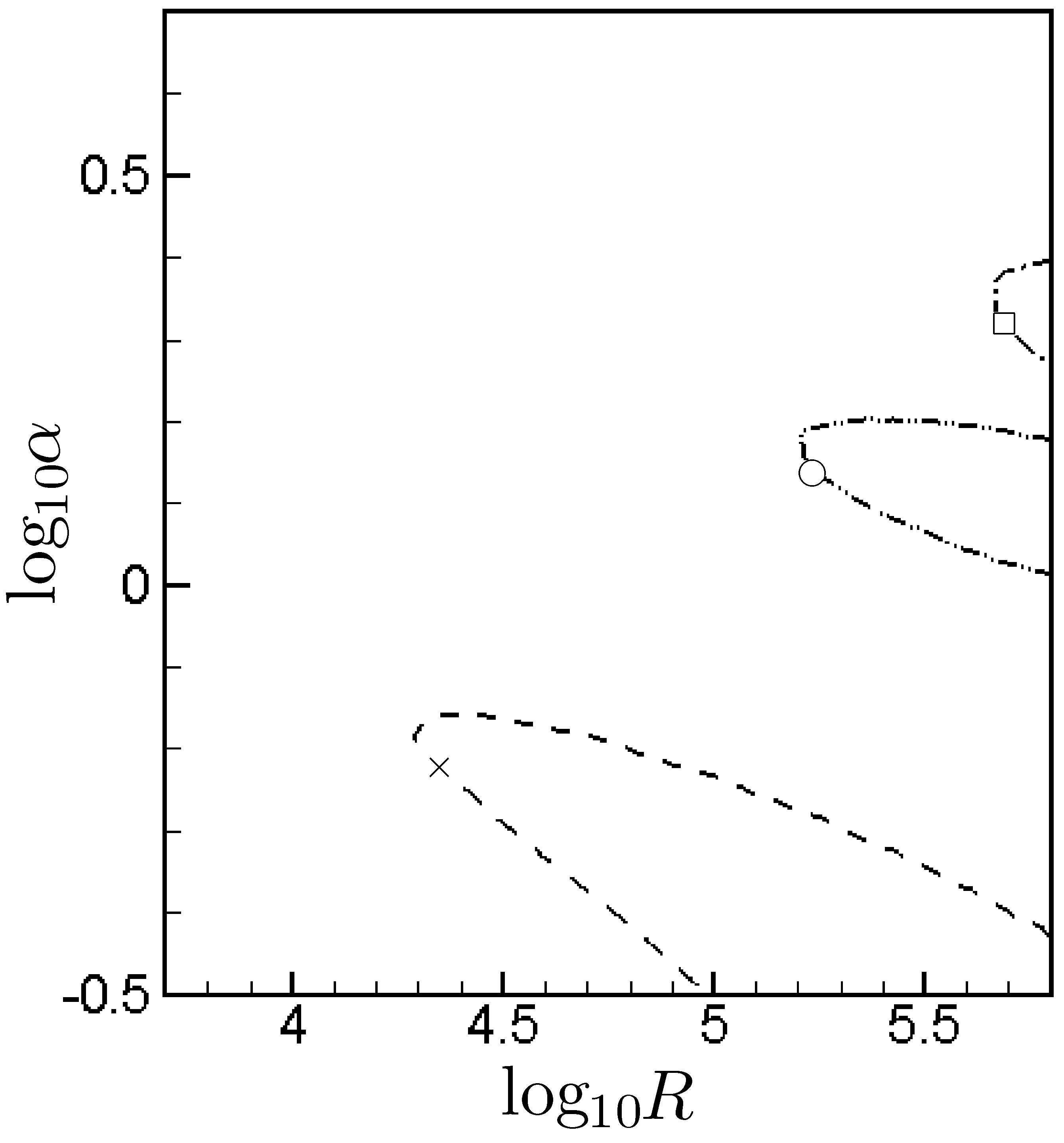}

}

\caption{Neutral Stability Curves~(NSCs)
for (a) $k=0$ and (b) $0.18$ at different $R_{inj}$. The symbols indicate different values of $R_{inj}$ and are as follows: $\times\rightarrow R_{inj}=0$,~$\circ\rightarrow R_{inj}=6$~in (a) and $4$ in (b),~$\square\rightarrow R_{inj}=12$~in (a) and $8$ in (b)
\label{fig:Neutral-Stability-Curves}
}
\end{figure}

We now consider the range $0\leq k\leq c_{r,crit}$, in which the critical layer at the upper wall is still present. We investigate its stability characteristics by adding cross-flow of intermediate strength ($0\leq R_{inj} \lesssim 21$), avoiding for the moment the second transition.
It is intuitive that the presence of the critical layer will affect the stability behavior. To verify this we have studied the two extremities of the range of $k$ considered, i.e.~$k=0$ (PP flow) and $k=0.18$. The respective NSCs are shown in Fig.~\ref{fig:Neutral-Stability-Curves}. It is evident that the presence of the critical layers render shorter wavelength modes unstable. Yet, it is also observed that with $R_{inj}$ in this intermediate range, the  stability increases dramatically.

We have been unable to make any advance analytically in this range of $R_{inj}$, and therefore have proceeded numerically. First we note that when we have considered $k \gtrsim 0.19$ for the range of $1.3 < R_{inj} <21$, we have found that the least stable modes are long wavelength modes and that these are linearly stable. Thus, $k \gtrsim 0.19$ appears to represent an absolute cut-off in this range of $R_{inj}$.

\begin{figure}
~~~~~~~~~~~~~~~~~~~~~~~~~~~~\includegraphics[scale=0.08]{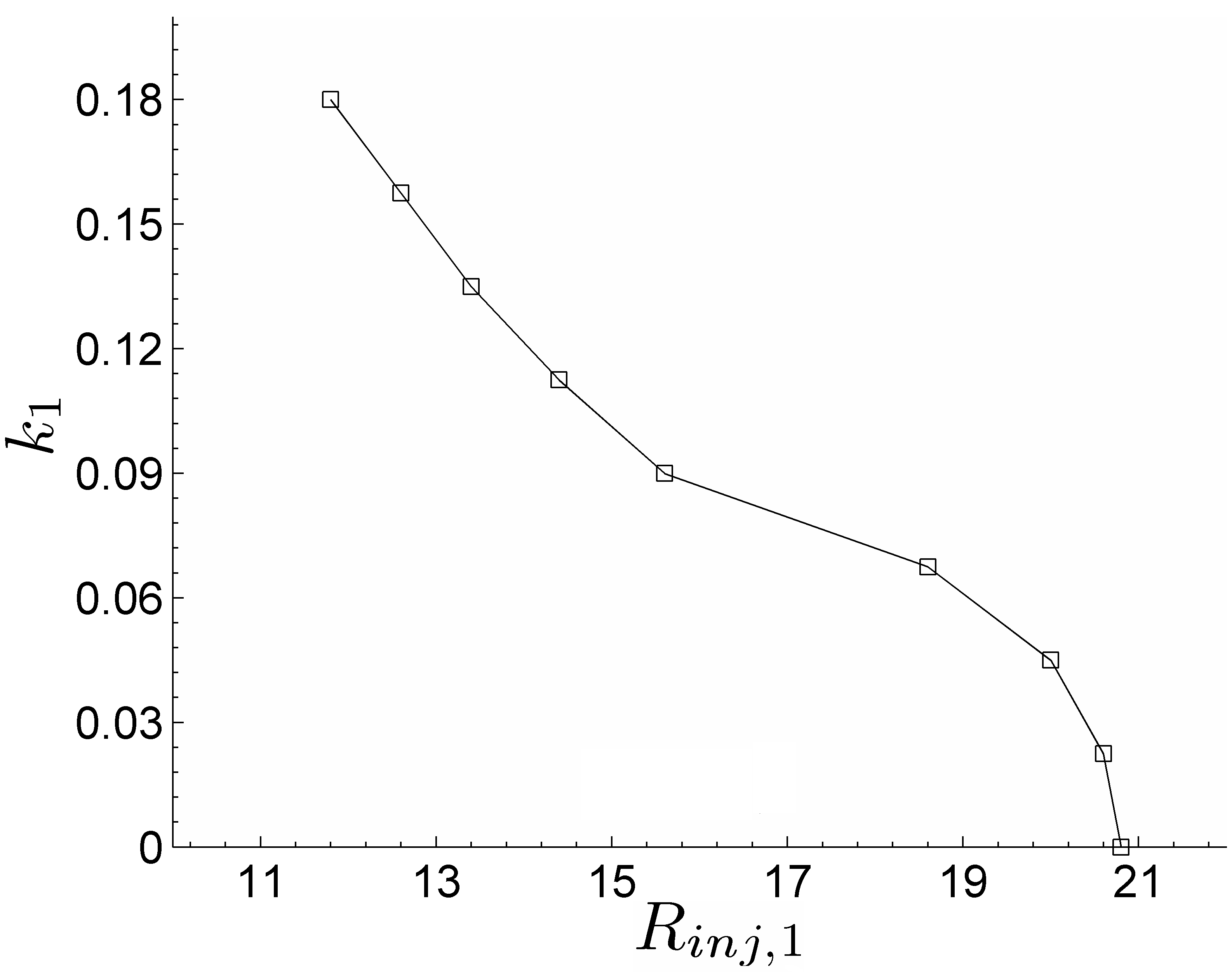}

\caption{~Shorter wavelength cut-off showing $k_{1}$ as a function of $R_{inj,1}$.~The flow is  linearly stable for $R\le 10^{6}$ above the curve.~The values in Table~\ref{tab:shortwave} are marked by $\square$.}
\label{fig:shortwave}

\end{figure}

For smaller $k$ we have seen that the NSC's occur with wavenumbers that are $O(1)$ and apparently increasing with $R_{inj}$. Unlike the long wavelength problem, the asymptotic behaviour along the branches of the NSC's is not easily treated. At fixed large $R$, we are able to compute numerically a cut-off value of $k$ for increasing $R_{inj}$, i.e.~$k = k_1(R_{inj},R)$. These cut-off curves do lie below $k \sim 0.19$, but are not wholly independent of $R$, at least within the range of $R$ up to which our numerical code is reliable, i.e.~it is quite possible that these asymptote to a cut-off curve as $R \to \infty$, but we cannot reliably evaluate this limit numerically.
As an example of this numerical cut-off, (at $R = 10^{6}$), we have computed the cut-off values $R_{inj,1}$, as listed in Table~\ref{tab:shortwave} and shown in Fig.~\ref{fig:shortwave}(a). For the range $1.4 < R_{inj} <11.8$, the cut-off is close to $k \sim 0.19$.

\begin{table}
~~~~~~~~~~~~~~~~~~~~~~~~~~~~~~~~~~~~~~~~~~\begin{tabular}{|c|c|c|c|}
\hline
$k_{1}$ & $R_{inj,1}$ & $\alpha_{crit}$\tabularnewline
\hline
\hline
$0$ & $20.8$ & $3.5227$\tabularnewline
\hline
$0.0225$ & $20.6$ & $3.7458$\tabularnewline
\hline
$0.0450$ & $20.0$  & $3.9381$\tabularnewline
\hline
$0.0675$ & $18.6$  & $3.9831$\tabularnewline
\hline
$0.0900$ & $15.6$  & $3.4146$\tabularnewline
\hline
$0.1125$ & $14.4$  & $3.2112$\tabularnewline
\hline
$0.1350$ & $13.4$  & $3.2112$\tabularnewline
\hline
$0.1575$ & $12.6$  & $3.2112$\tabularnewline
\hline
$0.1800$ & $11.8$ & $3.2112$\tabularnewline
\hline
\end{tabular}\caption{~Cut-off values evaluated for shorter wavelength instabilities for $R=10^{6}$.\label{tab:shortwave}}

\end{table}

Although we see that the unstable wavenumbers increase with $R_{inj}$ in Fig.~\ref{fig:Neutral-Stability-Curves}, note that asymptotically as $\alpha \to \infty$ the short wavelengths are stable. To see this, from (\ref{ci-identity}) we bound
\[ \langle (\phi_r D\phi_i - \phi_i D\phi_r ) Du \rangle \leq |Du|_{max} I_0 I_1 \leq 0.5 |Du|_{max} [\alpha I_0^2 +  I_1^2/\alpha], \]
so that $c_i < 0$ provided that:
\begin{equation}
R < \frac{ |Du|_{max} }{2 \alpha^2} ,
\end{equation}
(and better bounds are certainly possible). In Table~\ref{tab:shortwave} we note that the maximal critical wavenumber is in fact attained at an intermediate $R_{inj}$.

\subsection{Behaviour of preferred modes for intermediate $R_{inj}$.}

In our preliminary results, (\S \ref{sec:preliminary}), we saw that at fixed values of $(R,k,\alpha)$, increasing the $R_{inj}$ led to regimes of stabilisation, then destabilisation, and then finally stabilisation. For $k\geq c_{r,crit}$, only long wavelengths appear unstable and how the cut-off values of $k$ and $R_{inj}$ vary in this regime are illustrated in Fig.~\ref{fig:Cut-off-Velocity}. For the lower range of $k$, our results are primarily numerical, indicating a cut-off value $k \approx 0.19$ for $1.3 \lesssim R_{inj} \lesssim 11.8$ and then with decaying cut-off $k$ for $11.8 \lesssim R_{inj} \lesssim 20.8$, as illustrated in Fig.~\ref{fig:shortwave}. Therefore, we have linear stability as we cross some cut-off frontier, $k > k_1(R_{inj})$ in the $(R_{inj},k)$-plane, (alternatively for $R_{inj} > R_{inj,1}$).

\begin{figure}
\subfloat[]{\includegraphics[scale=0.055]{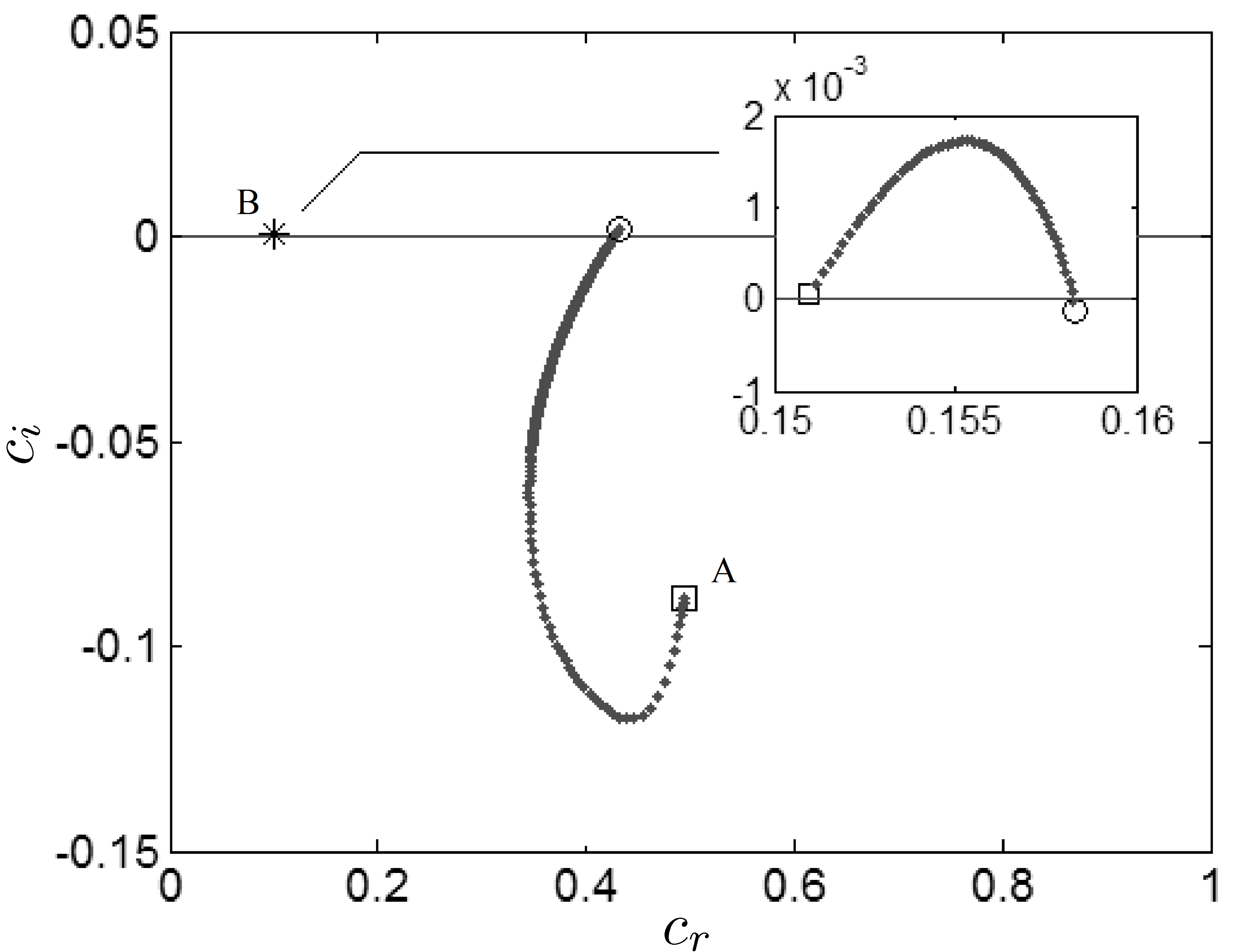}

}~\subfloat[]{\includegraphics[scale=0.055]{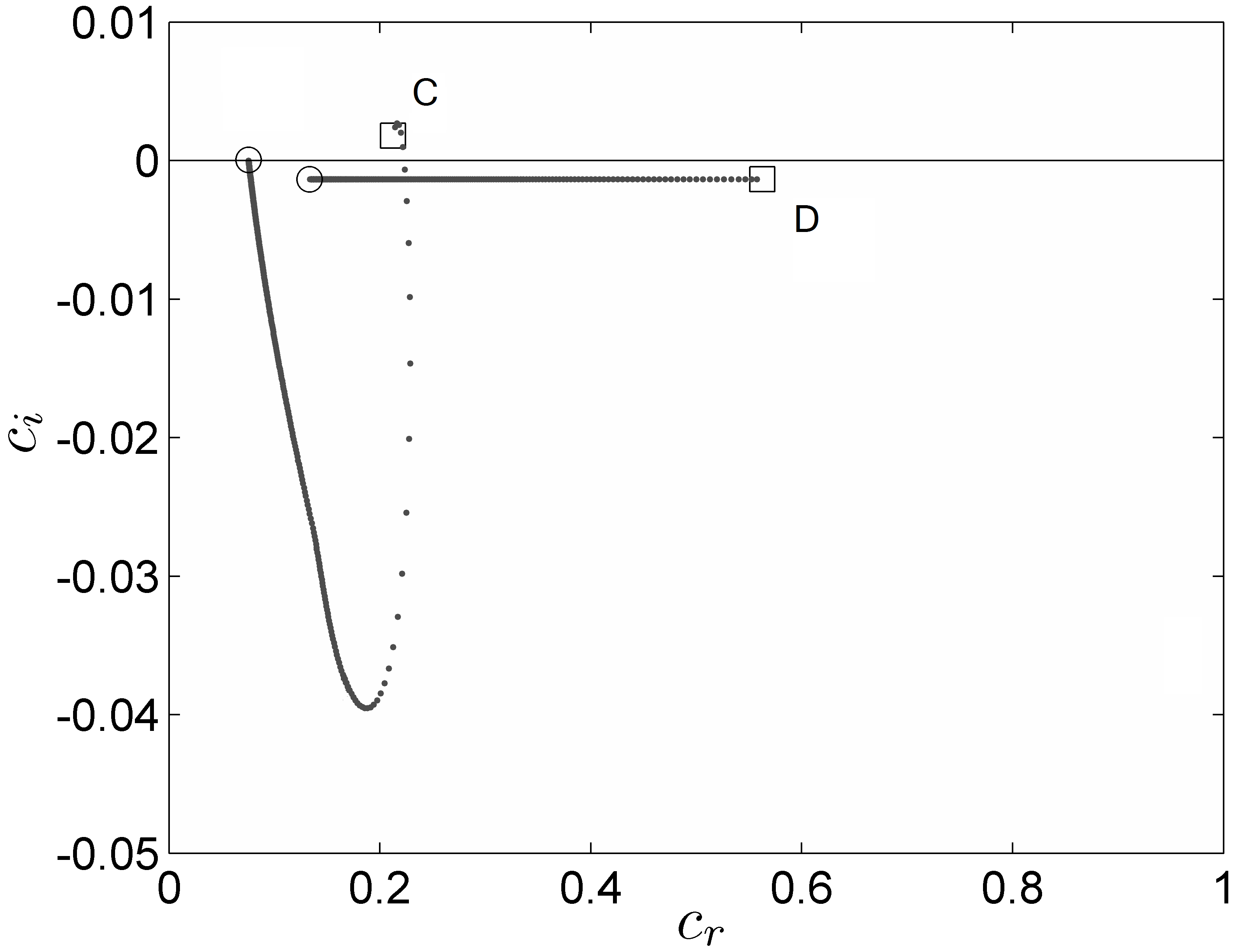}

}

\caption{~Behavior of preferred modes (belonging to different wavelengths and denoted by alphabets {}`A'-{}`D') under the influence of cross-flow with $R=10^{6}$.~Symbols $\square$ and $\circ$ respectively imply the starting and the ending position of the preferred mode in the $c_{i},c_{r}$ plane, whereas the dots~({}`.') trace the locus. The difference in $R_{inj}$ between consecutive dots is $0.1$.~(a)~$k=0$. Mode {}`A' has $\alpha=0.001$ and is traced for  $R_{inj}$=[$0$,$25$].~Mode {}`B' has $\alpha=3.5227$ and is traced for $R_{inj}$=[$15$,$21$]~(shown in the inset), the position at $R_{inj}=15$ is marked by {}`\textasteriskcentered{}'.~(b)~$k=0.5$.~Mode {}`C' has $\alpha=0.01$ and is traced for  $R_{inj}$=[$0$,$30$].~Mode {}`D' has $\alpha=2.5$ and is traced for  $R_{inj}$=[$7$,$30$].}
\label{fig:preferredemodes}
\end{figure}

We now consider what happens to the certain eigenmodes (preferred modes) as we extend the injection cross-flow up until the second critical $R_{inj}$. Our analysis up to now suggests that the behaviour may be different depending on whether we consider small or moderate $k$. In Fig.~\ref{fig:preferredemodes}, we have plotted the locations of certain eigenmodes as $R_{inj}$ is increased, by keeping the Reynolds number $R$ constant at $10^{6}$. This gives us some idea of how cut-off behaviour changes with $R_{inj}$. Although the ``preferred modes'' are simply those we have selected, we implicitly mean modes that are involved in the transition from stable to unstable as one of our dimensionless parameters is varied (here $R_{inj}$), i.e.~at some point a preferred mode becomes the least stable mode and then unstable.

Figure~\ref{fig:preferredemodes}(a) shows two eigenmodes corresponding to $k=0$, (PP flow). A least stable long wavelength mode is tracked for $\alpha=0.001$, denoted by `A'. This mode is stable at $R_{inj} = 0$ and its stability increases further as $R_{inj}$ increases up to around $1.7$. However, further increases in $R_{inj}$ destabilise this mode progressively until it becomes unstable at $R_{inj}=25$. In the inset of Fig.~\ref{fig:preferredemodes}(a) we have also plotted the least stable short wavelength mode at $\alpha=3.5227$. Such modes become unstable only under the influence of cross-flow of intermediate strength. This particular mode, (denoted by `B'), starts becoming unstable approximately when $R_{inj}>15$, but recovers stability later for $R_{inj}\geq 20.8$. This behavior is  a direct consequence of the trajectory of the NSCs observed in  Fig.~\ref{fig:Neutral-Stability-Curves}(a). The preferred mode `B' is the critical mode at cut-off, (see Table~\ref{tab:shortwave}). Thus PP flow with cross-flow is unconditionally linearly stable in the range $ 20.8\leq R_{inj}\lesssim 25$.

For larger $k$, the stability behavior is primarily governed by the long wavelength modes, as shown in Fig.~\ref{fig:preferredemodes}(b) for $k=0.5$. The least stable mode corresponding to $\alpha=0.01$ is unstable for $R_{inj} = 0$, denoted mode `C'. This viscous mode becomes stable when $R_{inj}$ increases to $0.6$, which is indeed  the cut-off value, i.e.~$R_{inj,1}$. This is expected, according to Table~\ref{tab:1}. Mode `C' is weakly damped and its stability increases for $R_{inj} \lesssim 3$, after which it starts destabilising. The mechanism of this destabilisation can probably be analysed along the lines of resonant interactions of the Tollmien-Schlichting~(T-S) waves; see \cite{Baines1996}. To show this interaction, we have traced the locus, (for $R_{inj}=[7,30]$), of the least stable inviscid short wavelength mode `D', at $\alpha=2.5$. This mode, being inviscid, remains stable but has $c_i$ very close to zero as $R_{inj}$ increases. The wave speed $c_{r}$ decreases continuously with $R_{inj}$ for mode `D'. The resonant interaction takes place when its wave speed matches with that of mode `C', which signals the destabilisation of mode `C'. This destabilisation continues until mode `C' becomes unstable when $R_{inj}\gtrsim30$.

\begin{figure}
\includegraphics[scale=0.09]{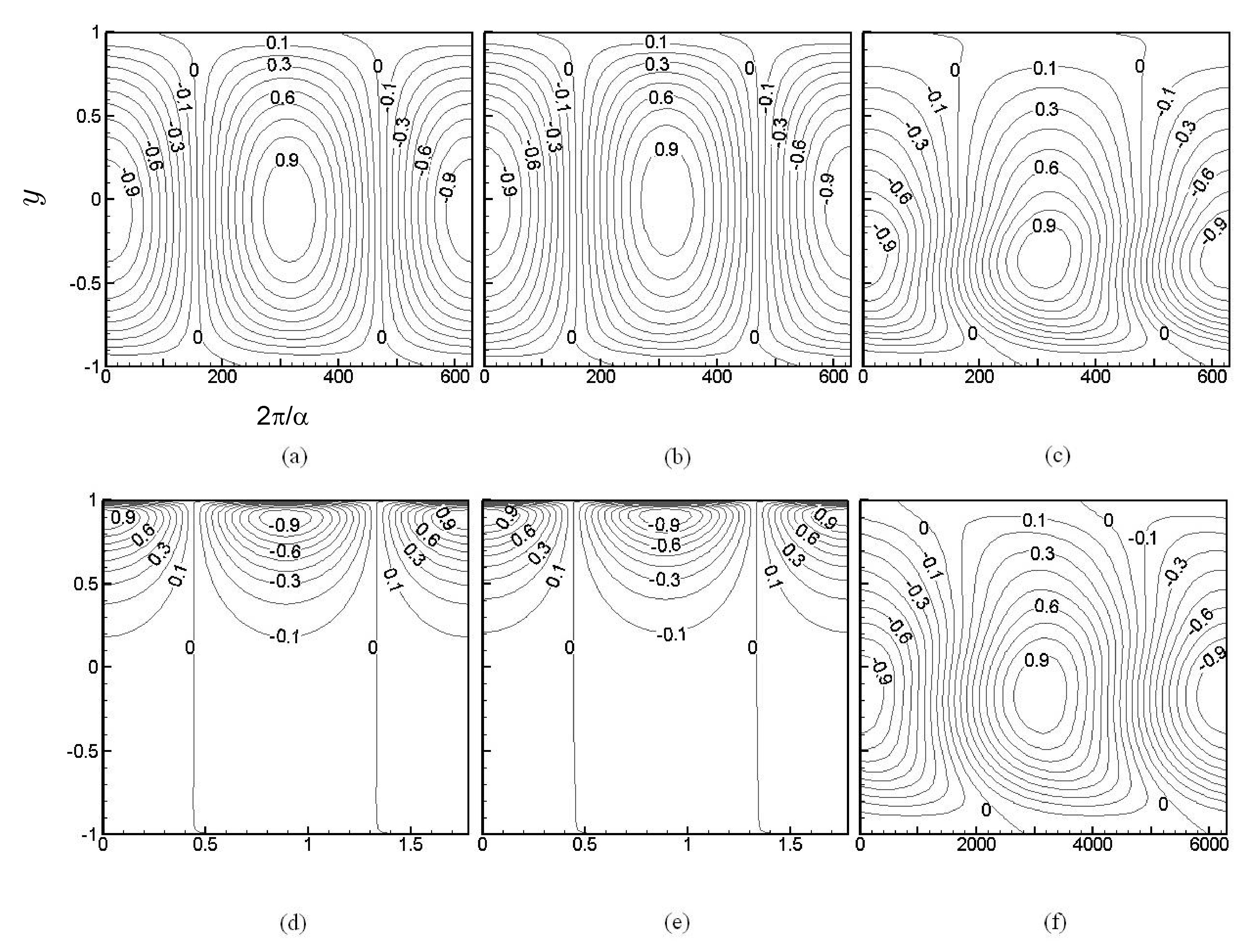}
\caption{~Isovalues of the normalised perturbation stream functions ($\hat{\psi}$) for the preferred modes at $R=10^{6}$ under different $R_{inj}$. The streamwise extent of the domain is one wavelength. Corresponding to $k=0.5$, the long wavelength mode {}`C' is shown for (a) $R_{inj}=0.1$ (unstable), (b) $R_{inj}=1$ (stable) and (c) $R_{inj}=30$ (unstable). 
Corresponding to $k=0$, $\hat{\psi}$ for two different preferred modes, viz. {}`A' and {}`B' are shown.  The  shorter wavelength mode {}`B' ($\alpha=3.5227$) is shown for (d) $R_{inj}=15$ (unstable) and (e) $R_{inj}=21$ (stable). The longer wavelength mode {}`A' ($\alpha=0.001$) is shown for (f) $R_{inj}=25$ (unstable).}
\label{fig:newfig}
\end{figure}

In Fig.~\ref{fig:newfig} we show examples of the stream function for the preferred modes, corresponding to various $k$ and $R_{inj}$ in the transitions of Fig.~\ref{fig:preferredemodes}. For the long wavelength mode {}`C', Figs.~\ref{fig:newfig}(a)-(c) show that strong $R_{inj}$ appears to skew the streamlines towards the lower wall. The same is true for the long wavelength mode {}`A'  under strong injection; see Fig.~\ref{fig:newfig}(f). On the contrary, Figs.~\ref{fig:newfig}(d)-(e) show that for large $R_{inj}$, the streamlines of the shorter wavelength  mode {}`B'  are skewed and localised towards the upper wall.

\section{Stability and instability at large $R_{inj}$.}
\label{sec:high}

We turn now to the transition to instability at $R_{inj,2}$ and then later to stabilising effects at very large $R_{inj}$. As observed in \S \ref{sec:preliminary}, the transition at $R_{inj,2}$ appears to be independent of streamwise  Reynolds number $R$ (see Fig.~\ref{fig:4}(b)) and occurs for all $k$. Although there is sensitivity to $k$, it is not very significant. Values of $R_{inj,2}$ are found for all  $k\in[0,1]$ and are in a fairly tight range  of $R_{inj} \sim 22-25$.

\begin{figure}
\subfloat[]{\includegraphics[scale=0.056]{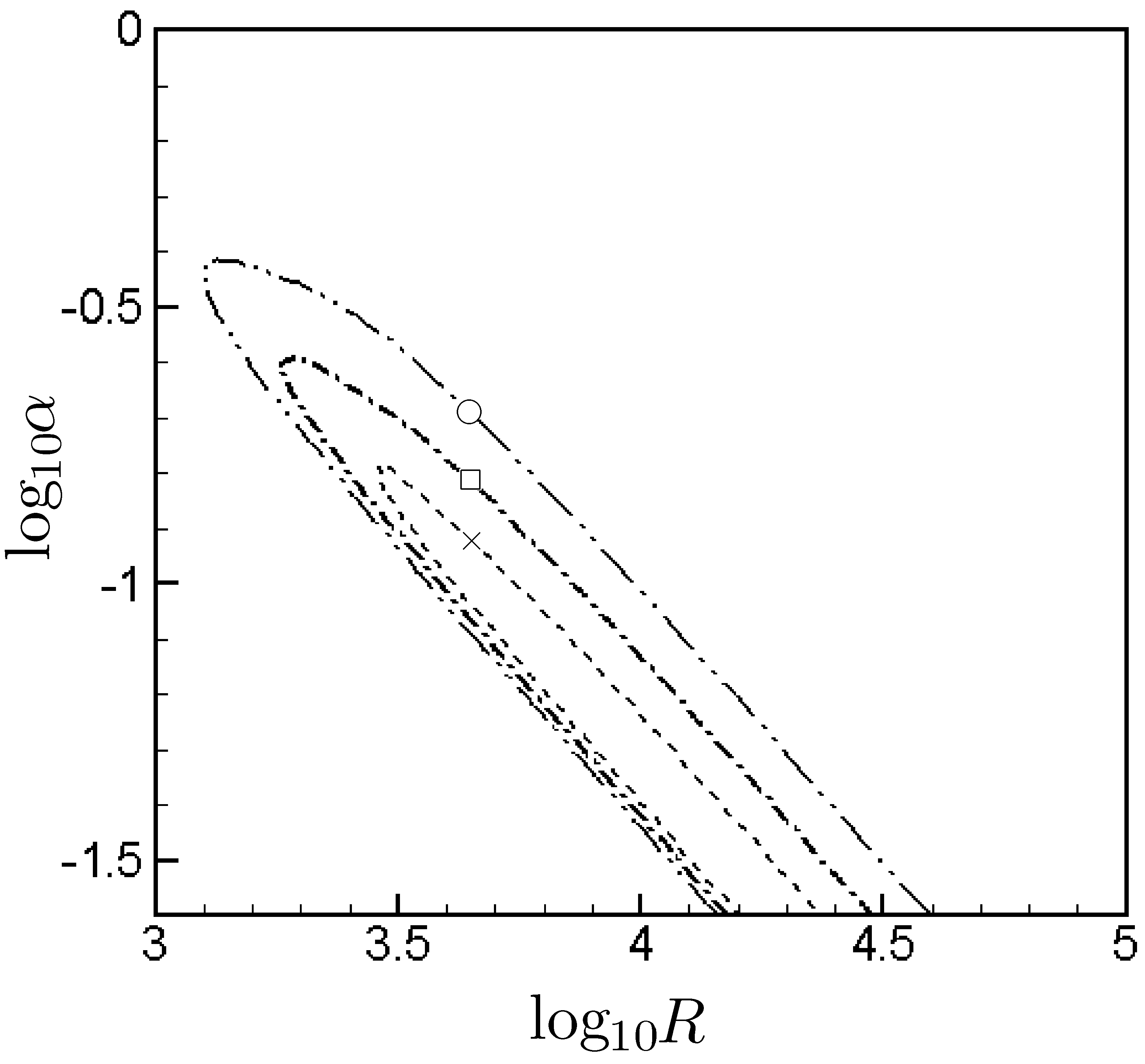}

}~\subfloat[]{\includegraphics[scale=0.055]{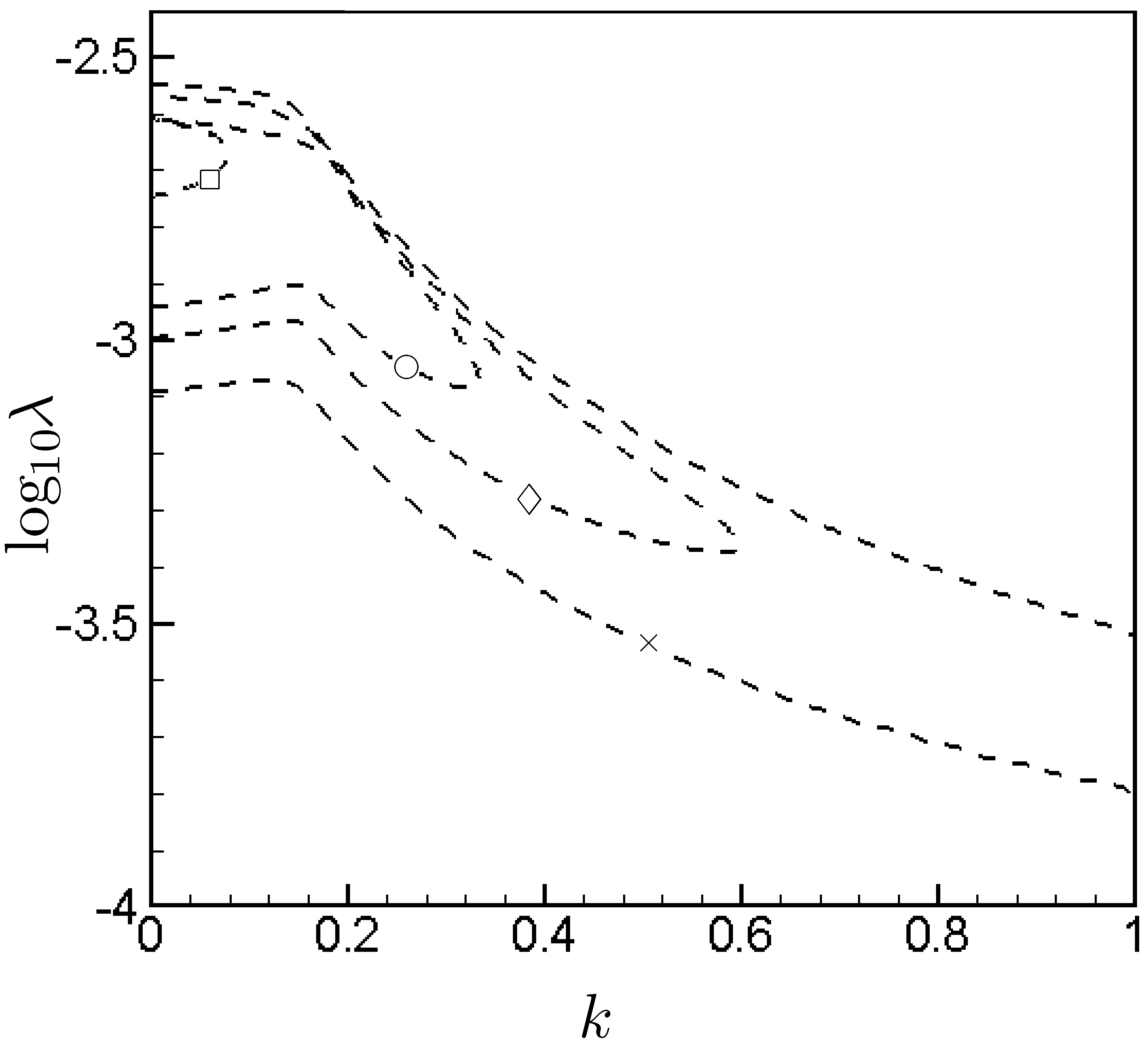}

}

\caption{(a)~NSC of PP flow ($k=0$) when $R_{inj}\rightarrow R_{inj,2}^{-}$.~The different values of $R_{inj}$ are $22.5$~(dashed line with $\times$),~$23$~(dash-dot line with $\square$)~and~$24$~(dash-dot-dot line with $\circ$).~Near cut-off, $\alpha R$ ~is constant along the upper and lower branches. (b)~Long wave NSCs showing the dependence
of $\log_{10}\lambda$ on $k$.~The different values of $R_{inj}$ are~$22.4$~($\square$),~$23.5$~($\circ$),~$24$~($\diamond$)~and~$25$~($\times$).~Cut-off is achieved over the entire range of $k$,
i.e. $[0,1]$.}
\label{fig:Rinj2}
\end{figure}

As suggested in the previous subsection, although instability at moderate $R_{inj}$ may be either short wavelength or long wavelength, according to $(k - c_{r,crit})$, as we approach $R_{inj,2}$ from below it is the long wavelengths that are unstable. Figure~\ref{fig:Rinj2}a shows the neutral stability curves corresponding to PP flow for $R_{inj}$ just above $R_{inj,2}$. The NSC's are nested with decreasing $R_{inj}$ and as we approach $R_{inj,2}$ the upper and lower branches of the NSC are seen to coalesce. The slope of the two branches suggests that $\alpha\sim R^{-1}$ in the limit of cut-off, and hence the previous long wavelength approximation, leading equation (\ref{eq:9}), should be effective for predicting the cut-off in the $(k,\lambda)$-plane; (recall $\lambda = (\alpha R)^{-1}$).

Figure~\ref{fig:Rinj2}b shows the NSC's obtained from long wave approximation. The cut-off velocity $k_{2}$ is the maximum value of $k$ encountered along the NSC for a given $R_{inj} = R_{inj,2}$. Unlike Fig.~\ref{fig:Long-wave-results}, the entire range of $k$ becomes unconditionally stable. For $R_{inj,1} < R_{inj}<22.2$, $c_{i}<0 ~\forall~ k\in[0,1]$. Another significant difference with Fig.~\ref{fig:Long-wave-results} and the results of \cite{CowleySmith1985} is that ``bifurcation from infinity'' is not observed as $k\rightarrow0$. This is possibly because the curves bifurcate from infinity for negative values of $k$, but we have not studied this range.
Finally, we mention that for cross-flow rates slightly greater than $R_{inj,2}$, the $R_{crit}$ is relatively low for the entire range of $k$. For example, $R_{inj,2} \approx23.8$ for $k=0.5$, (implying $R_{crit}\to \infty$ as $R_{inj} \to R_{inj,2}^-$). Increasing $R_{inj}$ to $25$ decreases $R_{crit}$ to around $6000$. Thus, on crossing $R_{inj,2}$ we find a dramatic decrease in the flow stability.

\subsection{Linear energy balance at $R_{inj,2}$}

\begin{figure}
\subfloat[]{\includegraphics[scale=0.055]{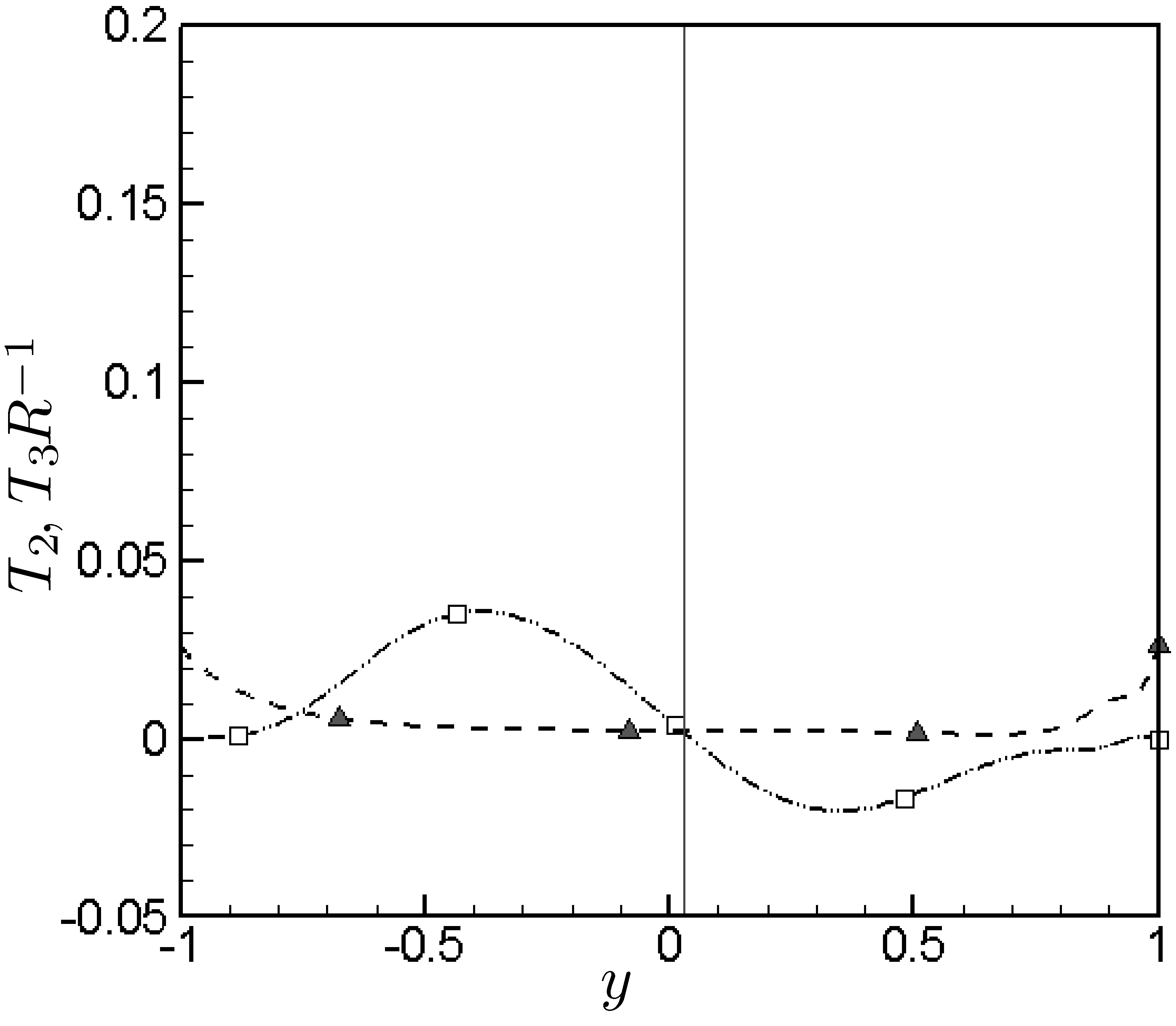}

}~\subfloat[]{\includegraphics[scale=0.055]{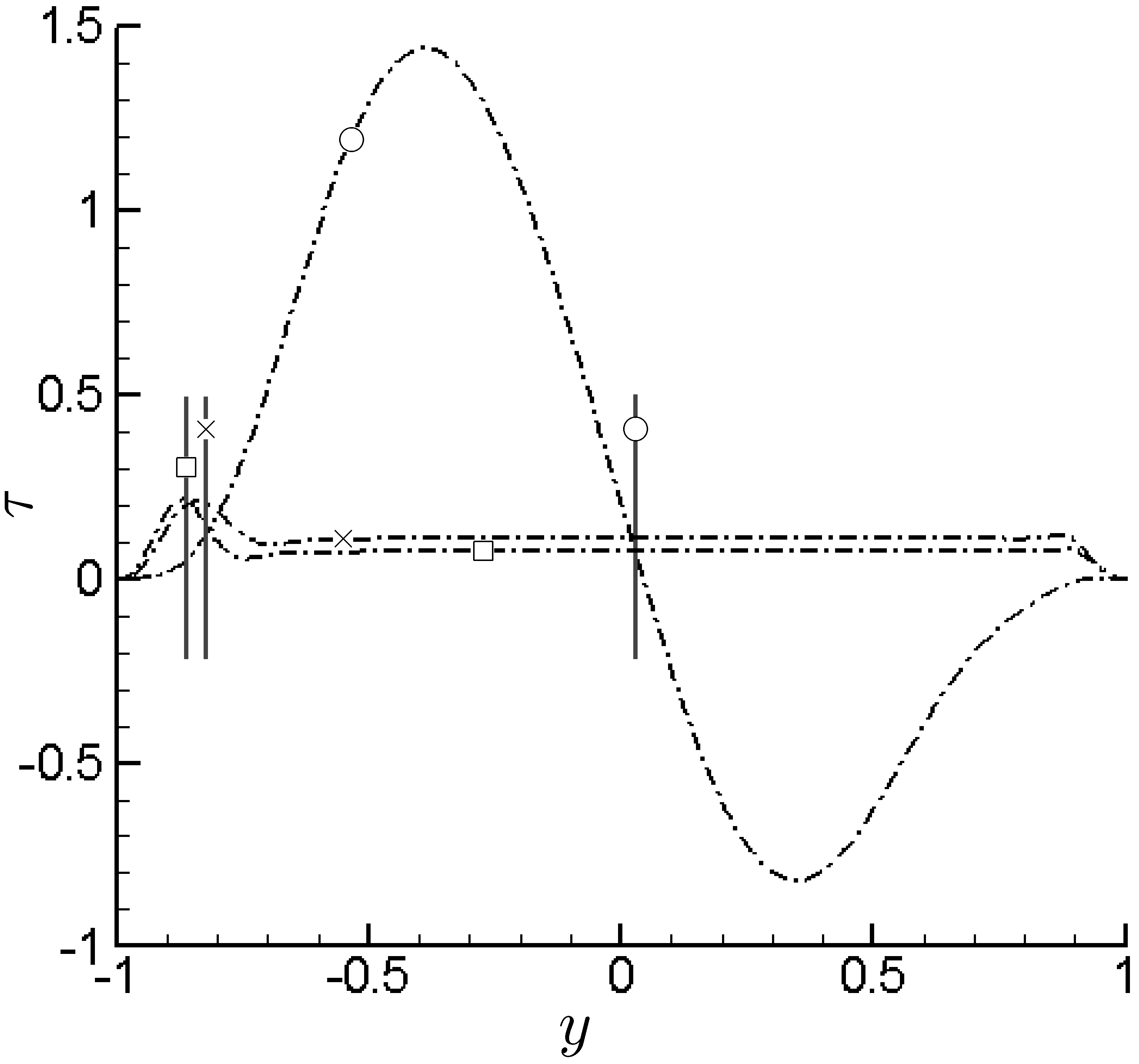}

}

\caption{~(a)~Distribution of energy production~$(T_{2})$~and dissipation~$(\frac{1}{R}T_{3})$~terms
across the domain corresponding to criticality at  $R_{inj}=25$.
Dash-dot-dot line with symbol $\square$ represents $T_{2}$,~dashed line with  filled $\triangle$ represents~$\frac{1}{R}T_{3}$
and solid vertical line represents the location of the critical layer.~(b)~Reynolds Stress $\tau$ distribution at criticality  for $R_{inj}=0$~(denoted by $\square$ symbol),~$R_{inj}=0.6$~(denoted by $\times$)~and~$R_{inj}=25$~(denoted by $\circ$). The location of the critical layers are shown by solid lines with corrosponding symbols.}
\label{fig:neutralmodesatRinj2}
\end{figure}

An interesting feature of transition at $R_{inj,2}$ is the independence with respect to $R$. With reference to the energy equation (\ref{eq:ReynoldsOrr}), this insensitivity implies that in this range $|T_2|$ is much larger than the viscous dissipation, $\frac{1}{R}T_3$. In other words, at criticality $c_i = 0$ is achieved by a balance of energy production and dissipation within $T_2$, more so than via balance with the viscous dissipation. Figure~\ref{fig:neutralmodesatRinj2}(a) investigates the energy budget at criticality for $k=0.5$ at $R_{inj}=25$. The critical parameters are observed to be $\left(\alpha_{crit},R_{crit}\right)=\left(0.31,6000\right)$. This implies that crossing the cut-off $R_{inj,2}$, there is a  transition from unconditional stability~($R_{crit}\rightarrow\infty$) to high instability($R_{crit}=6000$). Comparing with Fig.~\ref{fig:Critenergy}~(which shows energy distribution corresponding to criticality for $k=0.5$ and $R_{inj}\leq R_{inj,1}$), it is obvious that $T_{2}$ has a higher amplitude while the viscous dissipation $\frac{1}{R}T_{3}$ is weaker.

This behavior is due to  the generation of larger Reynolds stresses $\tau$, as $R_{inj}$ increases, as illustrated in Fig.~\ref{fig:neutralmodesatRinj2}(b). The dominance of $T_2$ over the viscous dissipation suggests that the critical layers have little to do with instability in this range. Note that $\tau$ is small in the critical layer, which has now moved towards the channel centre, and hence $T_{2}$ is also small. Referring to Fig.~\ref{fig:2}(b), the vanishing vorticity gradient ($D^{2}u$) found in the bulk of the flow domain at high values of $R_{inj}$ removes/diminishes the singular effects associated with the critical layer.

The growth of $\tau$ is probably not responsible for the spreading of the spectrum along the real axis, that we have observed in Fig.~\ref{fig:BasicSol}(b). Equation (\ref{cr-identity}) may be rewritten as:
\begin{eqnarray}
c_r &=& \frac{\langle (\alpha^2 |\phi|^2 + |D\phi|^2 ) u \rangle + \displaystyle{\frac{R_{inj}}{\alpha R}}
\langle \alpha^2 \tau - \phi_r D^3\phi_i + \phi_i D^3\phi_r  \rangle}
{I_1^2 + \alpha^2 I_0^2 } . \nonumber \\[-1ex] \label{cr-identity2}
\end{eqnarray}
The first term leads simply to values of $c_r$ in the range of $u$. The second term does contain $\alpha^2 \tau$, i.e. longitudinal gradients of the Reynolds stresses. However, note that even for the shorter wavelengths we have $\alpha \sim O(1)$, and if we consider long wavelengths, we have typically found instability only for $\alpha R \gg 1$. Thus, even for these larger $R_{inj}$, the term involving  $\alpha^2 \tau$ is likely to be insignificant.

The extension of $c_r$ beyond the usual bounds of the base flow velocity is therefore due to the 3rd derivative terms in (\ref{cr-identity2}), which cannot be bounded by the denominator. Interestingly therefore, the larger values of $c_r$, which indicate less regular eigenmodes, also lead to larger viscous dissipation, and hence more stable modes. This explains the shape of the spectrum in Fig.~\ref{fig:BasicSol}(b).

\begin{figure}
\subfloat[]{\includegraphics[scale=0.0382]{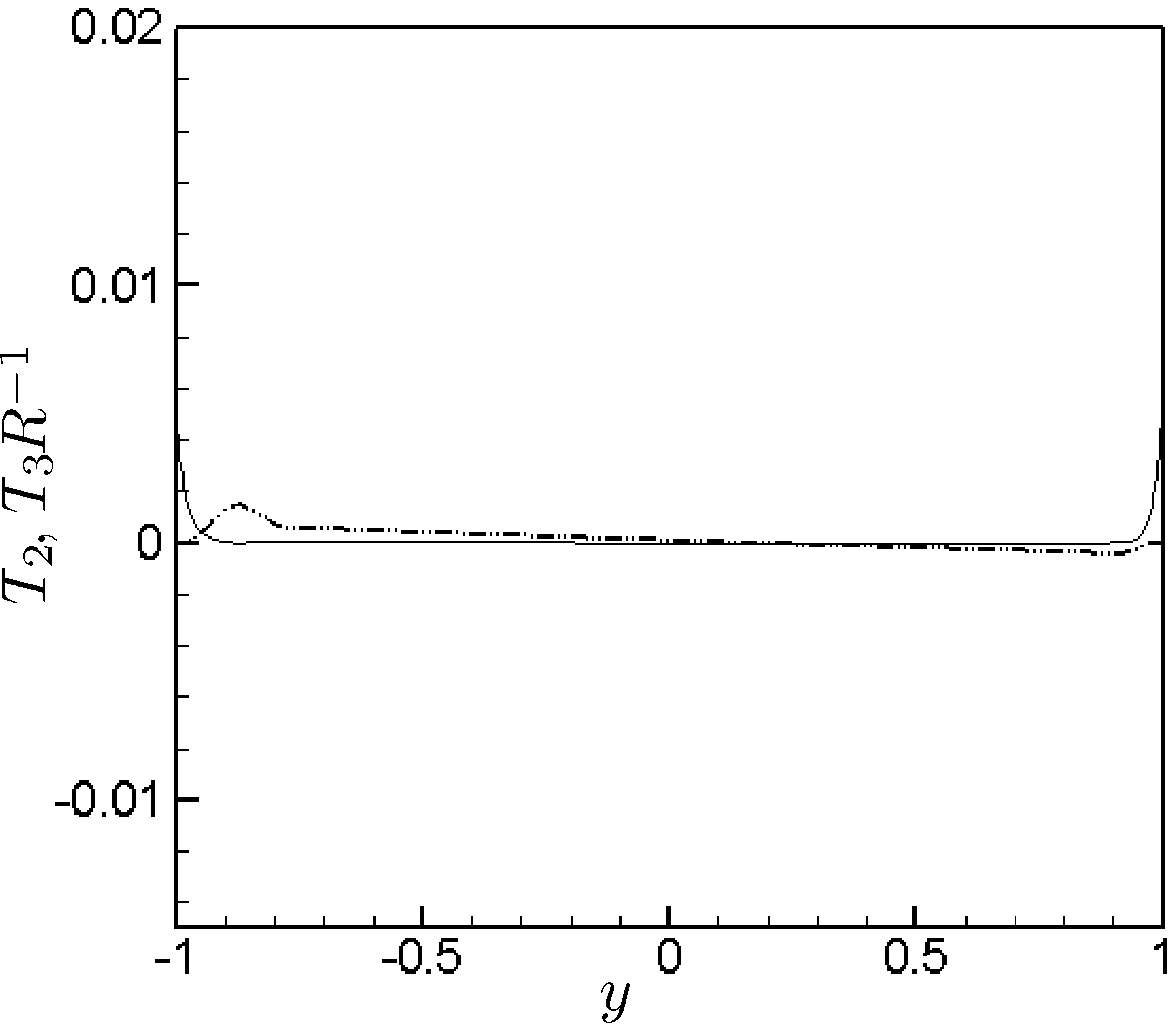}
}~\subfloat[]{\includegraphics[scale=0.037]{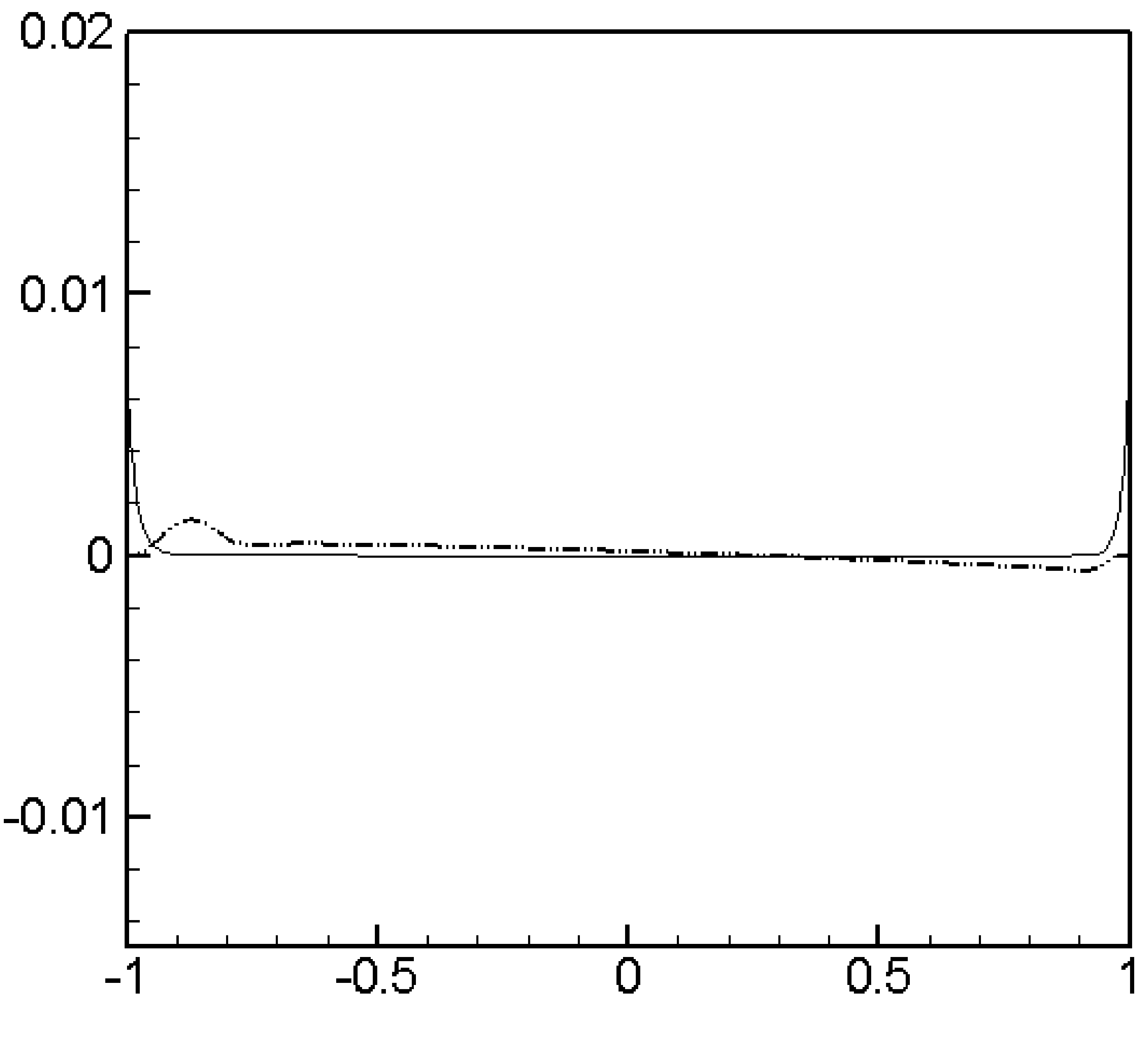}
}\subfloat[]{\includegraphics[scale=0.037]{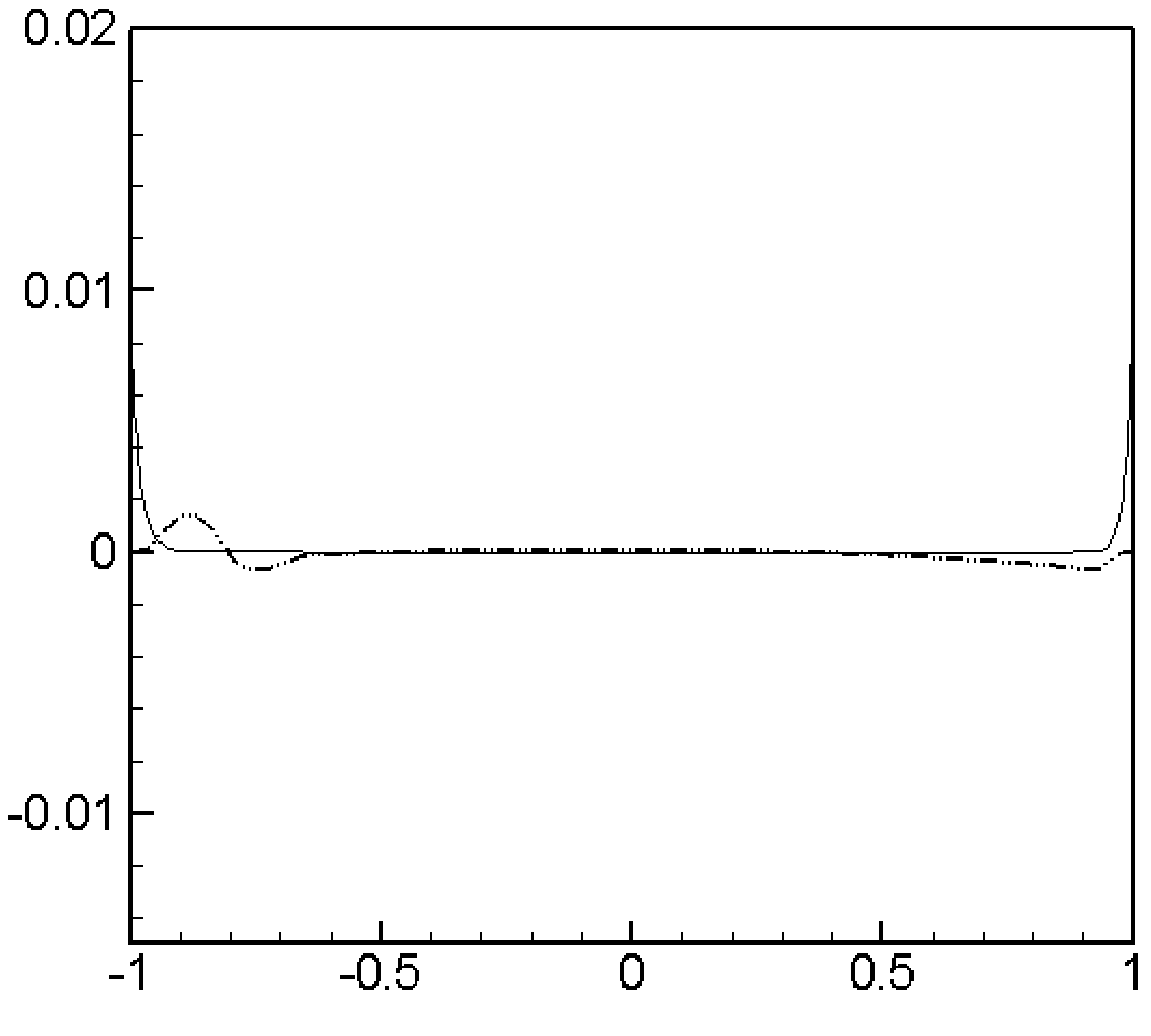}
}~~~\\\subfloat[]{\includegraphics[scale=0.037]{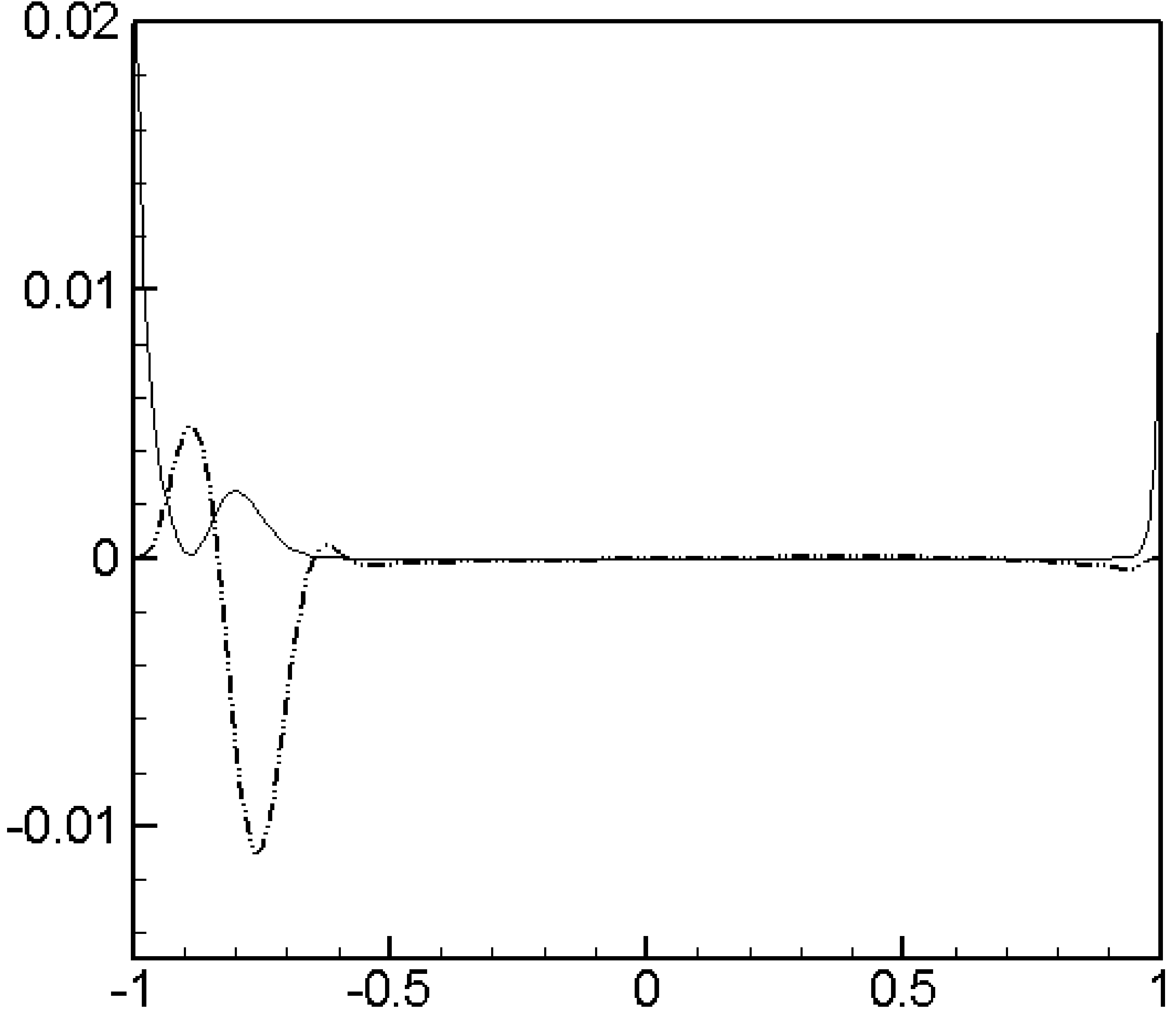}
}~\subfloat[]{\includegraphics[scale=0.037]{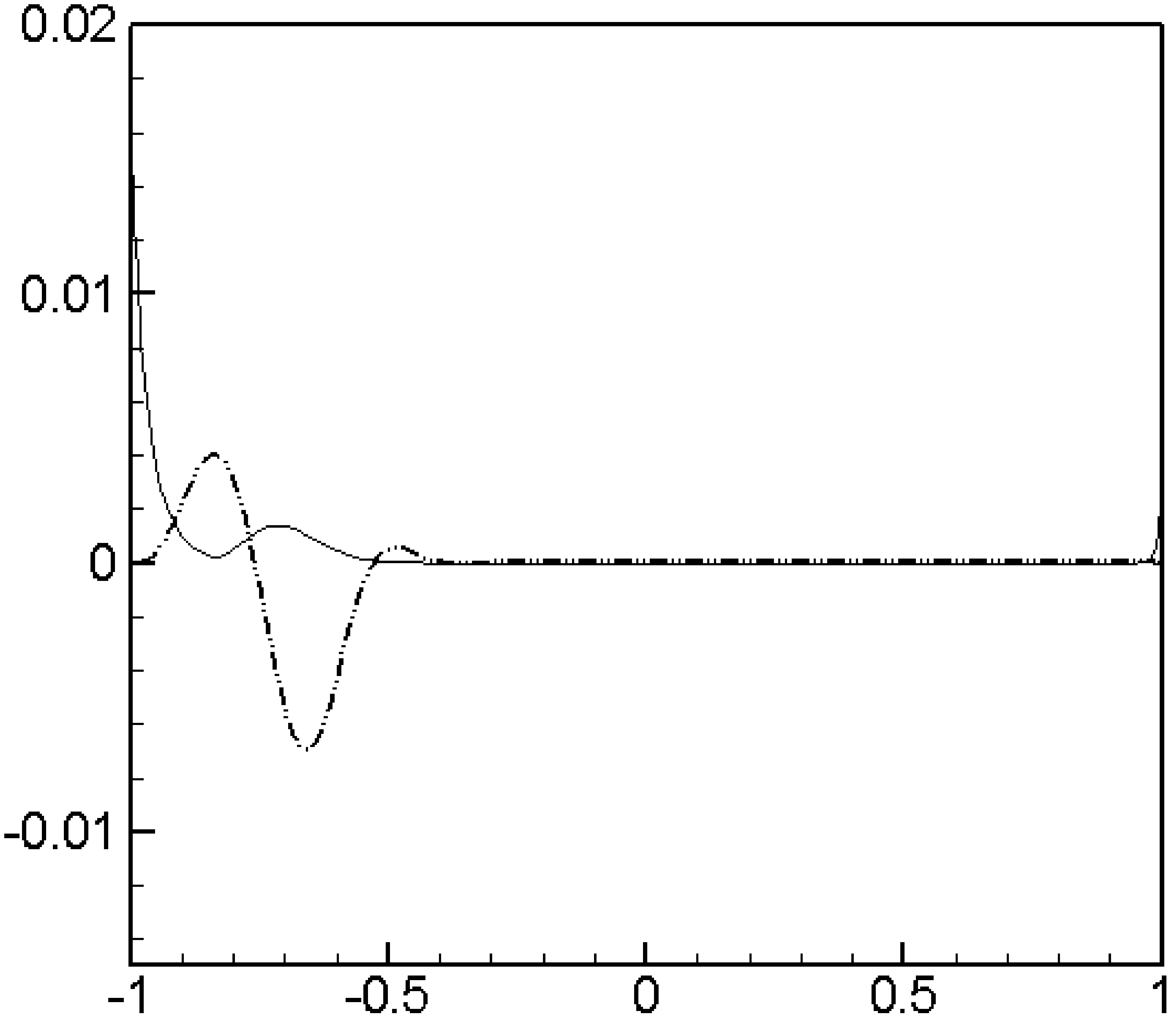}
}~\subfloat[]{\includegraphics[scale=0.037]{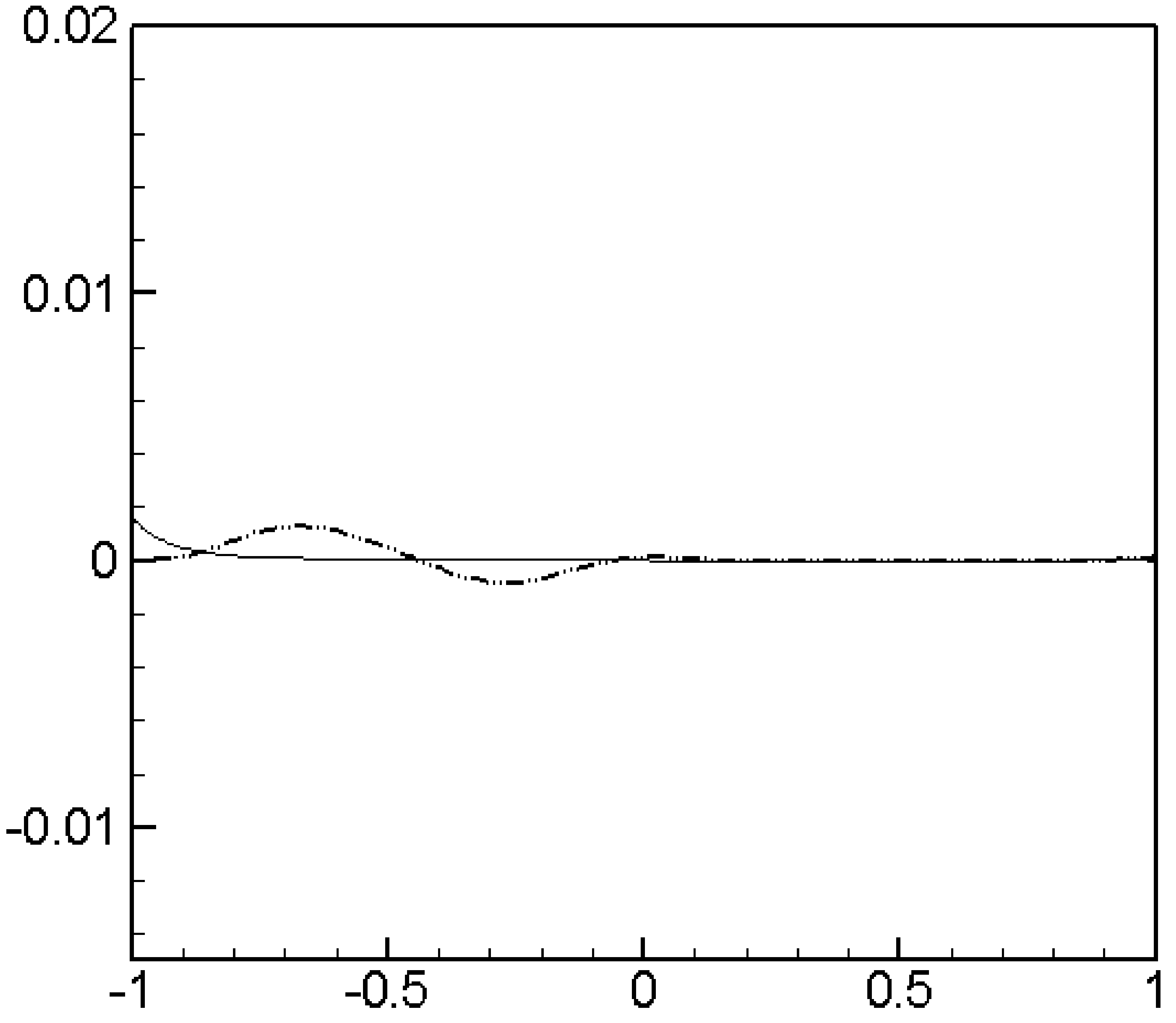}
}
\caption{Distribution of energy production~$(T_{2})$~and dissipation~$(\frac{1}{R}T_{3})$~terms
across the domain corresponding to mode {}`C' at  $R_{inj}$=~(a)~$0$,~(b)~$0.6$,~(c)~$1$,~(d)~$3$,~(e)~$10$~and~(f)~$30$.
Dash-dot-dot line represents $T_{2}$,~solid line represents~$\frac{1}{R}T_{3}$.}
\label{fig:Distribution-of-energy}
\end{figure}

\begin{figure}
~~~~~~~~~~~~~~~~~~~~~~~~~~~~\includegraphics[scale=0.06]{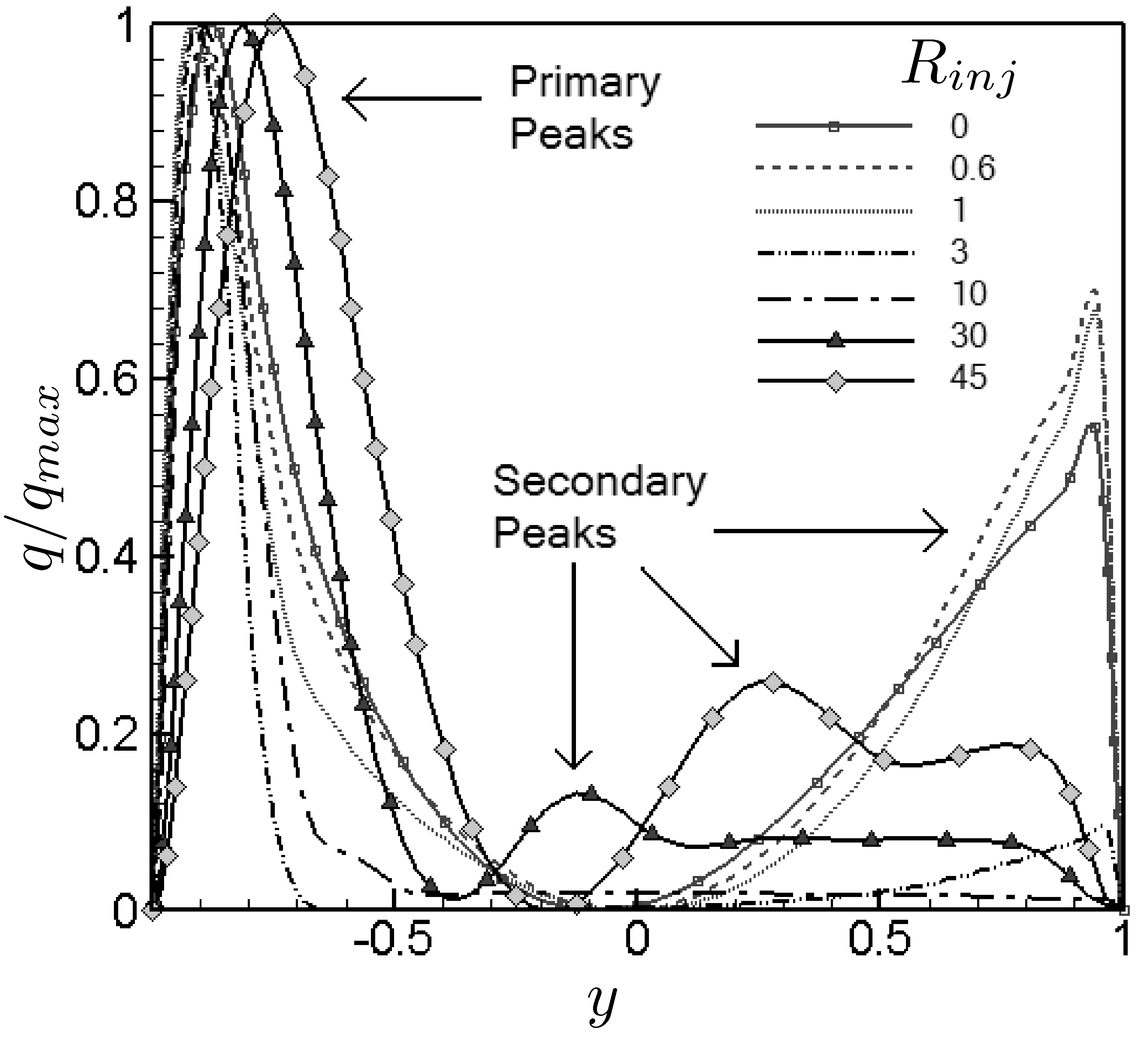}
\caption{~Non-dimensional mean perturbation kinetic energy profiles for mode~{}`C' at different $R_{inj}$. Solid lines with symbols denote the unstable modes. For each $R_{inj}$, $q$ has been scaled by its maximum value.}
\label{fig:Non-dimensional-mean-perturbation}
\end{figure}

In Fig.~\ref{fig:preferredemodes}(b), we tracked the behaviour of mode {}`C' as $R_{inj}$ increased. This mode becomes unstable for $R_{inj}\geq R_{inj,2}$, implying that it governs the transition behavior.  In Fig.~\ref{fig:Distribution-of-energy} we show the evolution of the energy balance terms for this mode as $R_{inj}$ increases from zero. This mode is stable from $R_{inj}=0.6$ to $30$. The cut-off achieved at $R_{inj}=0.6$ is primarily due to the increased viscous dissipation at both walls. This phenomenon continues until $R_{inj}\approx 3$, see Fig.~\ref{fig:Distribution-of-energy}(d). At this point, the {}`viscous hump' observed near the lower wall gets amplified. This mechanism is probably due to the resonant interaction between mode {}`C' and an (approximately) neutrally stable inviscid mode, for example mode {}`D'. Further increase in $R_{inj}$ thins out the viscous layer at the suction wall faster than that at the injection wall. Suction negates both the exchange as well as the dissipation of energy, and the viscous hump is localised within the lower half of the channel, i.e. injection side. $\frac{1}{R}T_{3}$ reduces faster than $T_{2}$ and finally the mode becomes unstable when $R_{inj}$ increases to $30$. The condition at this point is $\left\langle T_{2}\right\rangle > \frac{1}{R}\left\langle T_{3}\right\rangle $; see Fig.~\ref{fig:Distribution-of-energy}(f). It is interesting to observe that the mode becomes unstable when the viscous hump reaches the centre of the channel. Further increase of $R_{inj}$ results in a gradual reduction of $T_{2}$ and the mode becomes stable again.
The mean perturbation kinetic energy $q(y)$ distribution provides further insight into the instability mechanism. It is defined modally to be:
\begin{equation}
q=\frac{1}{4}\left(\left|D\phi\right|^{2}+\alpha^{2}\left|\phi\right|^{2}\right)
\end{equation}
Figure~\ref{fig:Non-dimensional-mean-perturbation} shows the mean perturbation kinetic energy profiles for mode `C' at different  $R_{inj}$. Each distribution of $q$ has been normalised by its maximum value.

Without any cross-flow, the amount of energy in the two halves of the domain are comparable, the suction half having $\sim43\%$ of the energy, (note $k=0.5$). Increasing cross-flow up to $R_{inj}\thickapprox R_{inj,1}=0.6$ increases the secondary peak until the cut-off is achieved. The energy in the suction half at this point is~$46.7\%$. The primary peak moves toward the lower wall but cannot reach it because of the no-slip conditions. At $Rinj>1$, the primary peak starts moving away from the suction wall. At $Rinj=3$, the mode is at its maximum stability (see Fig. \ref{fig:preferredemodes}(b)). At this point, the perturbation energy is highly localised within the lower $\frac{1}{8}$th of the channel, along with a small secondary peak at the upper quarter. Further increase of $R_{inj}$ to $10$ causes the secondary peak to vanish; the energy content in the suction half being only $\sim7.6\%$. The resonant interactions of T-S waves  result in the development of a secondary peak from the primary peak itself. During this process, the secondary peak slowly separates from the primary peak and moves in the direction of the upper wall. For $R_{inj}=30$, the perturbation reaches the channel centre and the mode becomes unstable. The amount of energy in the suction half increases to $18.1\%$. For even higher values of $R_{inj}$, for example $45$, the upper half holds $\sim32\%$ of the energy.

Thus it appears that the onset of the cut-off at $R_{inj,1}$ occurs when the secondary peak holds maximum energy. Increasing injection decays this peak until it reaches a minimum and then starts to grow out from the primary peak. The end of the cut-off regime, marked by $R_{inj}>R_{inj,2}$, occurs when the secondary peak reaches the channel centre and holds sufficient energy.

\subsection{Eventual stabilisation at $R_{inj,3}$}

We have not studied in detail the eventual stabilisation of the flow at very large $R_{inj}$ (i.e.  $R_{inj}\sim R_{inj,3}$), but we believe the energetics of this stabilisation are due to a decay in the energy production. This can be seen most clearly from the identity (\ref{ci-identity}), which is in the same form as that for any parallel shear flow, i.e.~cross-flow only influences (\ref{cr-identity}) directly. Joseph has used this expression to derive general bounds that depend on $|Du|_{max}$, and various functional inequalities; see \cite{Joseph1968,Joseph1969}. For example, we have linear stability provided that:
\begin{equation}
\alpha R|Du|_{max} < \max(\xi^{2}\pi+2^{3/2}\alpha^{3},\xi^{2}\pi+\alpha^{2}\pi)
\label{eq:7-3}
\end{equation}
where $\xi = 2.36502$ is the least eigenvalue of a vibrating rod with clamped ends at $y=\pm1$.

The condition (\ref{eq:7-3}) evidently holds for the flows we consider, but is very conservative and especially so in the limit of large $R_{inj}$. This conservatism at large $R_{inj}$ stems directly from the simplistic treatment of $Du$ in bounding the energy production term:
\[ \langle (\phi_r D\phi_i - \phi_i D\phi_r ) Du \rangle < |Du|_{max} I_0 I_1 .\]
With reference to Figs.~\ref{fig:1} \& \ref{fig:2} and to (\ref{Dumax}), we see that at large $R_{inj}$ the base velocity profile consists of a thin layer near the upper suction wall, within which $Du \sim |Du|_{max} \sim R_{inj} $, which has thickness of $O(R_{inj}^{-1})$. Away from this thin boundary layer, the velocity gradients are of size $Du \sim 2(k R_{inj})^{-1} + O(R_{inj} \ee^{-R_{inj}(1-y)})$.
Note however, that within this suction layer, we have $\phi \sim (1-y)^2$ due to the boundary conditions on the perturbation. Therefore, taking a nominal suction layer boundary at $y=y_s$, we may estimate as follows:
\begin{eqnarray}
\langle (\phi_r D\phi_i - \phi_i D\phi_r ) Du \rangle
&=&
\int_{-1}^{1-y_s} (\phi_r D\phi_i - \phi_i D\phi_r ) Du~\dd y + \int^{1}_{1-y_s} (\phi_r D\phi_i - \phi_i D\phi_r ) Du~\dd y \nonumber \\
&\leq& \frac{2}{k R_{inj}} \int_{-1}^{1-y_s} |\phi_r D\phi_i - \phi_i D\phi_r |~\dd y + O(|Du|_{max} (1-y_s)^4) \nonumber \\
&\leq& \frac{2}{k R_{inj}} I_0 I_1 + O(R_{inj}^{-3})
\end{eqnarray}
Following \cite{Joseph1969}, this leads directly to the bound
\begin{equation}
\frac{2 \alpha R }{k R_{inj}} \lesssim \max(\xi^{2}\pi+2^{3/2}\alpha^{3},\xi^{2}\pi+\alpha^{2}\pi),
\label{eq:7-4}
\end{equation}
sufficient for linear stability at large $R_{inj}$, (with asymptotically $k R_{inj} \gtrsim 4$ required).
In other words, at large $R_{inj}$, the energy production $T_2$ will decay like $(kR_{inj})^{-1}$ at leading order, so that the viscous dissipation need only be of this order to stabilise the flow.

\section{Summary}
\label{sec:discuss}

\begin{figure}
~~~~~~~~~~~~~~~~~~~~~~~~~~~~\includegraphics[scale=0.06]{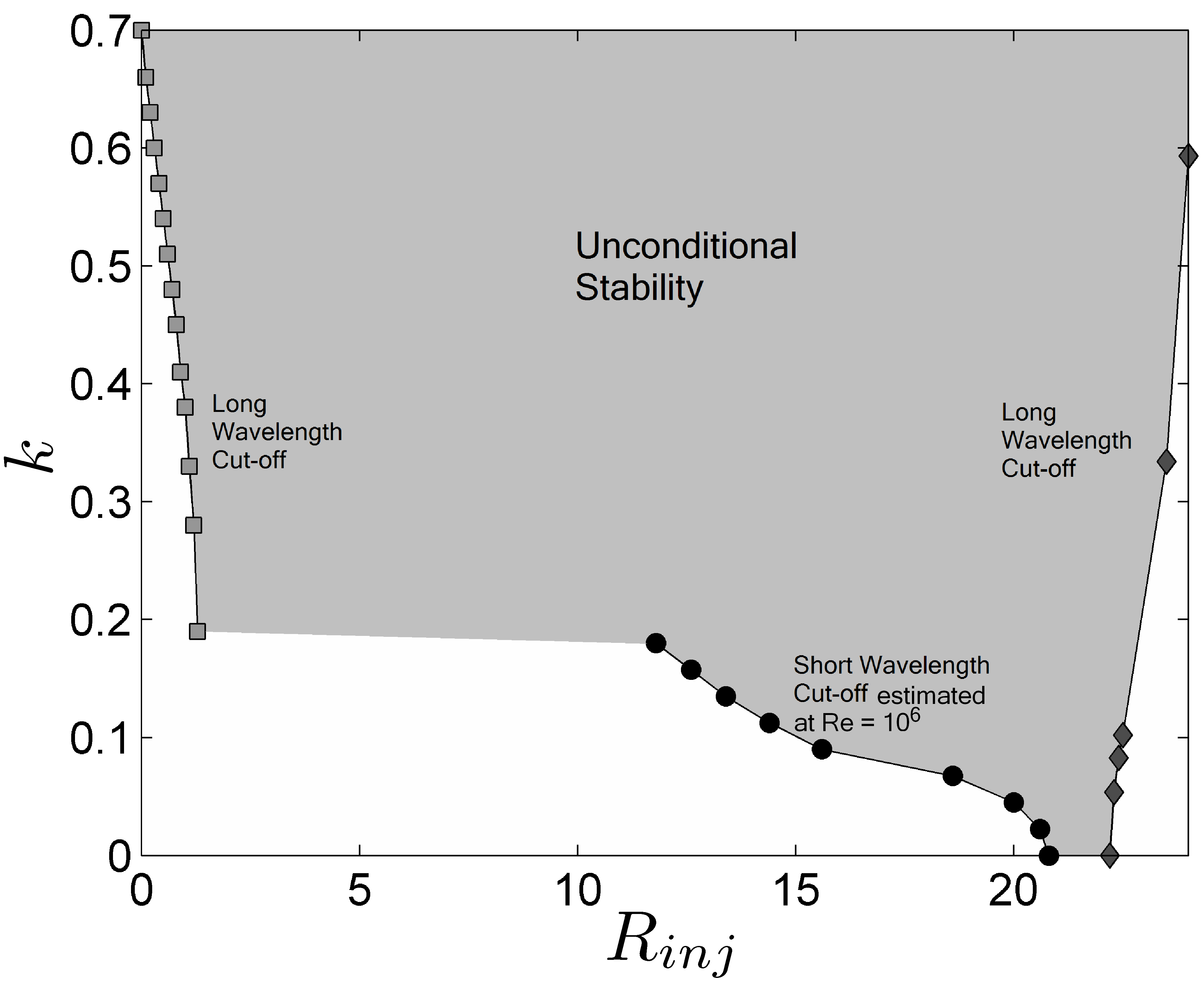}
\caption{~Variation of $k$ with $R_{inj}$. The filled $\square$ symbols show the long wavelength cut-off achieved for $0.7 \geq k \geq 0.19$. The filled $\circ$ symbols show the shorter wavelength cut-off for  $0.19 > k \geq 0$ evaluated numerically for $R=10^{6}$. The filled $\diamond$ symbols imply the second long wavelength cut-off. The shaded region depicts the entire zone of unconditional linear stability.}
\label{fig:summary}
\end{figure}

To summarise, we have presented a detailed analysis of linear stability and instability in the $(R_{inj},k)$-plane, for PCP flow with cross-flow. The most complete analysis concerns the important range of low $R_{inj}$ and modest $k$. In this range we have demonstrated that the stabilisation mechanism, due to either injection or wall motion, is essentially the same. Long wavelengths dominate. Skewing of the velocity profile shifts the critical layer and at the same time the energy production is diminished until viscous dissipation dominates at cut-off. In Fig.~\ref{fig:Cut-off-Velocity}, we have also shown an interesting quantitative analogy with the cut-off behavior of ACP flows; see \cite{Sadeghi1991}.

This lower range of $R_{inj}$ and modest $k$ is probably that which is most important practically. Essentially, this range allows one to compensate cross-flow by wall-motion and vice-versa, achieving unconditional linear stability via either mechanism. With reference to Fig.~\ref{fig:1}, it is  the range of $R_{inj}$ in which the cross-flow and wall motion are modifications of a base Poiseuille flow. Due to the scaling, the peak velocity is always 1, but at larger $R_{inj}$ with modest $k$ the Poiseuille component is completely dominated by cross-flow and wall motion.

Globally, the cut-off regimes in the $(R_{inj},k)$-plane are as illustrated in Fig.~\ref{fig:summary}. The shaded area shows the region of unconditional linear stability. In the intermediate range of approximately $1.3 \leq R_{inj} \leq 20.8$ values of $k \gtrsim 0.19$ are dominated by long wavelengths and are stable. Below this value, we are able to compute numerical cut-off curves for fixed $R$. With the limits of our computations, we cannot determine if these cut-off curves asymptote to an unconditional cut-off curve as $R \to \infty$.

There appears to be a short band of unconditional linear stability for all computed values of $k$ around approximately $20.8 \leq R_{inj} \leq 22$, before the destabilisation occurs at larger $R_{inj} = R_{inj,2}$. Since this band can make PP flow unconditionally stable, it could be effectively used in applications where wall motion is not feasible, e.g.~cross-flow filtration, medical dialysis. From the practical perspective, it is worth noting that the transition across $R_{inj,2}$, is from unconditional stability to critical values of $R$ which are relatively modest (e.g.~in the range $10^3-10^4$) just a short distance beyond $R_{inj,2}$. Assuming that the PP flow is linearly unstable, this means stabilisation can be achieved with cross-flow velocities of the order of 1\% of the mean axial flow velocity. 

This destabilisation at $R_{inj,2}$ is again a long wavelength mechanism, which we have analysed using the long wavelength approximation of \cite{CowleySmith1985}. A possible cause of this instability has been found to be resonant interactions of the T-S waves. Study of the linear  energetics of the upper limit, $R_{inj,2}$, has shown that neither viscous dissipation, nor the involvement of a critical layer are significant. Rather, the balance of energy production and dissipation within $T_2$ keeps the mode neutrally stable. Energy analysis of the preferred mode {}`C' has revealed that the precursor of the transition to instability from unconditional stability is the amplification of disturbances near the injection wall. The mean perturbation kinetic energy has also been analyzed. It has shown that the lower limit occurs when the secondary peak holds maximum energy. Increasing injection decreases the secondary peak until it reaches a minimum and then it starts to grow from the primary peak. When the secondary peak reaches the channel centre and holds a sufficient amount of energy, the unconditional stability mechanism breaks down.

The final stabilisation occurring at large $R_{inj}\ge R_{inj,3}$ has been analyzed using linear energy bounds. By careful treatment of the energy production term, we are able to show that the energy production terms decreases asymptotically like $R_{inj}^{-1}$ as $R_{inj} \to \infty$. We believe that this mechanism leads to the eventual domination of the viscous dissipation at large enough $R_{inj}$.

In terms of the spatial structure of the perturbations, we note that the stabilisation at small and moderate $R_{inj}$ are both long wavelength phenomena for which the approximation of Cowley \& Smith has been shown effective. Implicitly therefore, the critical wavenumbers scale like $R^{-1}$ in these limits. For the shorter wavelength instabilities we have not analysed the asymptotic behaviour of the wavenumber with $R$. A more detailed look at the spatial structure of certain eigenmodes has been presented in Fig.~\ref{fig:newfig}. This shows a skewing of the streamline recirculatory regimes towards the lower wall for long wavelengths as $R_{inj}$ is increased, and towards the upper wall at shorter wavelengths as $R_{inj}$ is increased.

\appendix

\end{document}